\documentclass[preprint,12pt]{common/elsarticle}

\usepackage{amssymb}
\usepackage{amsmath,amsthm,amsfonts,amscd,amsmath,amsfonts} % Some packages to write mathematics.
\usepackage{upgreek}
\usepackage{mdframed}
\usepackage{multicol}
\usepackage{graphicx}
\usepackage{nomencl}
\usepackage[nottoc]{tocbibind}
\usepackage{setspace}
\usepackage[caption=false]{subfig}  % don't load caption or captcont
\usepackage{verbatim}   	% Allows quoting source with commands.
\usepackage{makeidx}    	% Package to make an index.
\usepackage{color}
\usepackage{xcolor}
\usepackage{enumerate}
\usepackage{url}	
\usepackage{tcolorbox}
\usepackage{xcolor}
\usepackage{multirow}
\usepackage[linesnumbered,ruled,vlined]{algorithm2e}
\usepackage{ulem}
\usepackage{multirow}
\usepackage{bbm, dsfont}
\usepackage[percent]{overpic}
\newcommand{\blue}[1]{\textcolor{black}{#1}}

\newcommand{\thetab}{\boldsymbol\theta}
\newcommand{\bs}{\boldsymbol}

\journal{Computer Methods in Applied Mechanics and Engineering}

\begin{document}

%========================================================================
% First Page
%========================================================================
\begin{frontmatter}



\title{A Framework for Strategic Discovery of Credible Neural Network Surrogate Models under Uncertainty}


\author[inst1]{Pratyush Kumar Singh}

\affiliation[inst1]{organization={Department of Mechanical and Aerospace Engineering, \\University at Buffalo},%Department and Organization
            %addressline={Address One}, 
            city={Buffalo},
            %postcode={00000}, 
            state={NY},
            country={USA}}

\author[inst2]{Kathryn A. Farrell-Maupin}

\affiliation[inst2]{organization={Sandia National Laboratories},%Department and Organization
            %addressline={Address One}, 
            city={Albuquerque},
            %postcode={00000}, 
            state={NM},
            country={USA}}

\author[inst1]{Danial Faghihi\corref{cor1}}

\cortext[cor1]{Corresponding Author, \texttt{danialfa@buffalo.edu} (D. Faghihi)}            

%========================================================================
% Abstract
%========================================================================
\begin{abstract}

The widespread integration of deep neural networks in developing data-driven surrogate models for high-fidelity simulations of complex physical systems highlights the critical necessity for robust uncertainty quantification techniques and credibility assessment methodologies, ensuring the reliable deployment of surrogate models in consequential decision-making.
This study presents the Occam Plausibility Algorithm for surrogate models (OPAL-surrogate), providing a systematic framework to uncover predictive neural network-based surrogate models within the large space of potential models, including various neural network classes and choices of architecture and hyperparameters.
The framework is grounded in hierarchical Bayesian inferences and employs model validation tests to evaluate the credibility and prediction reliability of the surrogate models under uncertainty. Leveraging these principles, OPAL-surrogate introduces a systematic and efficient strategy for balancing the trade-off between model complexity, accuracy, and prediction uncertainty.
The effectiveness of OPAL-surrogate is demonstrated through two modeling problems, including the deformation of porous materials for building insulation and turbulent combustion flow for ablation of solid fuels within hybrid rocket motors.

\end{abstract}


\end{frontmatter}

%%\linenumbers

%%\tableofcontents

%\newpage
%========================================================================
% Document body
%========================================================================

\section{Introduction}
\label{sec:introduction}

%------------------------------------------------------------------------------------------------
% 1. Identify the problem and say why it is important
%------------------------------------------------------------------------------------------------
\noindent
The remarkable advancements in scientific machine learning (SciML) techniques, specifically deep neural networks, ignited an extraordinary revolution in the creation of data-driven surrogate models. These models aim to approximate solutions of high-fidelity physics-based scientific simulations and \blue{allowing} computational predictions at significantly lower costs.
Going beyond the enabler of once-computationally prohibitive outer-loop challenges such as uncertainty quantification, e.g., \cite{TRIPATHY2018565, ZHU2018415, Georgalis23, scarabosio2019goal}, Bayesian inference, e.g., \cite{li2023surrogate, cao2023bayesian}, design under uncertainty \cite{luo2023efficient, tan2024scalable}, digital twins, e.g., \cite{chattopadhyay2023oceannet,he2023hybrid, kapteyn2021probabilistic, mowlavi2023optimal, zohdi2022digital}, and optimal experimental design, e.g., \cite{wu2023large, stuckner2021optimal} for complex physical systems, neural network-based surrogate models hold transformative potential in reshaping the formulation and resolution of scientific problems across diverse domains of science, engineering, and medicine.

%------------------------------------------------------------------------------------------------
% 2. The current state of the field and current limitation(s) in the field
%------------------------------------------------------------------------------------------------

Despite notable advancements, applying machine learning techniques -- originally designed for large data regimes in domains like image processing, computer vision, and natural language processing -- encounters significant challenges when directly utilized to construct and establish trust in surrogate models. 
These obstacles emerge from the inherent spatiotemporal sparsity, limitation, and incompleteness of the scientific data extracted from high-fidelity physical simulations.
This intrinsic uncertainty presents substantial challenges to the credibility and prediction reliability of neural network-based surrogate models, as highlighted in works such as \cite{arzani2023interpreting, yuan2022towards, samek2021explaining, zhong2022explainable}.
The importance of verification, validation, and uncertainty quantification (VVUQ) for physics-based models against experimental measurements is well-established \cite{OdenMoserGhattas2010I,Babuska2008, odenbabuska2017,tan2022toward}. However, as SciML becomes increasingly integrated and deployed into high-consequence decision-making for complex physical systems, there arises a critical need for even more robust UQ techniques and rigorous methodologies to assess the credibility of neural network-based surrogate models.
In particular, the first shortcoming lies in the common training approach, based on maximum likelihood parameter estimation, which limits neural networks' robustness against data uncertainty (i.e., adversarial attack), resulting in overfitting and overconfident predictions.
The second challenge stems from the existing validation methodologies for neural networks, such as train-and-test approaches that rely on empirical performance assessments with asymptotic guarantees in large data \cite{yaseen2023quantification, twomey1997validation, arzani2023interpreting}. Moreover, the interpretability and explainability approaches employed by the machine learning community to build trust in neural network models often resort to heuristic and problem-dependent strategies \cite{arzani2023interpreting, samek2021explaining, zhong2022explainable}.
The third and perhaps most pivotal challenge lies in the uncertainty associated with selecting the surrogate model itself. Achieving a delicate balance in the model's complexity is crucial, as overly simplistic models may compromise predictive ability, while excessively complex ones are prone to overfitting the training data, resulting in poor generalization, especially when parameters are estimated via maximum likelihood methods \cite{mackay1995probable, neal2012bayesian, tan2022toward}. Consequently, determining the ``best" neural network model becomes challenging in the absence of predefined rules for network architecture and associated hyperparameters.
Current trial-and-error architecture selection approaches based on testing data performance are time-consuming and resource-intensive and may not effectively enhance the accuracy and reliability of the surrogate models.
Harnessing recent advancements in architecture optimization algorithms and software tools \cite{elsken2019neural, liu2021survey, wang2018hybrid, ghosh2022designing, mendoza2016towards, hutter2019automated, optuna_2019}, when adapted to align the objectives of surrogate modeling, holds the potential to address this challenge effectively.

%

% Bayesian NN
The Bayesian framework is the cornerstone for successful VVUQ methods, allowing the quantification of uncertainty in model parameters and choice of the model itself in small data regimes, e.g., \cite{faghihi2018fatigue, tan2021, tan2022predictive, oden2013selection, cao2023bayesian, jha2020bayesian, liang2023bayesian, lima2021bayesian, prudencio2015} and serving as a robust foundation for validating model predictions, e.g., \cite{OdenMoserGhattas2010I, Babuska2008, tan2022toward, odenbabuska2017, farrell2015jcp, oden2016reviewtumor}.
The seminal contributions of Mackay \textit{et al.} \cite{mackay1992bayesian, mackay1995probable} laid the groundwork for the adoption of the Bayesian framework in inferring neural network parameters and hyperparameters, inspiring subsequent developments \cite{bishop1995neural, neal2012bayesian}. Despite these early contributions, Bayesian neural networks (BayesNN) only recently gained recognition in the SciML community, e.g., \cite{psaros2023uncertainty, YANG2021109913, olivier2021bayesian, ZHU2018415, LINKA2022115346, MORA2023116207, kontolati2023influence, meng2021multi, MORA2023116207}. 
The delayed adoption can be attributed to the incomplete understanding of UQ methods for neural networks as well as the computational complexities associated with high-dimensional parameter spaces in these models.
BayesNN provides remarkable advantages, including alleviating overfitting, preventing overconfident parameter estimation in small and uncertain datasets, and the ability to quantify prediction uncertainty.
This study emphasizes an additional benefit of BayesNN in surrogate modeling by relaxing rigid constraints on model complexity. This flexibility enables the retention of a sufficient number of parameters, which is crucial for capturing the underlying multiscale structure inherent in physics-based simulations.
Despite these merits, BayesNN surrogate modeling faces formidable challenges, particularly in (i) selecting a specific network architecture and hyperparameters, as it is challenging to assert that the chosen models align with prior beliefs about the problem, and (ii) developing methodologies to rigorously assess the credibility and prediction reliability of the surrogate models under uncertainty.

%------------------------------------------------------------------------------------------------
% 3. How do you plan on filling this hole with a novel approach
%------------------------------------------------------------------------------------------------

This contribution introduces the Occam Plausibility Algorithm for surrogate models (OPAL-surrogate), a systematic framework designed to discover predictive neural network-based surrogate models for high-fidelity physical simulations. 
The name is inspired by the principle of Occam's Razor, advocating the preference for simpler models over unnecessary complex ones, guided by the notion of model plausibility as a basis for its effectiveness in explaining the given data.
OPAL-surrogate is grounded in hierarchical Bayesian inferences, enabling a systematic determination of the probability distributions of the network parameters and hyperparameters and measures for comparing various neural network models. Moreover, it adheres to the principles of Bayesian model validation to assess the credibility and prediction reliability of the surrogate model. 
Leveraging these methodologies, OPAL-surrogate presents a strategy to adaptively adjust model complexity, utilizing a combination of bottom-up and top-down approaches until predefined validation criteria are met.
Consequently, within the wide space of potential BayesNN models involving choices of architecture and hyperparameters, OPAL-surrogate identifies the ``best" predictive surrogate model by balancing the trade-off between model complexity and prediction uncertainty.
The effectiveness of OPAL-surrogate is demonstrated via two modeling problems in solid mechanics and computational fluid dynamics, identifying credible surrogate models for reliable prediction of quantities of interest (QoIs).

%------------------------------------------------------------------------------------------------
% 4. Summary of each section
%------------------------------------------------------------------------------------------------

Following this introduction, Section \ref{sec:bnn} provides a comprehensive overview of Bayesian learning for neural networks along with an efficient and scalable solution algorithm. Section \ref{sec:hierarchical} delves into the definition of the BayesNN model and hierarchical inference of various classes of neural network parameters. The steps involved in the OPAL-surrogate for discovering \blue{predictive} BayesNN surrogate models are outlined in Section \ref{sec:opal}. Section \ref{sec:results} shows numerical examples, including the identification of BayesNN surrogate models for the elastic deformation of porous materials with random microstructures and the direct numerical simulation of turbulent combustion flow for shear-induced ablation of solid fuels within hybrid rocket motors. Concluding remarks can be found in Section \ref{sec:conclusions}.

%++++++++++++++++++++++++++++++++++++++++++++++++++++++++++++++++++++++++
%++++++++++++++++++++++++++++++++++++++++++++++++++++++++++++++++++++++++
\section{Neural Networks Learning as Probabilistic Inference}\label{sec:bnn}
\noindent
% NN
A neural network is a nonlinear map $\bs g: \mathbb{R}^{d_i} \rightarrow \mathbb{R}^{d_o}$ from the inputs $\bs x\in \mathbb{R}^{d_i}$ to the outputs $\bs{u}_{\bs \theta} \in \mathbb{R}^{d_o}$, representing a continuously parameterized function.
The functional form of a feed-forward neural network with $D$ number of layers is expressed as follows,
\begin{equation}\label{eq:NNfunc}
    \bs{u}_{\bs \theta}(\bs x) = 
f^{(D)} (\bs w^{(D)} \cdots f^{(2)}(\bs w^{(2)} f^{(1)} (\bs w^{(1)} \bs x + \bs b^{(1)}) + \bs b^{(2)}) \cdots + \bs b^{(D)}).
\end{equation}
Here, $\bs w^{(\ell)}$ represents the weight vector connecting layer ${\ell}-1$ to layer ${\ell}$, 
$\bs b^{(\ell)}$ denotes the biases in layer ${\ell}$, and $f^{(\ell)}$ denotes activation functions applied element-wise. 
The set of weights and biases collectively forms the network parameters $\bs \theta \in \mathbb{R}^P$, and common activation functions include Hyperbolic Tangent (\textit{Tanh}), Rectified Linear Unit (\textit{ReLU}), Leaky Rectified Linear Unit (\textit{Leaky ReLU}), and Logistic (\textit{Sigmoid}) functions.
The output in each layer $\ell$ can be represented as
\begin{equation}\label{eq:NNfunc2}
\bs a^{(\ell)} = \bs w^{(\ell)} \bs z^{(\ell-1)} + \bs b^{(\ell)};
\quad 
\bs z^{(\ell)} = f^{(\ell)}(\bs a^{(\ell)});
\quad
1 \leq \ell < D,
\end{equation}
where $\bs a^{(\ell)}$ is the pre-activation and 
$\bs z^{(\ell)}$ denotes the activation values.
% Bayesian NN
Standard training based on the maximum likelihood estimate often leads to overconfidence in parameter values and ignoring inherent uncertainty in model predictions. 
\blue{
However, BayesNN accounts for this uncertainty by considering probability distribution functions (PDFs) over the parameters, inferred from data $\bs D = \{ \bs x_{\bs D}^j, \bs u_{\bs D}^j \}_{j=1}^{N_D}$ using Bayes' theorem},
\begin{equation}\label{eq:bayes}
 \pi_{post}(\boldsymbol{\theta}|\bs {D}) = 
 \frac{\pi_{like}(\bs {D}|\boldsymbol{\theta})
 \pi_{pr}(\boldsymbol{\theta})}
 {\pi_{evid}(\bs {D})}.
\end{equation}
Here, $ \pi_{post}(\boldsymbol{\theta}|\bs {D}) $ represents the posterior PDF, updating the prior PDF $ \pi_{pr}(\boldsymbol{\theta})$ given observational data and the evidence PDF $\pi_{evid}(\bs {D})$ serves as the normalization factor.
Additionally, the likelihood PDF $\pi_{like}(\bs {D}|\boldsymbol{\theta})$ is derived from a noise model depicting the discrepancy between the data and neural network output. 
Adhering to the maximum entropy principle with constraints on the mean and variance of parameters, we adopt a Gaussian prior $\pi_{pr}(\boldsymbol{\theta}) = \mathcal{N}(\bar{\boldsymbol{\theta}}, \bs \Gamma_{pr})$. Unless derived from pre-training or historical datasets, and without loss of generality, we assume $\bar{\boldsymbol{\theta}} = \bs 0$ and prior covariance is $\bs \Gamma_{pr} = (\sigma^{\text{pr}})^2\mathbf{I}$ where $\sigma^{\text{pr}}$ is the prior hyper-parameter. Additionally, we consider an additive noise model $\bs u_D = \bs{u}_{\bs \theta}(\bs x_{\bs D}) + \bs \eta$
where $\bs \eta$ is the total error, including both data uncertainty and network inadequacy in representing the data.
Assuming a zero-mean Gaussian distribution for the total error, $\boldsymbol{\eta} \sim \mathcal{N} \left(\mathbf{0}, (\sigma^{\text{noise}})^2\mathbf{I} \right)$ 
and considering independent and identically distributed (iid) data, the likelihood PDF is expressed as,
\begin{equation}\label{eq:sigma_noise}
    \pi_{\rm like} (\boldsymbol{D}|\boldsymbol{\theta}) = \prod_{j=1}^{N_D} \frac{1}{\sqrt{2\pi} \sigma^{\text{noise}}}
\exp \left[-\frac{(\boldsymbol{u}_{\bs D}^j - \boldsymbol{u}_{\boldsymbol{\theta}}(\boldsymbol{x}_{\bs D}^j))^2 }{2 (\sigma^{\text{noise}})^2} \right ],
\end{equation}
where $\sigma^{\text{noise}}$ is called noise hyper-parameter.
A point estimate of the most probable value for the parameter, considering both the observed data and the prior PDFs, is referred to as the Maximum A Posteriori (MAP) estimate such that,
\begin{equation}\label{eq:map}
    \boldsymbol{\theta}_{MAP} = \underset{\boldsymbol{\theta} \in \Theta}{\text{argmax}} \; 
    \pi_{post}(\boldsymbol{\theta}|\bs {D}).
\end{equation}
%

%++++++++++++++++++++++++++++++++++++++++++++++++++++++++++++++++++++++++
\subsection{Bayesian solution: Laplace approximation to the posterior}\label{sec:LA}
\noindent
Bayesian inference for neural networks poses substantial computational challenges, primarily due to the high-dimensional parameter space (number of weights) and the potential complexity of the posterior distribution geometry. 
To illustrate the proposed framework for BayesNN surrogate modeling in this work, we leverage an efficient Bayesian solution based on Laplace approximation (LA) to the posterior, offering scalability with respect to the parameter dimensions. 
\blue{The utilization of LA within BayesNN draws from seminal works of
\cite{mackay1992bayesian, bishop1995neural, buntine1991bayesian}, and recently, its computational efficiency has been improved to accommodate larger networks 
\cite{ritter2018scalable, deng2022accelerated, immer2021improving, immer2021scalable, humt2019laplace}.}
This approach involves approximating the posterior distribution by linearizing it around its dominant mode (the MAP estimate in \eqref{eq:map}), \blue{followed by the application of Laplace's method to evaluate the resulting normalization factor, yielding,}
\begin{equation}\label{eq:post_la}
\pi_{post} (\bs \theta | \bs D) 
\approx
\frac{1}{ \sqrt{(2\pi)^{P} \det (\bs H)}}
\exp \left[ 
-\frac{1}{2} (\bs \theta - \bs \theta_{MAP})^T \bs H (\bs \theta - \bs \theta_{MAP})
\right].
\end{equation}
Here, $\bs H = -\nabla^2_{\bs \theta \bs \theta} \ln \pi_{post} (\bs \theta | \bs D) |_{\bs \theta = \bs \theta_{MAP}} $ is the positive (semi-)definite Hessian of the negative log-posterior, describing its local curvature at the MAP point,
and
$P$ is the number of parameters $\bs \theta$.
The relation \eqref{eq:post_la} 
demonstrates that the posterior PDF \blue{in approximated} by a multivariate normal distribution $\mathcal{N}(\bs \theta_{MAP}, \bs \Gamma_{post})$. The mean of this distribution is the MAP point $\bs \theta_{MAP}$, and the posterior covariance matrix $\bs \Gamma_{post} = \bs H^{-1}$ is given by the inverse of the Hessian matrix evaluated at $\bs \theta_{MAP}$.
%-- MAP
The MAP estimate for the parameters can be computed by minimizing the negative log-posterior, which can be expressed equivalently through the likelihood and prior PDFs
as,
\begin{equation}\label{eq:map_opt}
\bs \theta_{MAP} = 
\underset{\bs \theta}{\text{argmin}} \;
\{- \ln \pi_{post} (\bs \theta | \bs D) \}= 
\underset{\bs \theta}{\text{argmin}} \;
\{- \ln \pi_{like} (\bs D| \bs \theta) - \ln\pi_{pr} (\bs \theta)\}.
\end{equation}
The solution to the above optimization problem is analogous to deterministic neural network training with an updated loss function. 
%-- Posterior covariance 
Furthermore, the posterior covariance can be expressed as,
\begin{equation}\label{eq:post_cov}
\bs \Gamma_{post} = \bs H^{-1} = 
\left( \bs H_{lnlike} + \bs \Gamma_{pr}^{-1} \right)^{-1},
\end{equation}
where
$\bs \Gamma_{pr}$ is the prior covariances, and 
$ \bs H_{lnlike} = -\nabla^2_{\bs \theta \bs \theta} \left.\ln \pi_{like} ( \bs D | \bs \theta)  \right\vert_{\bs \theta = \bs \theta_{MAP}} $ denotes the Hessian of log-likelihood evaluated at MAP point. 
To this end, the efficient computation of the log-likelihood Hessian is crucial for the calculation of the posterior covariance, as the prior terms are typically straightforward.

%+++++++++++++++++
\subsection{Scalable algorithm via Kronecker-factored Laplace approximation}\label{sec:kfac}
\noindent
The computation of the posterior covariance encounters a significant challenge due to the large size of the $\bs H_{lnlike} \in \mathbb{R}^{P \times P}$ matrix resulting from the high-dimensional parameter space in neural networks. This matrix is often non-diagonal, not guaranteed to be positive semi-definite, and may lack sparsity.
To develop a scalable solution algorithm for Bayesian inference, we adopt the Kronecker factored representation of the Hessian proposed by \cite{martens2015optimizing, botev2017practical}. This approach leverages the structure of neural network parameters, enabling explicit calculation, inversion, and storage of this matrix in a tractable manner.

Due to the intractability of the $\bs H_{lnlike}$, it is common practice to resort to its generalized Gauss-Newton approximation. The resulting symmetric positive and semi-definite approximated Hessian matrix is given by,
\begin{equation}\label{eq:ggn}
\bs H_{lnlike} \approx
\bs J^T \bs L \bs J,
\end{equation}
where the matrix $\bs J \in \mathbb{R}^{N_D d_o \times P}$ is formed by concatenating $N_D$ Jacobian matrices of the network with respect to the parameters 
$\bs j (\bs x_{\bs D}^j) = \nabla_{\bs \theta} \bs{u}_{\bs \theta}(\bs x_{\bs D}^j) |_{\bs \theta = \bs \theta_{MAP}}$ $\in \mathbb{R}^{d_o \times P}$ for all $\{ \bs x_{\bs D}^j \}_{j=1}^{N_D}$ in the data set,
and
$\bs L \in \mathbb{R}^{N_D d_o \times N_D d_o}$ is a block-diagonal matrix with blocks 
$\bs l (\bs u_{\bs D}^j, \bs u_{\bs \theta}(\bs x_{\bs D}^j)) = -\nabla^2_{\bs g \bs g} \ln \pi_{like} ( D^{\blue{j}} | \bs \theta) |_{\bs \theta = \bs \theta_{MAP}} \in \mathbb{R}^{d_o \times d_o}$ 
\blue{, where the likelihood of the $j$-th data is defined as $\pi_{\rm like} ({D}^j|\boldsymbol{\theta}) =$ $\frac{1}{\sqrt{2\pi} \sigma^{\text{noise}}}$
$\exp \left[-\frac{(\boldsymbol{u}_{\bs D}^j - \boldsymbol{u}_{\boldsymbol{\theta}}(\boldsymbol{x}_{\bs D}^j))^2 }{2 (\sigma^{\text{noise}})^2} \right ]$.
}
Despite the approximation in \eqref{eq:ggn}, the log-likelihood Hessian remains a full matrix that needs to be inverted. To address the associated computational challenge, we employ a block-diagonal approximation in which the
$\bs H_{lnlike}$ matrix is divided into blocks, each corresponding to all weights of one layer of the network. This approach neglects covariance between layers but retains covariance within each layer.
We further leverage the Kronecker factored representation of the layer-wise Hessian, denoted as $\bs H_{lnlike}^{(\ell)}$. 
Proposed by \cite{martens2015optimizing, botev2017practical}, the Kronecker-factored representation relies on approximating the sum of Kronecker products over all data points by a Kronecker product of sums.
Consequently, the log-likelihood Hessian for the $\ell$-th layer of the neural network can be represented as,
\begin{equation}\label{eq:kfac}
[\bs J^T \bs L \bs J]^{(\ell)} \approx 
\bs {Q}^{(\ell)} \otimes \bs {R}^{(\ell)},
\end{equation}
where
$\bs{Q}^{(\ell)} = \bs z^{(\ell-1)} (\bs z^{(\ell-1)})^T$ is the uncentered covariance of the activations
and
$\bs{R}^{(\ell)} = -\nabla^2_{\bs a^{(\ell)} \bs a^{(\ell)}} \ln \pi_{like} (\bs D| \bs \theta)$ 
and 
$\bs z^{(\ell)}$ and $\bs a^{(\ell)}$ are the activation values and the pre-activation for layer $\ell$ defined in \eqref{eq:NNfunc2}.
As detailed in \cite{botev2017practical}, computing $\bs{R}^{(\ell)}$ involves a recursive procedure that starts at the output layer, followed by backward propagation through the network in a single pass for each data point, and then averaging across all data points.
The matrices $\bs{Q}^{(\ell)} \in \mathbb{R}^{d_i^{(\ell)} \times d_i^{(\ell)}}$ and 
$\bs{R}^{(\ell)} \in \mathbb{R}^{d_o^{(\ell)} \times d_o^{(\ell)}}$ depends in the dimensionality of the $\ell$-th layer's input $d_i^{(\ell)}$ and output $d_o^{(\ell)}$, and both are positive semidefinite.
Consequently, the Kronecker-factored approximation offers a significant computational advantage by decomposing the layer-wise log-likelihood Hessian into smaller matrices \cite{immer2021scalable, ritter2018scalable}.
By substituting \eqref{eq:ggn} and \eqref{eq:kfac} into \eqref{eq:post_cov}, the posterior covariance at the $\ell$-th layer is approximated as,
\begin{equation}\label{eq:post_cov_kfac}
\bs \Gamma_{post}^{(\ell)} \approx 
\left( \bs {Q}^{(\ell)} \otimes \bs {R}^{(\ell)} + {\bs \Gamma_{pr}^{(\ell)}}^{-1} \right)^{-1},
\end{equation}
while the entire term under inversion does not necessarily allow a Kronecker-factored representation.
To preserve a Kronecker factored structure, \cite{ritter2018scalable} suggested approximating the posterior covariance by introducing the effect of the prior as damping factors to $\bs {Q}^{(\ell)}$ and $\bs {R}^{(\ell)}$.
While such dampening is common in optimization methods employing Kronecker-factored Hessian approximations \cite{martens2015optimizing}, applying it to a posterior approximation artificially increases the posterior concentration.
To prevent the introduction of additional approximations that might overshadow the impact of the prior PDF on inference, we utilize the eigendecompositions 
$\bs {Q}^{(\ell)} = \bs M_{Q^{(\ell)}} \bs \Lambda_{Q^{(\ell)}} \bs M_{Q^{(\ell)}}^T$ and
$\bs {R}^{(\ell)} = \bs M_{R^{(\ell)}} \bs \Lambda_{R^{(\ell)}} \bs M_{R^{(\ell)}}^T$.
Here, $\bs \Lambda_{Q^{(\ell)}} = \text{diag}(\bs q^{(\ell)})$ and 
$\bs \Lambda_{R^{(\ell)}} = \text{diag}(\bs r^{(\ell)})$ where 
$\bs q^{(\ell)}$ and $\bs r^{(\ell)}$ are the eigenvalues of 
$\bs {Q}^{(\ell)}$ and $\bs {R}^{(\ell)}$, respectively.
Substituting the eigendecompositions into \eqref{eq:post_cov_kfac} and 
considering the $\ell$-th block-diagonal entry of prior covariance as 
$\bs \Gamma_{pr}^{(\ell)} = (\sigma^{\text{pr}(\ell)})^2 \bs I^{(\ell)}$, results in \cite{immer2021improving},
\begin{equation}\label{eq:post_cov_kfac_eig}
\bs \Gamma_{post}^{(\ell)} \approx 
\left( 
(\bs M_{Q^{(\ell)}} \otimes \bs M_{R^{(\ell)}}) \,
(\bs \Lambda_{Q^{(\ell)}} \otimes \bs \Lambda_{R^{(\ell)}} + (\sigma^{\text{pr}(\ell)})^{-2} \bs I^{(\ell)}) \,
(\bs M_{Q^{(\ell)}}^T \otimes \bs M_{R^{(\ell)}}^T)
\right)^{-1},
\end{equation}
where $\bs I^{(\ell)}$ is identity matrix with dimensions corresponding to the number of parameters in the $\ell$-th layer
and $\bs M_{Q^{(\ell)}} \otimes \bs M_{R^{(\ell)}}$ constitutes the eigenvectors of 
$\bs {Q}^{(\ell)} \otimes \bs {R}^{(\ell)} $, making it unitary and facilitating the representation of this identity.
The LA and approximated posterior covariance via generalized Gauss-Newton and Kronecker-factored approximations, along with the eigendecomposition of the individual Kronecker factors, results in an efficient and scalable algorithm for Bayesian inference of neural networks.

%++++++++++++++++++++++++++++++++++++++++++++++++++++++++++++++++++++++++
%++++++++++++++++++++++++++++++++++++++++++++++++++++++++++++++++++++++++
\section{Hierarchical Bayesian inferences and model plausibility}\label{sec:hierarchical}

%++++++++++++++++++++++++++++++++++++++++++++++++++++++++++++++++++++++++
\subsection{Bayesian neural networks model}\label{sec:bnn_model}
\noindent
We introduce the \textit{BayesNN model} denoted by the abstract form $M(\bs \phi)$. This representation encapsulates not only the network architecture but also the specifications of the prior of the parameters and the noise model utilized for constructing the likelihood.
We thus categorize the BayesNN model parameters $\bs \phi$ into four distinct subsets, as follows:

%--------
\paragraph{1) Network parameters $\bs \theta \in \mathbb{R}^P$} 
These encompass the set of weights and biases constituting a BayesNN, represented collectively as, 
\begin{equation}
    \bs \theta = [\bs w^{(1)}, \cdots, \bs w^{(D)}, \bs b^{(1)}, \cdots, \bs b^{(D)}].
\end{equation}
As discussed earlier, the primary objective of BayesNN training is to determine the probability distributions of the network parameters based on data.

%--------
\paragraph{2) Inference hyper-parameters $\bs \sigma \in \mathbb{R}^H$} 
These parameters characterize the Bayesian inference setup and stem from the specifications of the prior and likelihood. 
More precisely, the prior on the weights and biases, denoted as $\pi_{pr} (\bs \theta)$, can be parametrized by $\bs \sigma^{\text{pr}} \in \mathbb{R}^P$, such as mean and variance of a Gaussian prior. 
%In practice, one may opt for a single prior hyperparameter on all network parameters or consider distinct priors for each class of parameters (weights out of input units, biases of hidden units, and weights and biases of output units) \cite{neal2012bayesian, mackay1995probable}. 
Additionally, the likelihood function incorporates noise hyper-parameters $\bs \sigma^{\text{noise}} \in \mathbb{R}^{N_D}$ arising from the chosen noise model.
The inference hyper-parameters are a collection of prior and noise hyper-parameters,
\begin{equation}
    \bs \sigma = [\bs \sigma^{\text{pr}}, \bs \sigma^{\text{noise}}].
\end{equation}
%

%--------
\paragraph{3) Architecture hyper-parameters $\bs \xi \in \Xi$}
These parameters represent the structural and organizational characteristics of the network, specifying the functional mapping between inputs and outputs. In the case of fully connected networks, the architecture hyper-parameters include the number of layers $D$, the number of neurons $W$ and the activation function $f$ in each layer (i.e., the type of non-linearity applied between layers),
\begin{equation}
    \bs \xi = [W^{(1)}, \cdots, W^{(D)}, f^{(1)}, \cdots, f^{(D)}].
\end{equation}
Here, $D, W \in \mathbb{N}$ and $f \in \mathcal{F}$, where $\mathcal{F}$ denotes the function space adhering to properties expected of neural network activation functions, such as smoothness, continuity, and convexity.
For more advanced models, like convolutional neural networks, the space of architecture hyper-parameters expands significantly, including the number and kernel size of convolutions, channels per convolutional layer, and the number of fully-connected layers, e.g., \cite{qian2023biomimetic, ghosh2022designing}.

%
%--------
\paragraph{4) Solution hyper-parameters $\chi \in \mathbb{R}^X$}
Another class of parameters pertains to the selection of Bayesian inference solutions, encompassing choices like sampling algorithms and variational inference, along with the particulars of these algorithms, such as learning rate and weight initialization.
\blue{In this work, we do not investigate the influence of solution hyperparameters on the predictive capability of the models. Hence, throughout the rest of this manuscript, the BayesNN model explicitly specifies the network's functional form, as well as the likelihood and prior specifications in Bayesian inference.}

%++++++++++++++++++++++++++++++++++++++++++++++++++++++++++++++++++++++++
\subsection{Model evidence}\label{sec:evidence}
\noindent
A common method for neural network model selection is based on their prediction performance on unseen testing datasets. 
\blue{
However, this approach has been criticized for 
irreproducible outcomes \cite{pouchard2023rigorous},
lack of robustness to small and uncertain data \cite{olivier2021bayesian}, and
resulting in suboptimal architectures \cite{krishnanunni2022layerwise}.} 
Instead, we advocate a principled model selection method based on Bayesian model plausibility, leveraging the entire dataset for model construction.
Let $M(\bs \phi)$ be a BayesNN model with its own set of model parameters 
$\bs \phi =[\bs \theta, \bs \sigma, \bs \xi]$.
We rewrite the Bayesian inference \eqref{eq:bayes}, acknowledging known information on the inference and architecture hyper-parameters,
\begin{equation}\label{eq:bayes_M}
 \pi_{post}(\boldsymbol{\theta}|\bs {D}, \bs \sigma, \bs \xi) = 
 \frac{\pi_{like}(\bs {D}|\boldsymbol{\theta},\bs \sigma, \bs \xi)
 \pi_{pr}(\boldsymbol{\theta} | \bs \sigma, \bs \xi)}
 {\pi_{evid}(\bs {D}|\bs \sigma, \bs \xi)}.
\end{equation}
The denominator of Bayes theorem \eqref{eq:bayes_M} is the probability of observing data using model $M$, commonly referred to as the evidence PDF,
\begin{equation}\label{eq:evid}
 \pi_{evid}(\bs {D}|\bs \sigma, \bs \xi) = \int \pi_{like}(\bs {D}|\boldsymbol{\theta}, \bs \sigma, \bs \xi) \; \pi_{pr}(\boldsymbol{\theta} | \bs \sigma, \bs \xi) \; d\boldsymbol{\theta}.
\end{equation}
\blue{In this setting, the evidence extends beyond its conventional role as a mere normalization factor in classical Bayesian inference, emerging as a pivotal component in our model selection framework. Specifically, by rearranging \eqref{eq:bayes}, the logarithm of evidence can be formulated as the posterior mean of the log-likelihood and the relative entropy between the prior and posterior distributions \cite{muto2008}.}
In essence, the evidence reflects a trade-off between the model fitting the observed data and the extent to which network parameters $\boldsymbol{\theta}$ are learned from that data. This characteristic renders the evidence a valuable measure for model comparison \cite{odenbabuska2017}, particularly in BayesNN models, facilitating comparisons between different architectures, noise models, and prior formulations. 

%++++++++++++++++++++++++++++++++++++++++++++++++++++++++++++++++++++++++
\subsection{Hierarchical Bayesian inferences}\label{sec:plausibility}
\noindent
Building upon MacKay's evidence framework \cite{mackay1995probable}, the following {hierarchical Bayesian inferences} are postulated, treating the evidence PDF at each level as a new likelihood function for subsequent inference:

\begin{itemize}
    \item \textit{Level 1.} Infer the network parameters $\bs \theta$,
    \begin{equation}\label{eq:lev1}
    \pi_{post} (\bs \theta | \bs D, \bs \sigma, \bs \xi) = 
    \frac{\pi_{like} (\bs D| \bs \theta, \bs \sigma, \bs \xi) \,
    \pi_{pr} (\bs \theta| \bs \sigma, \bs \xi)}
    {\pi_{evid} (\bs D| \bs \sigma, \bs \xi)}.
    \end{equation}

    \item \textit{Level 2.} Infer the inference hyper-parameters $\bs \sigma$,
    \begin{equation}\label{eq:lev2}
    \pi_{post} (\bs \sigma | \bs D,  \bs \xi) = 
    \frac{\pi_{evid} (\bs D| \bs \sigma,  \bs \xi) \,
    \pi_{pr} (\bs \sigma |  \bs \xi)}
    {\pi_{evid} (\bs D|  \bs \xi)}.
    \end{equation}

    \item \textit{Level 3.} Infer the archtecture hyper-parameters $\bs \xi$,
    \begin{equation}\label{eq:lev3}
    \pi_{post} (\bs \xi | \bs D) =
    \frac{\pi_{evid} (\bs D| \bs \xi) \,
    \pi_{pr} (\bs \xi)}{\pi_{evid} (\bs D)}.
    \end{equation}

\end{itemize}
\blue{Although leveraging \textit{Level 1} and \textit{Level 2} inferences in BayesNN has gained recent attention \cite{psaros2023uncertainty}, the utilization of \textit{Level 3} for neural network architecture selection has been relatively understudied.}
In particular, \eqref{eq:lev3} facilitates the update of the modeler's prior belief in a BayesNN model $\pi_{pr} (\bs \xi)$ based on data, yielding the posterior model plausibility $\rho = \pi_{post} (\bs \xi | \bs D)$. 

\blue{With the posterior PDFs of model parameters and hyperparameters in hand, the next step is to make prediction using the updated model. For a fixed architecture, this entails marginalizing over uncertainty at the first two levels to derive the prediction distribution,
\begin{equation}\label{eq:prediction}
    \pi(\bs{u}_{\bs \theta}|\bs D, \bs \xi) = 
    \int \, \pi(\bs{u}_{\bs \theta}|\bs \theta, \bs \sigma, \bs \xi) \, \pi(\bs \theta, \bs \sigma|\bs {D}, \bs \xi) \, d \bs\sigma \, d \bs\theta,
\end{equation}
where $\pi(\bs \theta, \bs \sigma|\bs {D}, \bs \xi) = \pi_{post} (\bs \theta | \bs D, \bs \sigma, \bs \xi) \pi_{post} (\bs \sigma | \bs D,  \bs \xi)$ and the evaluation of $\pi(\bs{u}_{\bs \theta}|\bs \theta, \bs \sigma, \bs \xi)$ requires a single forward model evaluation at the desired inputs.
Similarly, 
the predictions of multiple architectures are obtained by computing a weighted sum of \eqref{eq:prediction}, where the model plausibility $\rho$ serves as the weight, known as Bayesian model averaging, e.g., \cite{vila2000bayesian, chitsazan2015prediction, olivier2021bayesian}.
}

%--------------
\begin{paragraph}{\blue{Remark 1}}
In the hierarchical Bayesian inference of \eqref{eq:lev1} and \eqref{eq:lev2}, there may appear to be a contradiction with Bayesian principles, as the prior for network parameters seems to be determined after observing the data. In other words, it might appear that the most probable value of the prior hyperparameters $\bs \sigma^{\rm pr}$ is chosen first, followed by the utilization of the corresponding prior to infer $\thetab$.
To clarify this apparent sequence, we recognize that the model prediction can be the integration over the ensemble of unknown network and inference hyperparameters collectively defining the prior.
The true posterior \blue{of the network parameters} used for model prediction in \eqref{eq:prediction} is expressed as:
\begin{eqnarray}
	\pi_{post} (\bs \theta | \bs D, \bs \xi) &=& 
	\int \pi (\bs \theta,  \bs \sigma | \bs D, \bs \xi) \; d\bs \sigma
  \nonumber\\
     &=& 
	 \int \pi_{post} (\bs \theta | \bs D, \bs \sigma, \bs \xi) \; \pi_{post} (\bs \sigma | \bs D, \bs \xi) \; d\bs \sigma.
\end{eqnarray}
This expression suggests that the posterior of $\bs \theta$ is obtained by integrating over the posteriors for all values of $\bs \sigma$, each weighted by the probability of $\bs \sigma$ given the data. This mirrors the concept of Bayesian model averaging, where predictions from multiple  BayesNN with different inference hyperparameters are combined, considering the associated uncertainty of each model.
Considering $\pi_{post} (\bs \sigma | \bs D, \bs \xi)$ exhibiting a dominant peak at $\bs \sigma_{MAP}$, then the true posterior $\pi_{post} (\bs \theta | \bs D, \bs \xi) $ will be primarily governed $\pi_{post} (\bs \theta | \bs D, \bs \sigma, \bs \xi)|_{\bs \sigma = \bs \sigma_{MAP}}$. Consequently, we are justified in using the following approximation \cite{mackay1992bayesian},
\begin{equation}
\pi_{post} (\bs \theta | \bs D, \bs \xi) \approx 
\pi_{post} (\bs \theta | \bs D, \bs \sigma, \bs \xi)|_{\bs \sigma = \bs \sigma_{MAP}}.
\end{equation}
Therefore, employing only the most probable prior hyperparameter is a valid approximation for posterior evaluation. 
We note that the Bayesian model averaging can be extended to combine predictions from multiple network architectures, with each model weighted according to its posterior probability based on the data according to \eqref{eq:lev3}.
\end{paragraph}

%--------------
\paragraph{Numerical solution}
The solution to hierarchical Bayesian inferences involves a multi-step procedure. 
Specifically, \textit{Level 1} and \textit{Level 2} inferences comprise an iterative scheme for the online determination of network parameters and inference hyperparameters.
In each step, the network parameters $\bs \theta$ are inferred with fixed values of the inference hyperparameters $\bs \sigma$ using \eqref{eq:lev1}. 
Subsequently, these parameters are utilized to update $\bs \sigma$ as per \eqref{eq:lev2}, maximizing the corresponding evidence $\pi_{evid} (\bs D| \bs \sigma, \bs \xi)$.
Upon completion of the network training and hyperparameters inference, the evidence $ \pi_{evid} (\bs D| \bs \xi) $ is employed to compute posterior model plausibility in \textit{Level 3}.
Using LA to the posterior and the prior and noise model considered in Section \ref{sec:bnn}, the process of network training and hyperparameter determination entails computing their corresponding MAP estimates and the posterior covariance. 
In the absence of prior information on the inference hyperparameters and the non-Gaussian form of likelihood, we adopt uniform priors for $\ln (\sigma^{\text{pr}^2})$ and $\ln (\sigma^{\text{noise}^2})$, in accordance with the recommendation of \cite{mackay1992bayesian, mackay1995probable}. This prior facilitates exploration across a broad spectrum of hyperparameters and ensures cancellation when comparing different models.
\blue{Taking $\bs \sigma = [\ln (\sigma^{\text{noise}^2}), \ln (\sigma^{\text{pr}^2})]$,}
the evidence for updating inference hyperparameters is expressed as, 
\begin{eqnarray}\label{eq:evid_lev2_la}
\ln \pi_{evid} \left(\bs D| \blue{\bs \sigma} , \bs \xi \right) 
 \approx 
\ln \pi_{like} \left(\bs D| \bs \theta, \blue{\bs \sigma}, \bs \xi\right)|_{\bs \theta = \bs \theta_{MAP}}
 \nonumber\\
 + 
\ln \pi_{pr} (\bs \theta | \blue{\bs \sigma}, \bs \xi)|_{\bs \theta = \bs \theta_{MAP}} 
 - 
 \ln \pi_{post} (\bs \theta | \bs D, \blue{\bs \sigma}, \bs \xi)|_{\bs \theta = \bs \theta_{MAP}} 
	\nonumber\\
 = 
- \frac{1}{2 \, \blue{\ln (\sigma^{\text{noise}^2})} }\sum_{j=1}^{N_D} (\bs u_D^j - \bs{u}_{\bs \theta_{MAP}}(\bs x_D^j))^2 
-\frac{N_D}{2}\ln (2\pi \, \blue{\ln (\sigma^{\text{noise}^2})} )
	\nonumber\\ 
- \frac{1}{2 \, \blue{\ln (\sigma^{\text{pr}^2})} } \| \bs \theta_{MAP} \|^2 
-\frac{\blue{P}}{2}\ln (2\pi \, \blue{\ln (\sigma^{\text{pr}^2})}  )
- \frac{1}{2} \ln \det\left(\frac{1}{2\pi}\bs H \right),
\end{eqnarray}
where $\| \cdot \|$ denotes the Euclidean norm.
\blue{
Utilizing a Gaussian approximation of $\pi_{post} (\bs \sigma | \bs D, \bs \xi)$, the MAP estimate of the inference hyperparameters in \eqref{eq:lev2} is obtained by maximizing the evidence in \eqref{eq:evid_lev2_la}. The posterior variances of $\ln (\sigma^{\text{noise}^2})$ and $\ln (\sigma^{\text{pr}^2})$ are derived by differentiating \eqref{eq:evid_lev2_la} twice, following the derivations provided in \cite{mackay1992bayesian}, as 
\begin{equation}\label{eq:var_hyper}
    \sigma^2_{\ln (\sigma^{\text{pr}^2})} \approx \frac{2 \left(\ln(\sigma^{\text{pr}^2}_{MAP}) \right)^2}
    {P - \frac{1}{\ln(\sigma^{\text{pr}^2}_{MAP})}\text{tr}(\bs H^{-1})}, \;
    \sigma^2_{\ln (\sigma^{\text{noise}^2})} \approx \frac{2 \left(\ln(\sigma^{\text{pr}^2}_{MAP}) \right)^2}
    {N_D - P + \frac{1}{\ln(\sigma^{\text{pr}^2}_{MAP})}\text{tr}(\bs H^{-1})}.
\end{equation}
}
Using \eqref{eq:var_hyper},
the model evidence for evaluating the posterior model plausibility in \textit{Level 3} is expressed as,
\begin{eqnarray}\label{eq:evid_lev3_la}
\blue{
 \pi_{evid} (\bs D| \bs \xi)
 }
& = &
\blue{
\int \pi_{evid} (\bs D| \bs \sigma,  \bs \xi) \,
    \pi_{pr} (\bs \sigma |  \bs \xi) \; d\bs \sigma
    }
    \nonumber\\
& \approx & 
\blue{
    \pi_{evid} (\bs D| \bs \sigma, \bs \xi)|_{\bs \sigma = \bs \sigma_{MAP}} \,
 \pi_{pr} \left(\bs \sigma | \bs \xi \right)|_{\bs \sigma = \bs \sigma_{MAP}}
    }
     \nonumber\\
 &  &
\blue{
  \left( 2\pi \sigma_{\ln (\sigma^{\text{pr}^2})} \sigma_{\ln (\sigma^{\text{noise}^2})} \right),
  }
\end{eqnarray}
\blue{where $\pi_{evid} (\bs D| \bs \sigma, \bs \xi)|_{\bs \sigma = \bs \sigma_{MAP}}$ is evaluated using \eqref{eq:evid_lev2_la}.}

The iterative updating network parameters, inference hyperparameters, and computing model plausibility is computationally expensive due to calculations involving the determinant of the Hessian in \eqref{eq:evid_lev2_la} and the trace of its inverse in \eqref{eq:evid_lev3_la} for high-dimensional parameters.
However, we leverage the available eigenvalues of the Kronecker-factored matrices described in Section \ref{sec:kfac} for efficient computation of the determinant and trace,
\begin{equation}
    \det( \bs H ) = \prod_{\ell} \prod_{ij} q_i^{(\ell)} r_j^{(\ell)} + (\sigma^{\text{pr}(\ell)})^2, \quad
    \text{tr}( \bs H ) = \sum_{\ell} \sum_{ij} q_i^{(\ell)} r_j^{(\ell)} + (\sigma^{\text{pr}(\ell)})^2. 
\end{equation}
%

%++++++++++++++++++++++++++++++++++++++++++++++++++++++++++++++++++++++++
%++++++++++++++++++++++++++++++++++++++++++++++++++++++++++++++++++++++++
\section{A Framework for Surrogate Model Discovery}\label{sec:opal}
\noindent
The main focus of this study is on surrogate modeling, specifically utilizing BayesNN models as low-computational-cost approximations for solutions of high-fidelity physics-based simulations. Beyond addressing uncertainties arising from limited and sparse data in high-fidelity simulations, credible surrogate models must capture the overall hierarchy of interactions across multiple scales inherent in physics-based simulations \cite{shekhar2020hierarchical}.
We argue that Bayesian training of neural networks is well-suited for this purpose, as it allows for more complex models with an adequate number of weights to capture the multiscale structure of physics-based simulation. This stands in contrast to maximum likelihood training, where an excessive number of parameters may lead to overfitting and compromise the models’ generalization.
Despite these merits, challenges persist in surrogate modeling using BayesNN, particularly in identifying the right model within an enormous space of potential architectures and rigorous methodologies for assessing the accuracy and reliability of surrogates under uncertainty.

In formulating the surrogate modeling problem, let $\mathcal{M}$ be a set of different BayesNN models,
\begin{equation}\label{eq:set}
\mathcal{M} = \Big\{ M_1(\bs \phi_1), M_2(\bs \phi_2), \cdots, M_K(\bs \phi_K) \Big\},
\end{equation}
where each model $M_i, \; i=1, 2, \cdots, K$ has its own set of model parameters $\bs \phi_i =[\bs \theta_i, \bs \sigma_i, \bs \xi_i]$. 
The high-fidelity data $\bs {D}$ is obtained from the scenario $\mathcal{S}$ that encapsulates features of the high-fidelity simulation such as domain, boundary conditions, and spatial location and frequency from which data is gathered. Later we will argue the importance of considering a hierarchy of scenarios to rigorously assess the validity and predictive reliability of surrogate models.
The training process for each model involves hierarchical Bayesian inferences \eqref{eq:lev1}-\eqref{eq:lev3}, where we recognize known information about the set $\mathcal{M}$ and scenario $\mathcal{S}$ in the probability distributions, i.e., $ \pi_{post} (\bs \theta | \bs D, \bs \sigma, \bs \xi, \mathcal{M}, \mathcal{S})$. In this context, the denominator of \eqref{eq:lev3} serves to normalize the discrete probability components over the models in the set $\mathcal{M}$,
\begin{equation}
    \pi (\bs D| \mathcal{M}, \mathcal{S}) = 
    \sum_{i=1}^K \pi_{evid}(\bs {D} | \bs \xi_i, \mathcal{M}, \mathcal{S})
\pi_{pr}(\bs \xi_i | \mathcal{M}),
\end{equation}
resulting in 
$
    \sum_{i=1}^K \rho_i = 1,
$
with $\rho_i = \pi_{post} (\bs \xi_i | \bs D)$.
Hence, within the set of competing models in $\mathcal{M}$ and for a given dataset $\bs {D}$, the model boasting the highest $\rho_i$ is deemed the most plausible model \cite{odenbabuska2017, muto2008}.
However, the model evidence alone does not ensure the generalizability and prediction reliability of the model, indicating the need for additional steps for credibility assessment and model validation.

%++++++++++++++++++++++++++++++++++++++++++++++++++++++++++++++++++++++++
\subsection{Occam-Plausibility Algorithm for Surrogates}\label{sec:opal_steps}
\noindent
Building on our previous framework for validation and selection of physics-based models \cite{farrell2015jcp, tan2022toward, odenbabuska2017},
we proposed Occam-Plausibility Algorithm for Surrogate models (OPAL-surrogate),
a systematic strategy for discovering the simplest (``best'') credible surrogate model shown in Figure \ref{fig:opalsur}. 
The key enabler to overcome the enormous space of potential BayesNN models is adaptive modeling by replacing discrete choices with continuous parameterization.
The OPAL-surrogate involves the following steps:

\begin{enumerate}

%------------------------------
\item \textbf{Initialization.}
Construct an initial set of possible surrogate models $\mathcal{M}$ in \eqref{eq:set}
and acquired training data $\bs D$ from a scenario $\mathcal{S}^{(s)}$ with setting $s=1$. 

%------------------------------
\item \textbf{Occam Categories.}
Partition the models in $\mathcal{M}$ based on model complexity measures into \textit{Occam categories}. The simplest models are placed in Category 1, and the most complex models are designated in the last category. 
We, therefore, produce a collection of subsets, 
\begin{equation}
\hat{\mathcal{M}}^{(l)} = 
\Big\{ M_1^l(\bs\phi_1^l), M_2^l (\bs\phi_2^l), \cdots, M_{K_l}^l(\bs\phi_{K_l}^l) \Big\}, \quad l=1,2 \cdots, H,
\end{equation}
where $l$ is the category and ${K_l}<K$.
A commonly used complexity measure is the number of model parameters $\bs \phi$, although other measures can also be considered, especially when implementing OPAL-surrogate to multiple classes of surrogate models. The categorization of models depends on factors such as the size of the initial model set and the available computational resources.

%------------------------------
\item \textbf{Occam Step and Model Plausibility.}
Start with the first Occam Category, $l=1$, employ $\bs D$
to identify the plausible model in Category $l$, denoted by $M_I^l(\boldsymbol{\phi}^l_I)$.
This process is formalized as,
\begin{equation}\label{eq:archopt}
    \bs \xi_I^l = \underset{\bs \xi_k^l \in \bar{\Xi}^l}{\text{argmax}} \; 
    \{\rho_k^l\}_{k=1}^{R_l},
\end{equation}
where $\rho_k^l = \pi_{post} (\bs \xi_k^l | \bs D, \hat{\mathcal{M}}^{(l)}, \{\mathcal{S}^{(s)}\}_{s=1}^{S})$ represents the Bayesian posterior plausibility for the $k$-th model in the set $\hat{\mathcal{M}}^{(l)}$, 
given the high-fidelity data $\bs D$ obtained from a collection of scenarios ${ \mathcal{S}^{(1)}, \cdots, \mathcal{S}^{(S)} }$.

Except for rare instances like model selection among a limited space of architectures $\bar{\Xi}^l$, such as in \cite{immer2021scalable}, discrete optimization in \eqref{eq:archopt} that requires Bayesian inference across all possible models to evaluate $\rho_i^l$, becomes computationally prohibitive. 
Instead, one can adopt a uniform prior on all (or a subset of) architecture hyper-parameters $\pi_{pr}(\bs \xi_i | \hat{\mathcal{M}}^{(l)})$, and employ adaptive modeling by replacing discrete model choices with continuous parameterization. This transformation turns the optimization problem in \eqref{eq:archopt} into a continuous maximization of the evidence $\pi_{evid}(\bs {D} | \bs \xi_i, \hat{\mathcal{M}}^{(l)}, \{\mathcal{S}^{(s)}\}_{s=1}^{S})$, facilitating efficient model discovery.
\blue{Solving the possibly non-convex optimization problem using pure machine learning techniques involves leveraging advanced neural search algorithms \cite{elsken2019neural, liu2021survey, wang2018hybrid, optuna_2019}.}
However, inspired by \cite{shekhar2022forward, shekhar2020hierarchical}, our adaptive surrogate modeling takes a distinctive perspective to ensure the BayesNN approximator effectively captures the underlying multiscale structure of the high-fidelity physics-based simulations.
To realize this objective, we propose a sequential addition of fully connected layers with sufficiently large widths, guided by the model evidence value, followed by the elimination of irrelevant weights. This strategic approach necessitates an effective network sparsification method to reveal sparsity patterns associated with the inherent multiscale structure encoded in high-fidelity data. One of such methods is described in Section \ref{sec:sparse}.

%------------------------------
\item \textbf{Credibility Assessment.}
Examine the validity of the most plausible model $M_I^l(\boldsymbol{\phi}^l_I)$, by subjecting it to \textit{leave-out cross-validation test}.
Accordingly, divide the training data $\bs D$ into $N_v$ leave-out subsets,
\begin{equation}
    \{\bs D_{LO}^{(n)}\}_{n=1}^{N_v}
	\quad \text{with} \quad
	\bs D_{LO}^{(n)} = \bs D \setminus \bs D^{(n)}.
\end{equation}
For each subset of data, train $M_I^l(\boldsymbol{\phi}^l_I)$ with $\bs D^{(n)}$ using the Bayesian inference \eqref{eq:lev1} and \eqref{eq:lev2}.
Use the posteriors of the network parameters and inference hyper-parameters to evaluate the model output $\bs u_{\bs \theta}(\bs x_{D_{LO}^{(n)}} )$ associated with $\bs D_{LO}^{(n)}$.
Compare a proper distance measure $\mathbbm{d}(\cdot , \cdot)$ between the model output and the leave-out data set with a given accuracy tolerance $TOL$.
\begin{figure}[htpb]
    \centering
    \includegraphics[width=0.85\textwidth]{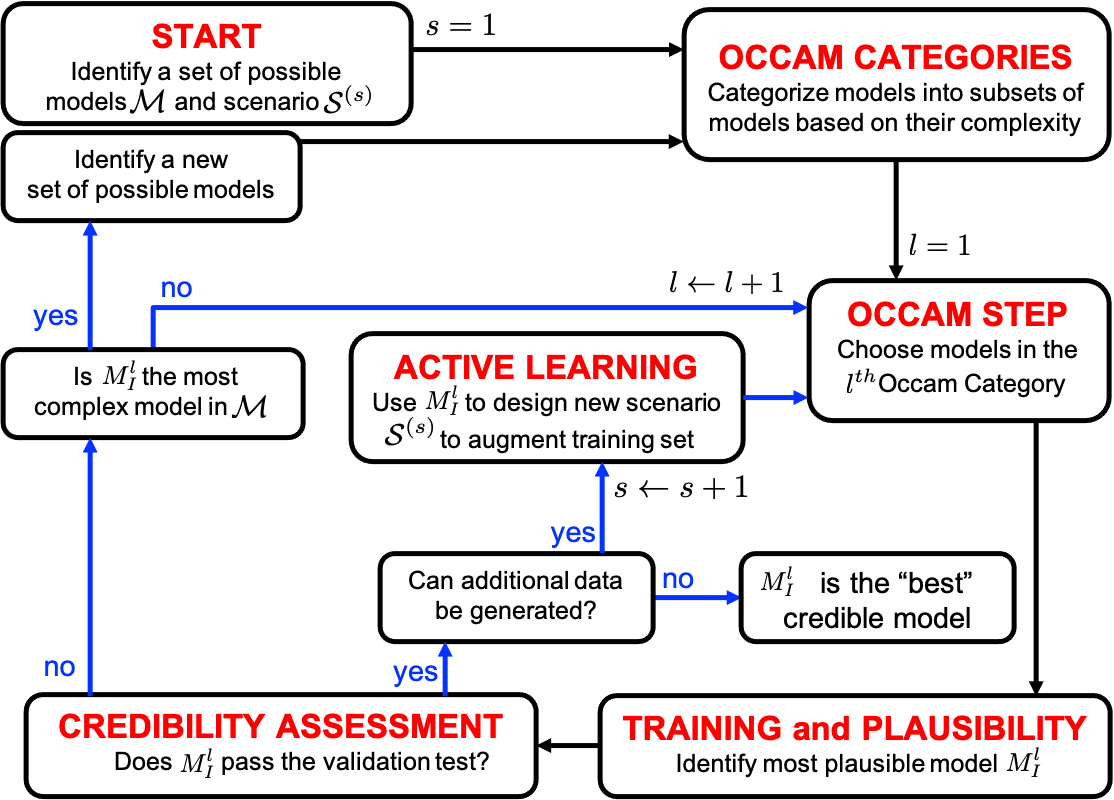}
    \vspace{-0.1in}
    \caption{
    The Occam Plausibility Algorithm for surrogate models (OPAL-surrogate): 
    Commencing with an initial set of potential models, the models are categorized based on measures of complexity. Starting with the first Occam Category, the most plausible model is determined and undergoes the validation test for credibility assessment. 
    If additional data can be acquired, the validated model guides the design of new scenarios to expand the training set and iteratively re-select new models in response to the updated dataset. The model that passes the validation test is considered the ``best" credible model.
    }
    \label{fig:opalsur}
\end{figure}
\\
For $n=1, \cdots, N_v$,
    \begin{eqnarray*}\label{eq:lev1_2}
    \pi_{post} (\bs \theta^l_I | \bs D^{(n)}, \bs \sigma^l_I, \bs \xi^l_I, \hat{\mathcal{M}}^{(l)}, \{\mathcal{S}^{(s)}\}_{s=1}^{S}) = \\
    \frac{\pi_{like} (\bs D^{(n)}| \bs \theta^l_I, \bs \sigma^l_I, \bs \xi^l_I, \hat{\mathcal{M}}^{(l)}, \{\mathcal{S}^{(s)}\}_{s=1}^{S}) 
    }
    {\pi_{evid} (\bs D^{(n)}| \bs \sigma^l_I, \bs \xi^l_I, \hat{\mathcal{M}}^{(l)}, \{\mathcal{S}^{(s)}\}_{s=1}^{S})}
    \pi_{pr} (\bs \theta^l_I| \bs \sigma^l_I, \bs \xi^l_I), \nonumber\\
    \pi_{post} (\bs \sigma^l_I | \bs D^{(n)},  \bs \xi^l_I, \hat{\mathcal{M}}^{(l)}, \{\mathcal{S}^{(s)}\}_{s=1}^{S})  =  \\
    \frac{\pi_{evid} (\bs D^{(n)}| \bs \sigma^l_I,  \bs \xi^l_I, \hat{\mathcal{M}}^{(l)}, \{\mathcal{S}^{(s)}\}_{s=1}^{S})
    }
    {\pi_{evid} (\bs D^{(n)}|  \bs \xi^l_I, \hat{\mathcal{M}}^{(l)}, \{\mathcal{S}^{(s)}\}_{s=1}^{S})}
    \pi_{pr} (\bs \sigma^l_I |  \bs \xi^l_I),
    \end{eqnarray*}
    \begin{equation}\label{eq:validtest}
		\mathbbm{d}^{(n)} \left(
        \pi(\bs u_{\bs D}(\bs x_{D_{LO}^{(n)}})) \, , \, 
        \pi(\bs u_{\bs \theta}(\bs x_{D_{LO}^{(n)}}))
        \right) 
        < TOL.
    \end{equation} 
If the inequality \eqref{eq:validtest}$_3$ holds for all $N_v$, the model $M_I^l$ is considered as ``not invalid'' and identified as the \textit{best credible surrogate model} for the given training data. This designation is made under the recognition that additional data and information could potentially falsify a model initially presumed to be valid.
Once the surrogate model is determined, the parameters are updated using the entire training set. This involves substituting $\bs D^{(n)}$ with $\bs D$ in \eqref{eq:validtest}$_{1,2}$, and utilizing the inferred network parameters $\pi_{post} (\bs \theta^l_I | \bs D, \bs \sigma^l_I, \bs \xi^l_I, \hat{\mathcal{M}}^{(l)}, \{\mathcal{S}^{(s)}\}_{s=1}^{S})$ for making predictions.

%------------------------------
\item \textbf{Scenario Design via Active Learning.} 
If additional computational resources allow for further data generation, the model $M_I^l(\boldsymbol{\phi}^l_I)$ is utilized to design the next scenario $s \leftarrow s+1$ to augment $\bs D$ with more effective data, following the active learning approach.
In cases where surrogate models are utilized for interpolation, this involves leveraging optimal experimental design methods, such as \cite{huan2013simulation, riis2022bayesian, dougherty2015}, which take a decision-theoretic approach to optimize features of the high-fidelity simulation scenario by maximizing an information gain metric as expected utility. %The choice of the utility function depends on the purpose of surrogate modeling and the cost and usefulness of the observational data in a specific problem.
However, complications arise in extrapolation when the surrogate model aims to predict \textit{unobservable QoI} beyond the scope of high-fidelity simulation. This poses the formidable challenge of designing a scenario to obtain observational data that reflects the structure of the QoI \cite{paquette2023optimal, tan2022toward} and further discussed in \textit{Remark 2} below.

%------------------------------
\item \textbf{Iteration and Refinements.} 
If the model $M_I^l$ is invalid, OPAL returns to the next Occam category $l \leftarrow l+1$ and repeats Steps 3 and 4 until identifying a ``not invalid'' model and possibly augmenting training data in Step 5. 
In case all BayesNN models in $\mathcal{M}$ are found to be invalid, it is necessary to enlarge the initial model set.

\end{enumerate}

\begin{paragraph}{Remark 2}
While the OPAL-surrogate provides an adaptive strategy for discovering neural network-based surrogate models, it emphasizes the vital role of integrating domain expert knowledge into the modeling process. The efficacy of this framework relies on various subjective decisions that the modeler must tailor to a specific problem. These decisions involve defining the initial model set, selecting appropriate complexity measures and categorization methods, establishing the prior distribution for architecture hyperparameters, defining metrics for credibility assessment, and choosing the utility function in scenario design.
\blue{
Of particular significance are the credibility assessment (step 4) and the choice of validation tolerance. In surrogate modeling scenarios, such as those illustrated in the numerical examples in Section \ref{sec:results}, where the objective is to identify architectures for a specific class of neural network models, a convergence study perspective may be considered. Instead of using a fixed tolerance, the model is selected within the Occam Category, where predictive performance diminishes with increasing complexity in the higher category.
However, when extending the application of the OPAL-surrogate, for instance, to identify predictive models among diverse classes of neural operators and varied architectures, the convergence perspective becomes highly sensitive to categorization. Therefore, it is more appropriate to consider a user-defined, problem-specific tolerance on the validation observables, indicating the acceptable level of prediction errors in the surrogate model. This approach also allows for the possibility of the OPAL-surrogate rejecting all models in the initial set, prompting the identification of new potential models to achieve acceptable prediction accuracy, rather than solely relying on models within the initial set.
}
\end{paragraph}

\begin{paragraph}{Remark 3}
A formidable challenge in scientific prediction is the extrapolation beyond the range of available data to predict unobservable QoIs. Section \ref{sec:results} presents a numerical example illustrating this situation, where the surrogate's objective is to make predictions on a larger domain inaccessible to high-fidelity simulation.
The formalization of surrogate modeling, grounded in the concept of a ``prediction pyramid" \cite{OdenMoserGhattas2010I,odenbabuska2017}, 
is pivotal in augmenting the reliability of extrapolation predictions.
At the base of the prediction pyramid is the pre-training scenario $\mathcal{S}_c$, with a lower computational cost of the high-fidelity simulation. It facilitates the generation of a large set of pre-training data $\bs {D}_c$, contributing to establishing a meaningful prior distribution on the network parameters.
Ascending the pyramid is the training scenario $\mathcal{S}$, involving more complex high-fidelity simulations that provide training dataset $\bs {D}$, used to inform and test the surrogate model's trustworthiness and prediction reliability.
At the top of the pyramid is the prediction scenario $\mathcal{S}_p$, the most complex scenario where conducting high-fidelity simulation is practically impossible. The surrogate model prediction relies on the extrapolation of training data and the design of a meaningful training scenario that accurately captures the features of the QoIs in the prediction scenario \cite{tan2022toward, paquette2023optimal}.
\end{paragraph}

%++++++++++++++++++++++++++++++++++++++++++++++++++++++++++++++++
\subsection{Network sparsification strategy}\label{sec:sparse}
\noindent
As outlined in Section \ref{sec:opal_steps}, an effective sparsification is needed to ensure the surrogate model captures the inherent multiscale structure within high-fidelity physical simulations.
The conventional method involves pruning by setting tentative network parameters to zero, followed by training the new model to accept the pruning based on a performance measure. However, explicitly pruning one parameter at a time becomes computationally prohibitive for large networks \cite{mackay1995probable}.
In contrast, we seek to automatically assess the relevance of network parameters, preserving those deemed relevant to enhance the surrogate model's predictive performance. 
%A suitable category of methods for this purpose is Automatic Relevance Determination (ARD), widely applied in sparse regression \cite{rudy2021sparse, zhang2018robust, fuentes2019efficient}. Despite the advantages of ARD over the pruning methods, practical challenges may arise in implementing ARD within BayesNN, as discussed by Mackay and Neal \cite{mackay1995probable, neal2012bayesian}. Notably, dealing with numerous relevance hyperparameters can hinder the calculation of evidence for different models, particularly when handling a large number of network parameters with limited data.
%
Here, we advocate exploiting sparsity-enforcing priors to eliminate irrelevant parameters of BayesNN, e.g., \cite{williams1995bayesian, rudy2021sparse}. From a Bayesian perspective, such a prior reflects our belief that certain parameters are less likely to be relevant than others. This leads us to anticipate that these parameters will be centered around a specific value, with penalization for deviations from this mode.
Employing the maximum entropy principle for constructing prior probability distributions \cite{jaynes2003}, we obtain a Laplace distribution by imposing constraints on both the mean of the network parameters and their absolute deviation (L1 norm) from the mean, expressed as,
\begin{equation}\label{eq:laplace}
\pi_{pr}(\boldsymbol{\theta}) = \prod_{i=1}^{P} \frac{1}{\beta} \exp \left[ -\frac{ \left | \theta_i - \bar{\theta_i} \right |}{\beta} \right],
\end{equation}
where $\beta>0$ is the scale parameter.
%It is worth noting that, Williams \cite{williams1995bayesian} justified the use of the Laplace prior by employing a transformation groups approach to prior construction \cite{jaynes1968prior}. He exploited a symmetry property inherent in neural network models, asserting the existence of a functionally equivalent network where the weight on a given connection has the same magnitude but opposite sign. This symmetry mandates that the prior for a given weight should depend on its absolute value.
%
The network sparsification method involves applying a Laplace prior to the network parameters, followed by magnitude-based thresholding. Parameters with sufficiently small magnitudes, denoted as $| \theta_i | < TOL_{\bs\theta}$, are considered irrelevant and subsequently removed from the model. The level of aggressiveness in parameter removal is controlled to maximize the model evidence, achieved by incrementally increasing the threshold $TOL_{\bs\theta}$ until the evidence value of the resulting sparsed network begins to decline. 

%This approach can be regarded as a soft implementation of the sparsification method proposed by Williams \cite{williams1995bayesian}.

\begin{paragraph}{Discovering the category-wise plausible model}
We propose the following strategy for implementing Step 3 of OPAL-surrogate across a wide range of possible BayesianNN models. In this context, we characterize fully connected neural network architectures by their depth $D$ (number of layers), width $W$ (number of neurons in each layer), and each layer's activation function, employing a uniform prior on all architecture hyper-parameters:

\begin{enumerate}	
\item Within each Occam category, identify a fully connected network architecture with the smallest depth and largest width. Among different choices of activation functions for the layer, select the one that yields the highest evidence $\pi_{evid} (\bs D | \bs \xi_{(\cdot)}^l, \hat{\mathcal{M}}^l, \mathcal{S})$.	

\item Sequentially add a fully connected layer and select the corresponding activation function based on evidence value.

\item Among all fully connected networks with different layers $M_{F(D,W)}^l$, choose the one with the largest model evidence and subject that model to appropriate network sparsification. The resulting network, with eliminated irrelevant neural connections (corresponding weight and bias parameters), is deemed the plausible network $M_I^l$ in this category.

\end{enumerate}

The proposed strategy combines bottom-up (adding layers) and top-down (removing connections) approaches, ensuring the retention of essential parameters and unveiling the sparsity pattern needed to capture multiscale interactions within a high-fidelity dataset.

\end{paragraph}

%++++++++++++++++++++++++++++++++++++++++++++++++++++++++++++++++++++++++
%++++++++++++++++++++++++++++++++++++++++++++++++++++++++++++++++++++++++
\section{Numerical Results}\label{sec:results}
\noindent
This section outlines numerical experiments on the OPAL-surrogate implementation for two high-fidelity physical simulations for problems in solid mechanics and computational fluid dynamics.
The first application focuses on the elastic deformation of porous materials through which we explore hierarchical Bayesian inference and the proposed network sparsification method. OPAL-surrogate is then applied to identify the BayesNN surrogate model, ensuring accuracy and reliability in predicting strain energy as QoI and facilitating forward uncertainty quantification for material systems with domain sizes beyond the capacity of high-fidelity physical simulations.
The second application focuses on the direct numerical simulation of turbulent combustion flow. Specifically, we apply OPAL-surrogate to determine BayesNN models for the interpolation prediction of combustion dynamics in shear-induced ablation of solid fuels within hybrid rocket motors.
\blue{
The implementation of Bayesian neural networks is built upon the Laplace Redux library \cite{daxberger2021laplace}, finite element solutions are conducted using the FEniCS library \cite{fenics}, and direct numerical simulations of turbulent combustion flow are performed using the ABLATE library \cite{ABLATE}.
}

%++++++++++++++++++++++++++++++++++++++++++++++++++++++++++++++++++++++++
\subsection{Elasticity in porous materials with random microstructure}\label{sec:porous}
\noindent
Our first application focuses on the deformation of porous silica aerogel, a high-performance insulation material for net-zero buildings, e.g., \cite{tan2022predictive, sarkar2024carbon, an2023flexible}.
\blue{The high-fidelity model is characterized by a stochastic partial differential equation (PDE) representing the elastic deformation of the two-phase material. The primary model parameter governing the deformation behavior is the Young's modulus of the solid aerogel phase $E_s$.
This model entails a stochastic microstructure indicator field, with samples derived from microstructural images of silica aerogel obtained from a lattice Boltzmann simulation of the foaming process; for further details, refer to \cite{bhattacharjee2023integrating}.
The finite element solution of the PDE utilizes a uniformly fine mesh to resolve the microstructure patterns accurately. The model output is defined as the stochastic strain energy of the material system,
\begin{equation}\label{eq:strain_energy}
    u_{\bs D} = \int_\Omega \bs T : \bs E \; d\bs y,  
\end{equation}
where $\bs T$ is the Cauchy stress and $\bs E$ is the strain tensors.
}

\begin{figure}[htpb]
    \centering
    \includegraphics[trim=0in 0in 0in 0in, clip, width=0.8\textwidth]{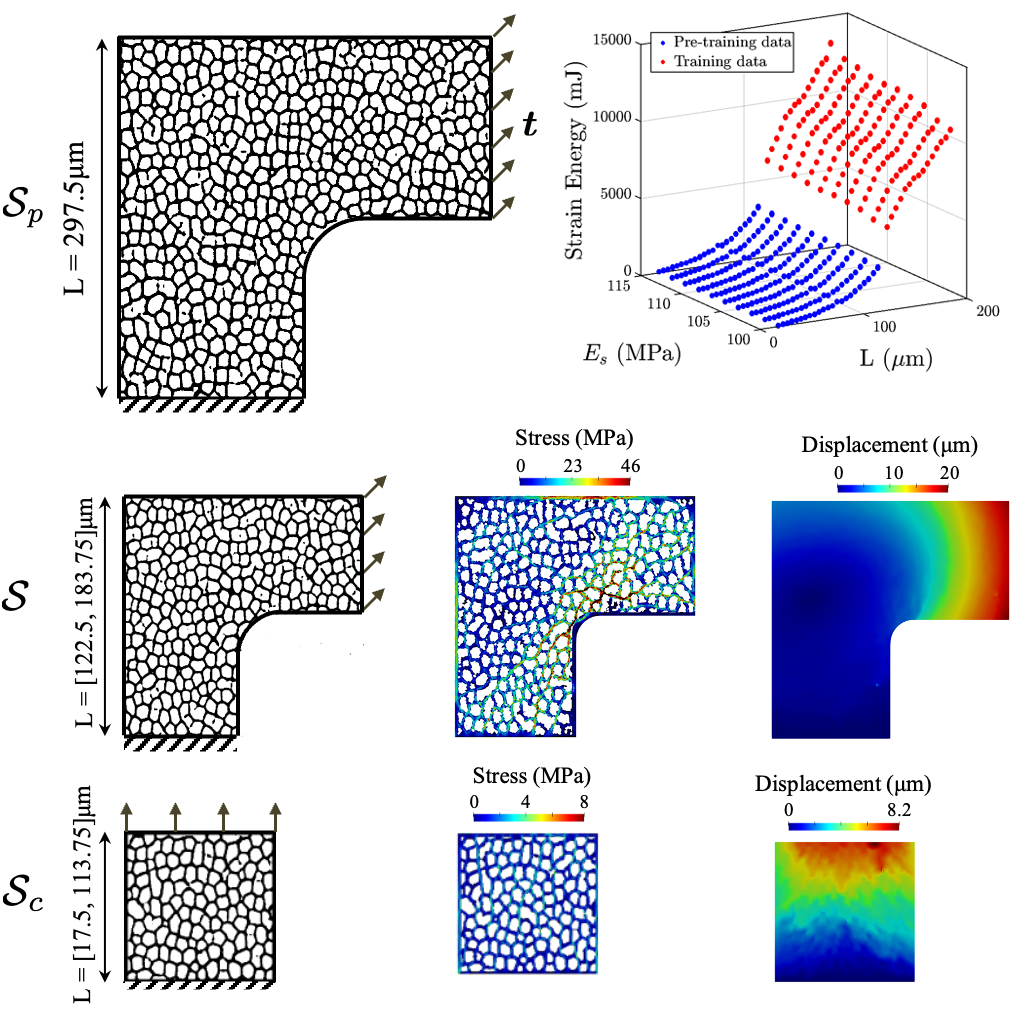}
    \vspace{-0.2in}
    \caption{
    Illustrations of the hierarchy of scenarios and high-fidelity data considered for the surrogate modeling of the elasticity problem:
    The pre-training scenario $\mathcal{S}_c$ consists of smaller domain sizes and under uniaxial loading conditions, while the training scenario $\mathcal{S}$ captures the mechanical features of the target prediction over domain sizes affordable for the high-fidelity model.
    The \blue{prediction scenario $\mathcal{S}_p$ } involves an aerogel insulation component with a size of $L = 297.5 \mu m$, and unobservable QoI is the strain energy.
    }
    \label{fig:aerogel}
\end{figure}

\blue{The prediction scenario $\mathcal{S}_p$ is illustrated at the top of Figure \ref{fig:aerogel} and involves an aerogel component with a characteristic length of $L = 297.5 \mu m$ and applied traction $\bs t$ = (70.71, 70.71) $N/m$ at the rightmost boundary.}
Given the computational demands associated with resolving microstructure patterns using a fine mesh, performing the high-fidelity simulation in this scenario is deemed computationally prohibitive. 
The objective of surrogate modeling is to approximate the solution of the high-fidelity simulation in $\mathcal{S}_p$ and enable the quantification of uncertainty in the strain energy, i.e., \textit{unobservable QoI}, stemming from the stochastic microstructure  and uncertainties in elasticity parameters.
We thus adhere to the notation of the prediction pyramid, as elucidated in \textit{Remark 3} in Section \ref{sec:opal_steps}, and adopt a hierarchy of scenarios for conducting high-fidelity simulations to generate data. As depicted in Figure \ref{fig:aerogel}, the pre-training scenario $\mathcal{S}_c$ entails small domains subjected to uniaxial loads, while the training scenario $\mathcal{S}$ encompasses larger domains specifically designed to capture physical features of the QoI in the prediction scenario, such as stress localization and loading conditions.
The pre-training dataset $\bs D_c$ comprises strain energies computed from 22 equidistant domain sizes within the range of $L = [17.5, 113.75] \mu m$, 10 equidistant elastic moduli within $E_s = [100, 115]$ MPa, and 10 realizations of the aerogel microstructure patterns.
The training set $\bs D$ includes strain energies from high-fidelity simulations obtained at 15 domain sizes within $L = [122.5, 183.75] \mu m$, 10 elastic moduli, and 10 microstructure realizations, resulting in a total of 3800 data points in the pre-training and training sets.
To assess extrapolation predictions, we consider the leave-out set $\bs D_{LO}$ containing data points with domain sizes $L_{LO} = \{179.4, 183.8\} \mu m$.
As the validation observables,
we take the product of the strain energy and Young's modulus at a specific domain length $L_{LO}$, evaluated using the high-fidelity simulations and surrogate models, expressed as,
\begin{equation}\label{eq:observ}
    \mathcal{Z}_D = \int_{E_s^{low}}^{E_s^{up}} u_{\bs D}|_{L=L_{LO}} \; d E, 
    \quad\quad
    \mathcal{Z}_M = \int_{E_s^{low}}^{E_s^{up}} u_{\bs \theta}|_{L=L_{LO}} \; d E,
\end{equation}
where $E_s^{low} = 100$ MPa and $E_s^{up} = 115$ MPa are the range of values for the elastic modulus within the training set.
Given the stochastic nature of both observables, we employ two validation measures \cite{maupin2018validation}, the normalized Kullback-Leibler divergence by the Shannon entropy,
\begin{equation}
    \mathbbm{d}_{DKL} = \frac{\mathcal{D}_{\rm KL} \left(\pi(\mathcal{Z}_D), \pi(\mathcal{Z}_M)\right)}{\mathcal{H}\left(\pi(\mathcal{Z}_D)\right)},
\end{equation}
and discrepancies between cumulative distribution functions,
\begin{equation}
    \mathbbm{d}_{CDF} = 
    \int_{-\infty}^{+\infty} | \Phi(\mathcal{Z}_D) - \Phi(\mathcal{Z}_M) | d\mathcal{Z}.
\end{equation}

%+++++++++++++++++++++++++++
\subsubsection{Illustrative 1D example}\label{sec:porous_evid}
\noindent
Prior to applying the OPAL-surrogate framework, we conduct a preliminary exploration of the methodologies detailed in Section \ref{sec:hierarchical} using a simple 1D example. In this example, the training data $\bs D$ consists of the strain energy calculations with $E_{s}$ = 100 MPa over the domain sizes of both pre-training and training sets outlined in the preceding section.

%-----------------
\begin{paragraph}{Model plausibility vs. network architecture}
\begin{figure}[htpb]
\vspace{-0.3in}
    \centering
    \includegraphics[trim=1in 0in 0in 0in, clip, width=0.48\textwidth]{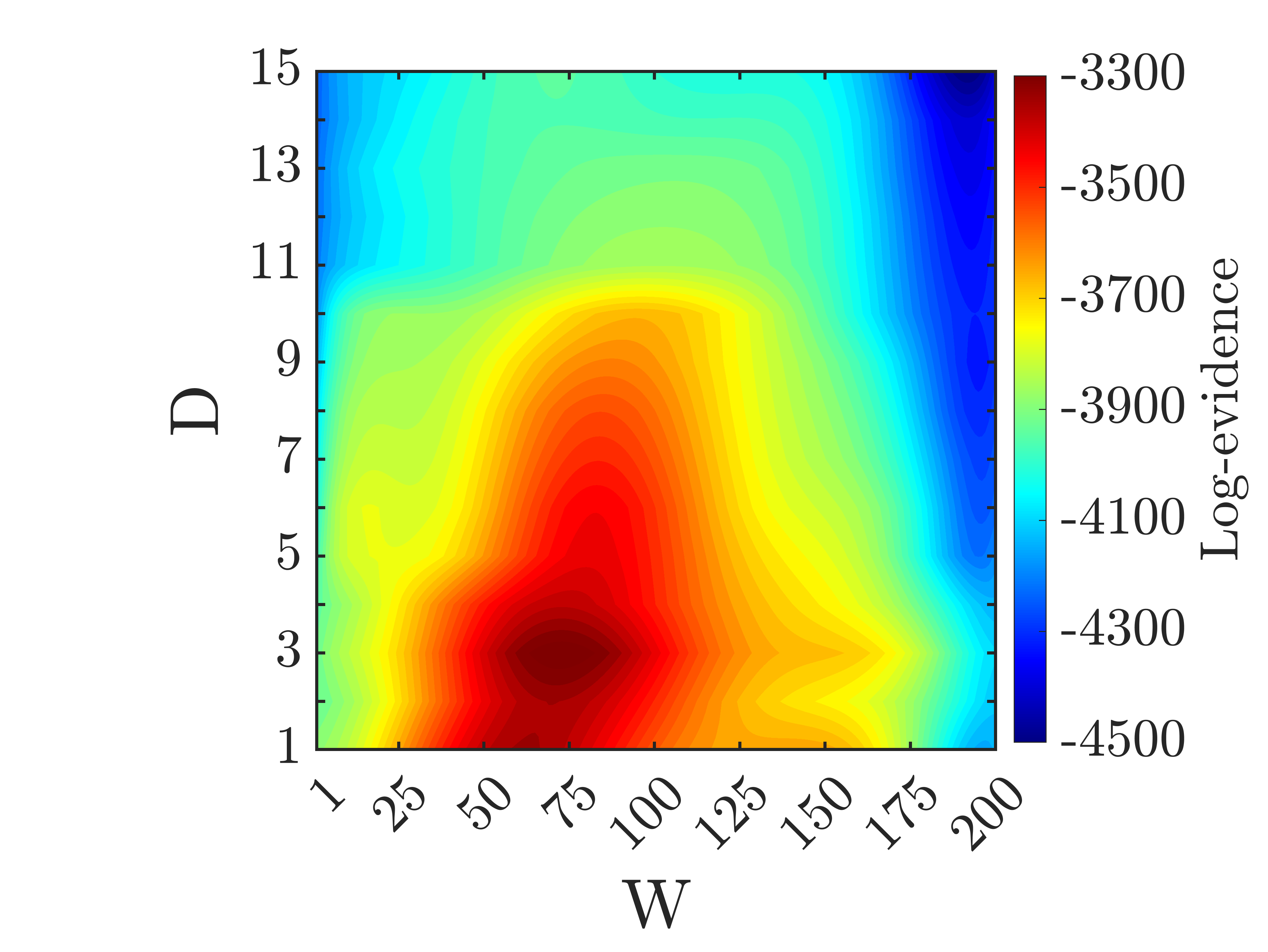}
    ~
    \includegraphics[trim=1in 0in 0in 0in, clip, width=0.48\textwidth]{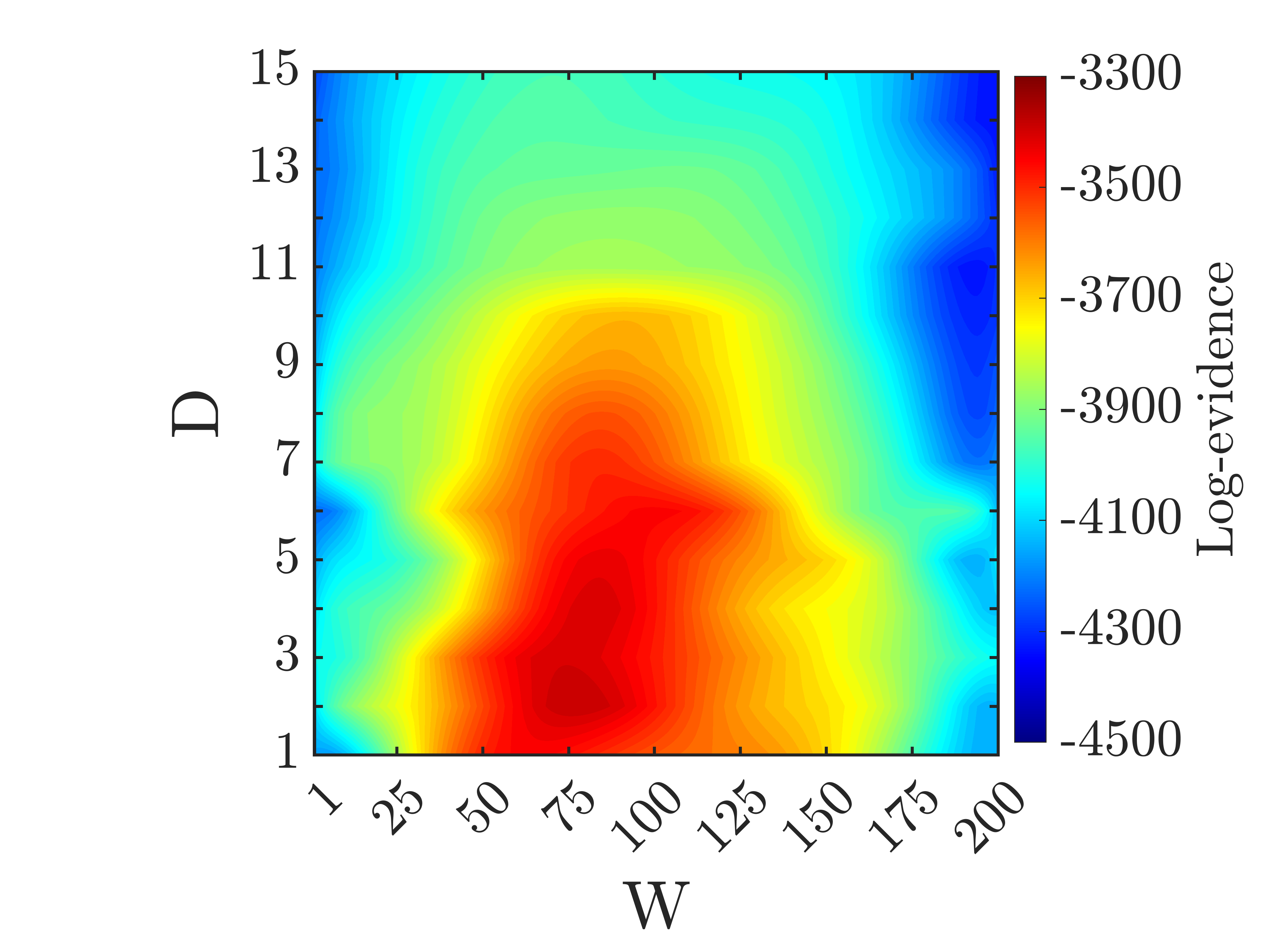}
    \\ (a) \hspace{2.2in} (b)\\
    \vspace{-0.05in}
    \includegraphics[trim=1in 0in 0in 0in, clip, width=0.48\textwidth]{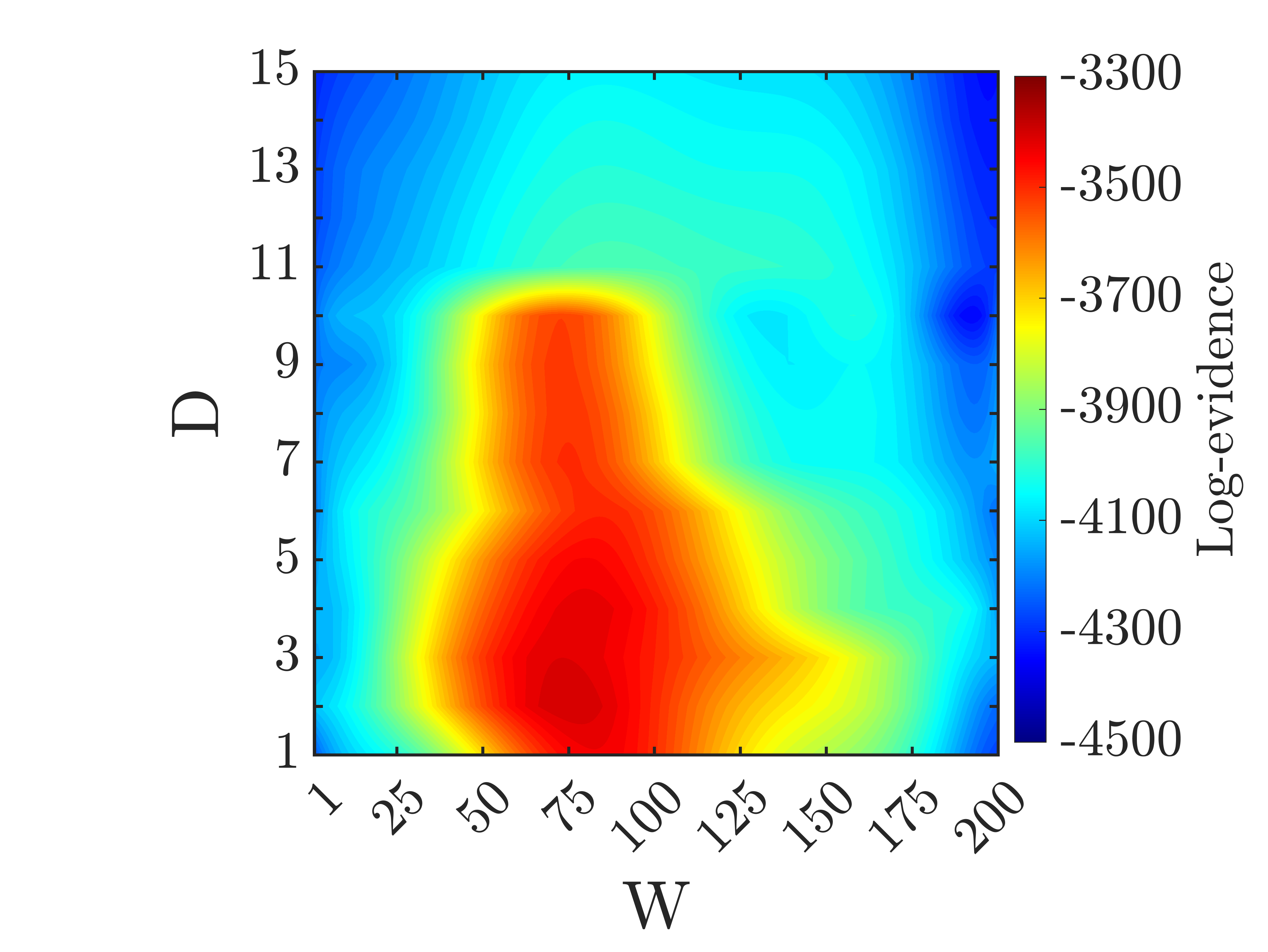}
    ~
    \includegraphics[trim=1in 0in 0in 0in, clip, width=0.48\textwidth]{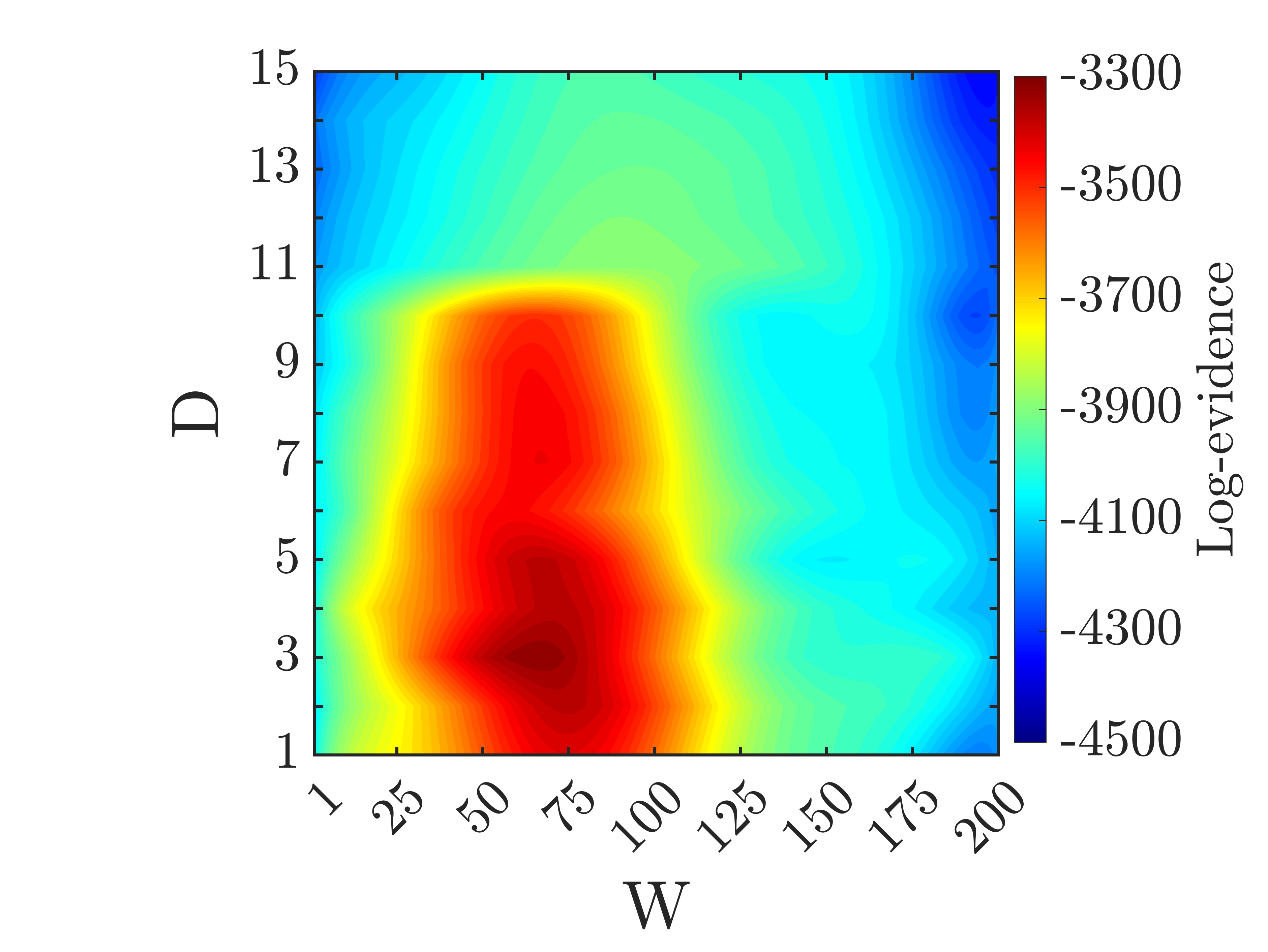}
    \\ (c) \hspace{2.2in} (d)
    \vspace{-0.05in}
    \caption{
    Illustrative 1D example:
    The log-evidence $\ln \pi_{evid} (\bs D| \bs \xi_i)$, corresponding to posterior model plausibilities $\rho_i$ under the uniform prior $\pi_{pr}(\bs \xi_i)$ assumption, for fully connected BayesNN models with varying number of layers $D$ and neurons in each layer $W$, and activation functions:
    (a) \textit{Tanh}, (b) \textit{Leaky ReLU}, (c) \textit{Sigmoid}, and (d) \textit{ReLU}.
    }
    \vspace{-0.1in}
    \label{fig:1d_evid_grid}
\end{figure}
Figure \ref{fig:1d_evid_grid} illustrates the log-evidence $\ln \pi_{evid} (\bs D| \bs \xi_i)$ for various fully connected neural networks with different depths $D$ (number of layers), widths $W$ (number of neurons in each layer), and activation functions. Considering uniform prior probabilities $\pi_{pr}(\bs \xi_i)$ for each model, the log-evidence values are equivalent to the posterior model plausibilities $\rho_i$.
As depicted in these plots, a specific range of architectural hyperparameters leads to higher plausibilities, implying that the corresponding models have a higher probability of accurately representing the dataset. However, overly simplistic models, characterized by smaller $D$ and $W$, exhibit limited predictive power within the dataset space. Conversely, excessively complex models with larger network parameters can capture a broader range of possible observations with low-confidence predictions within the dataset space.
These aspects are illustrated in Figure \ref{fig:1d_data_model}, depicting the mean and 95\% credible interval (CI) uncertainty in predictions from BayesNN models with \textit{Tanh} activation functions in comparison to the training data.
\blue{
As anticipated, owing to the extrapolation constraints inherent in data-driven surrogate models, model predictions consistently exhibit higher uncertainty beyond the data range compared to within the training data, across all architectures depicted in this figure. Models characterized by higher log evidence (Figure \ref{fig:1d_data_model}(a) and (b)) present more accurate and reliable approximations for both interpolation and extrapolation relative to those with lower log evidence (Figure \ref{fig:1d_data_model}(c) and (d)). For example, at $L=150 (\mu m)$ and $L=250 (\mu m)$, the standard deviations of model predictions for the model in Figure \ref{fig:1d_data_model}(b) are 153.65 and 230.25, respectively, whereas, for the model depicted in Figure \ref{fig:1d_data_model}(d), these values are 192.5 and 331.25.
}
\begin{figure}[htpb]
    \centering
    \includegraphics[trim=0.9in 0in 0.9in 0in, clip, width=0.48\textwidth]{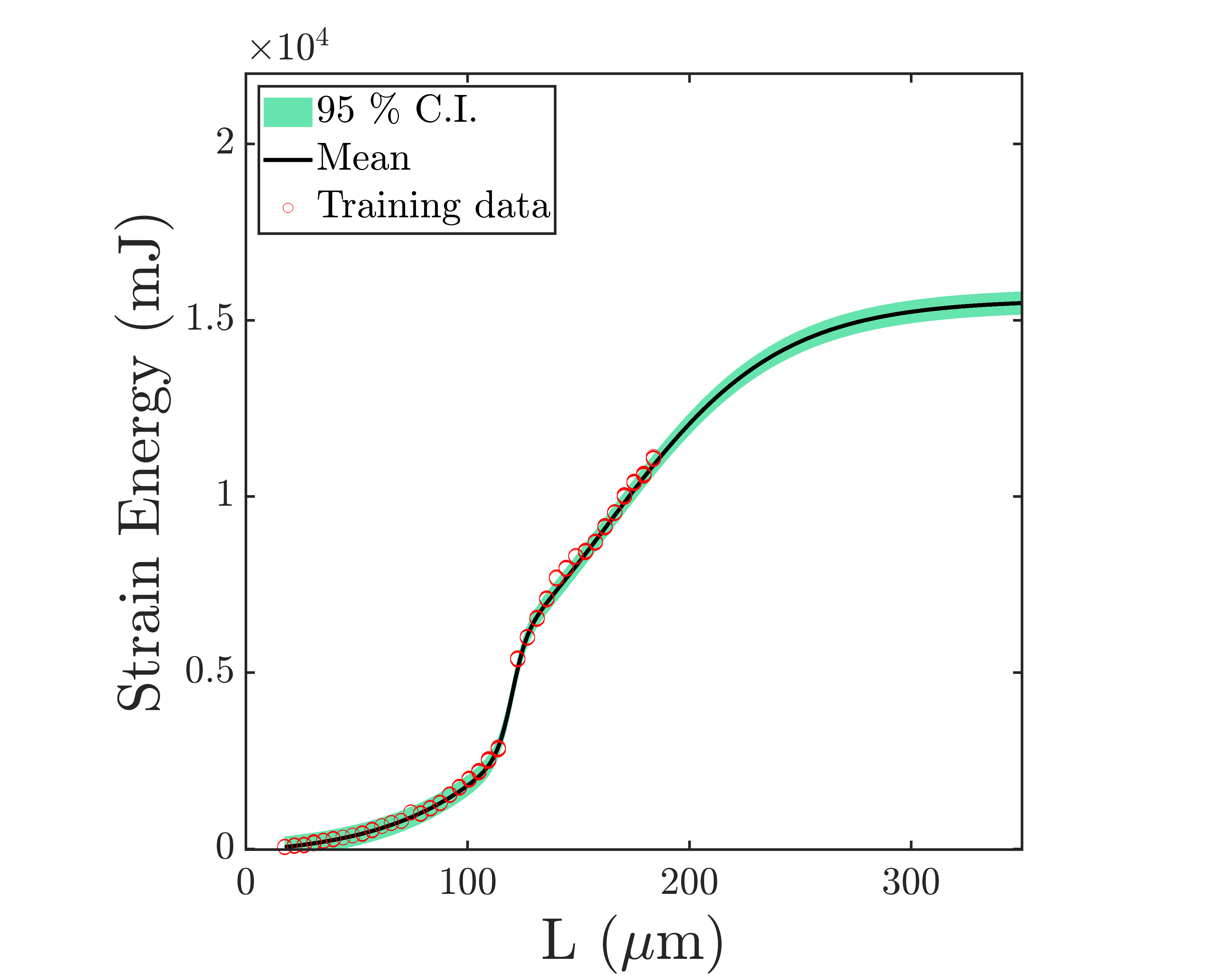}
    ~
    \includegraphics[trim=0.9in 0in 0.9in 0in, clip, width=0.48\textwidth]{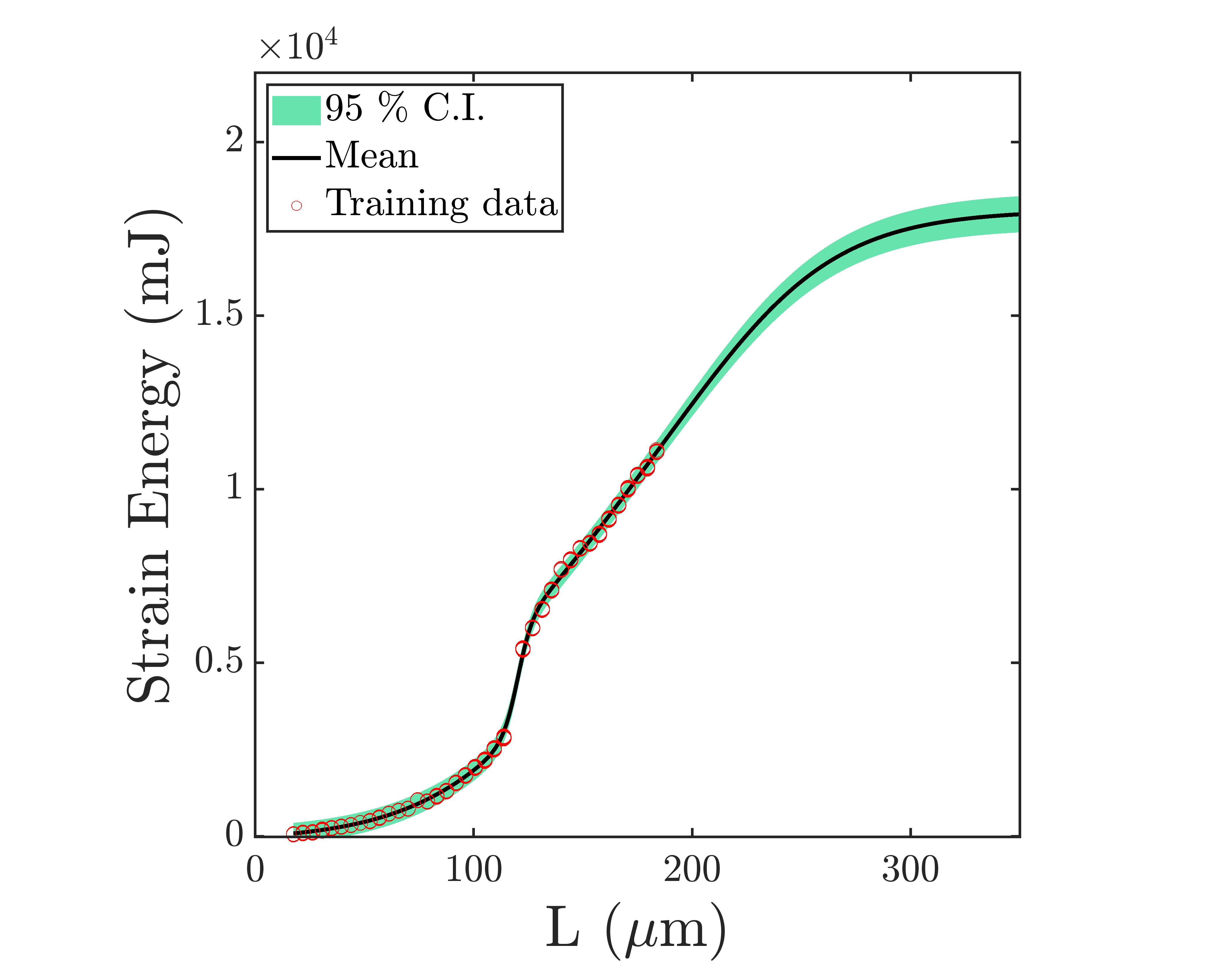}
    \\ (a) \hspace{2.2in} (b)\\
    \includegraphics[trim=0.9in 0in 0.9in 0in, clip, width=0.48\textwidth]{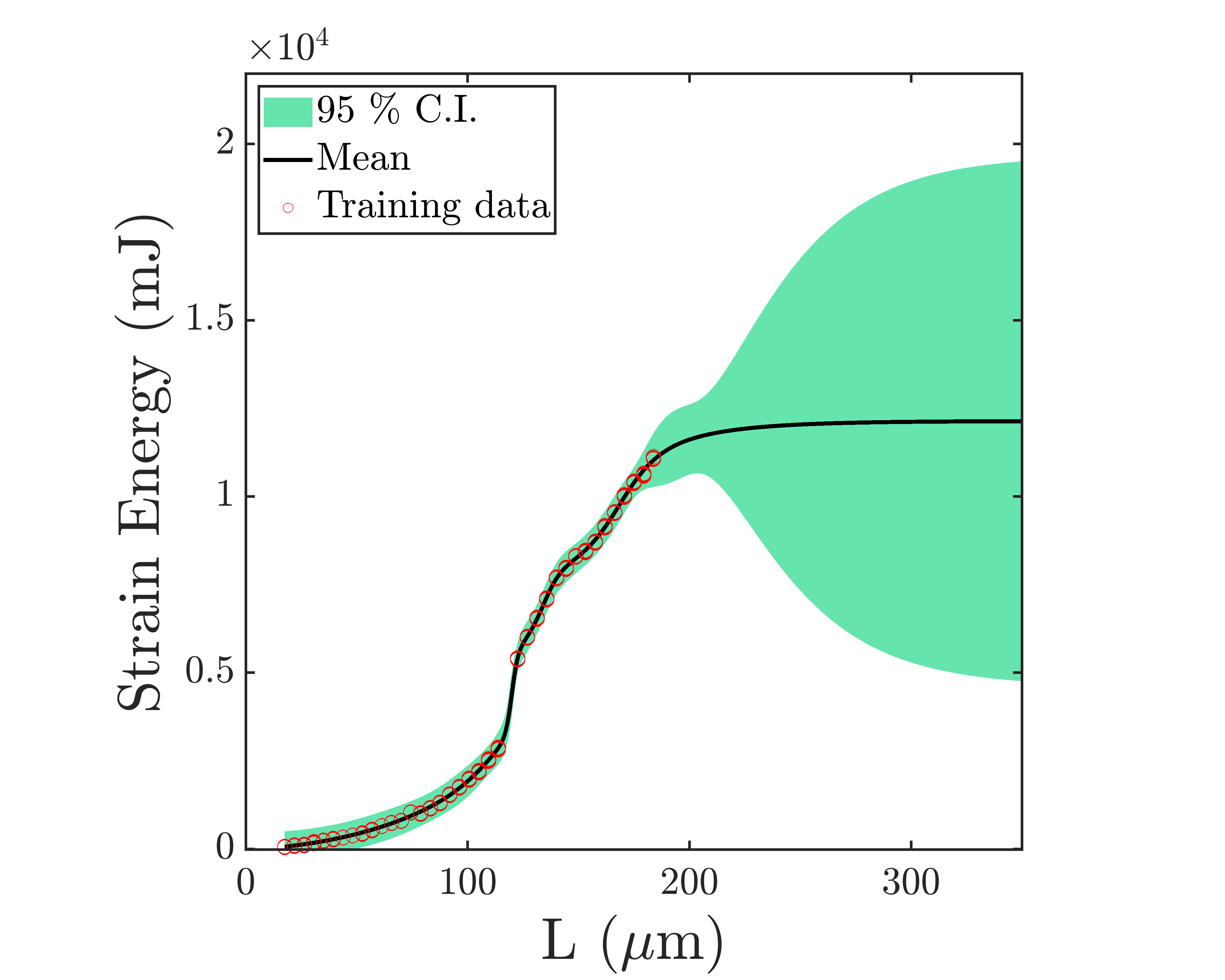}
    ~
    \includegraphics[trim=0.9in 0in 0.9in 0in, clip, width=0.48\textwidth]{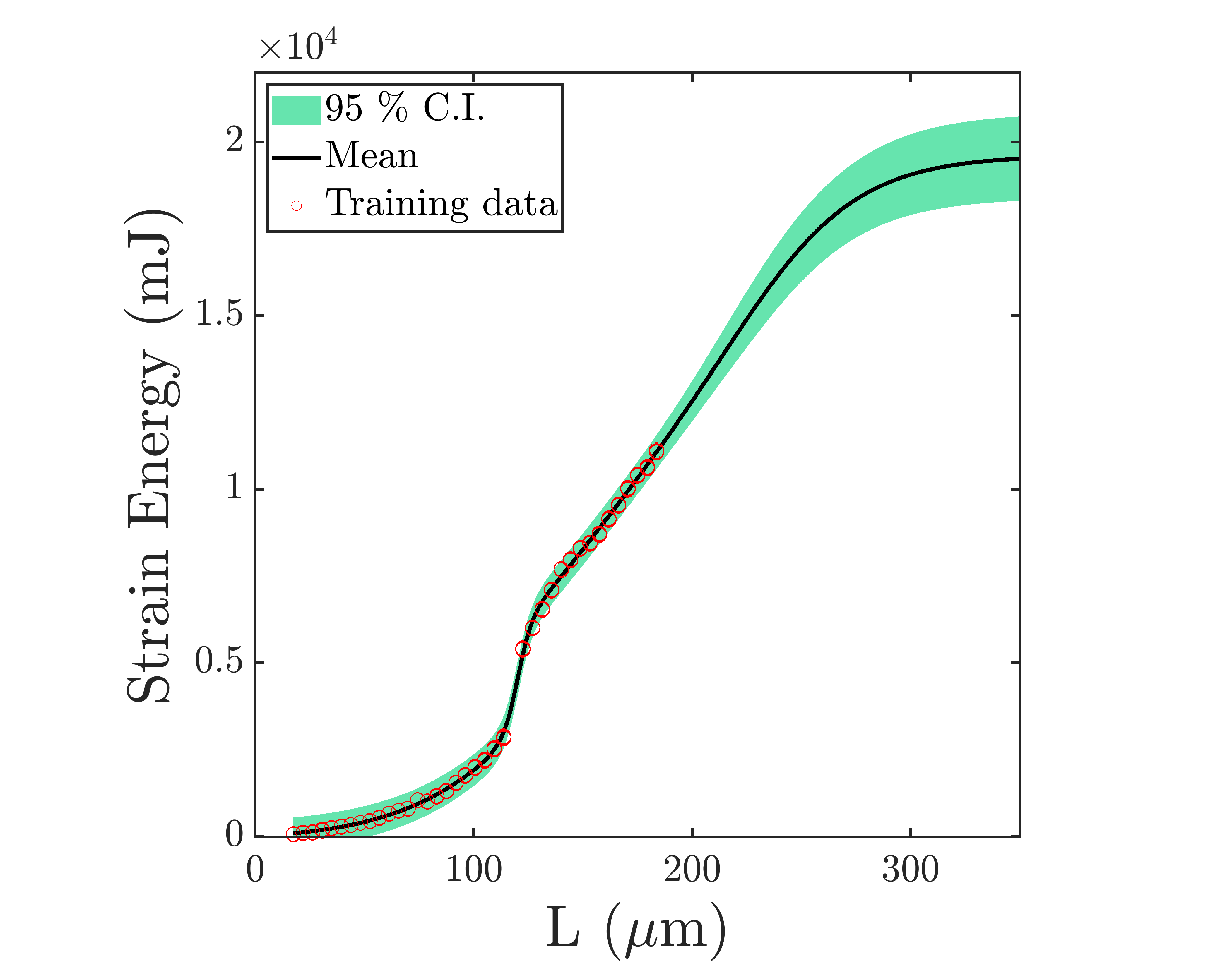}
    \\ (c) \hspace{2.2in} (d)
    \vspace{-0.1in}
    \caption{
    Illustrative 1D example:
    The mean and uncertainty predictions for different fully connected BayesNN models with \textit{Tanh} activation functions in Figure \ref{fig:1d_evid_grid}(a) compared to the training data:
    (a) $D=3$, $W=720$, $\ln \pi_{evid} (\bs D| \bs \xi) = -3354$ (highest posterior model plausibility);
    (b) $D=3$, $W=800$,  $\ln \pi_{evid} (\bs D| \bs \xi) = -3468$;
    (c) $D=2$, $W=1450$, $\ln \pi_{evid} (\bs D| \bs \xi) = -3877$;
    (d) $D=1$, $W=1840$, $\ln \pi_{evid} (\bs D| \bs \xi) = -3926$.
    }
    \label{fig:1d_data_model}
\end{figure}
\blue{It is crucial to emphasize that model plausibility alone does not guarantee prediction credibility, underscoring the importance of validation tests, as discussed in Section \ref{sec:opal_steps}, to rigorously assess the robustness of surrogate model predictions.}

\end{paragraph}

%-----------------
\begin{paragraph}{Network sparsification}
Figure \ref{fig:1d_sparse_D4} demonstrates the effectiveness of the sparsification method detailed in Section \ref{sec:sparse} in enhancing model plausibility by eliminating irrelevant network parameters. Gradually increasing $TOL_{\bs\theta}$ initially increases the model evidence, followed by a decline due to excessive sparsification. The reported threshold values correspond to the maximum model evidence.
In Figure \ref{fig:1d_sparse_D4} (a,b), the fully connected network exhibits a plateau in extrapolation with significant uncertainty. Sparsification reduces uncertainty (13.23\% in prediction variance) by eliminating irrelevant parameters, albeit with limited enhancement in prediction accuracy. Conversely, Figure \ref{fig:1d_sparse_D4} (c,d) shows that network sparsification improves accuracy in extrapolation predictions by maintaining the trend in training data for larger values of $L$. While these figures depict two representative cases, additional experiments indicate that the proposed network sparsification generally enhances both accuracy and reliability in BayesNN models.

\begin{figure}[htpb]
    \centering
    \includegraphics[trim=0.9in 0in 0.9in 0in, clip, width=0.48\textwidth]{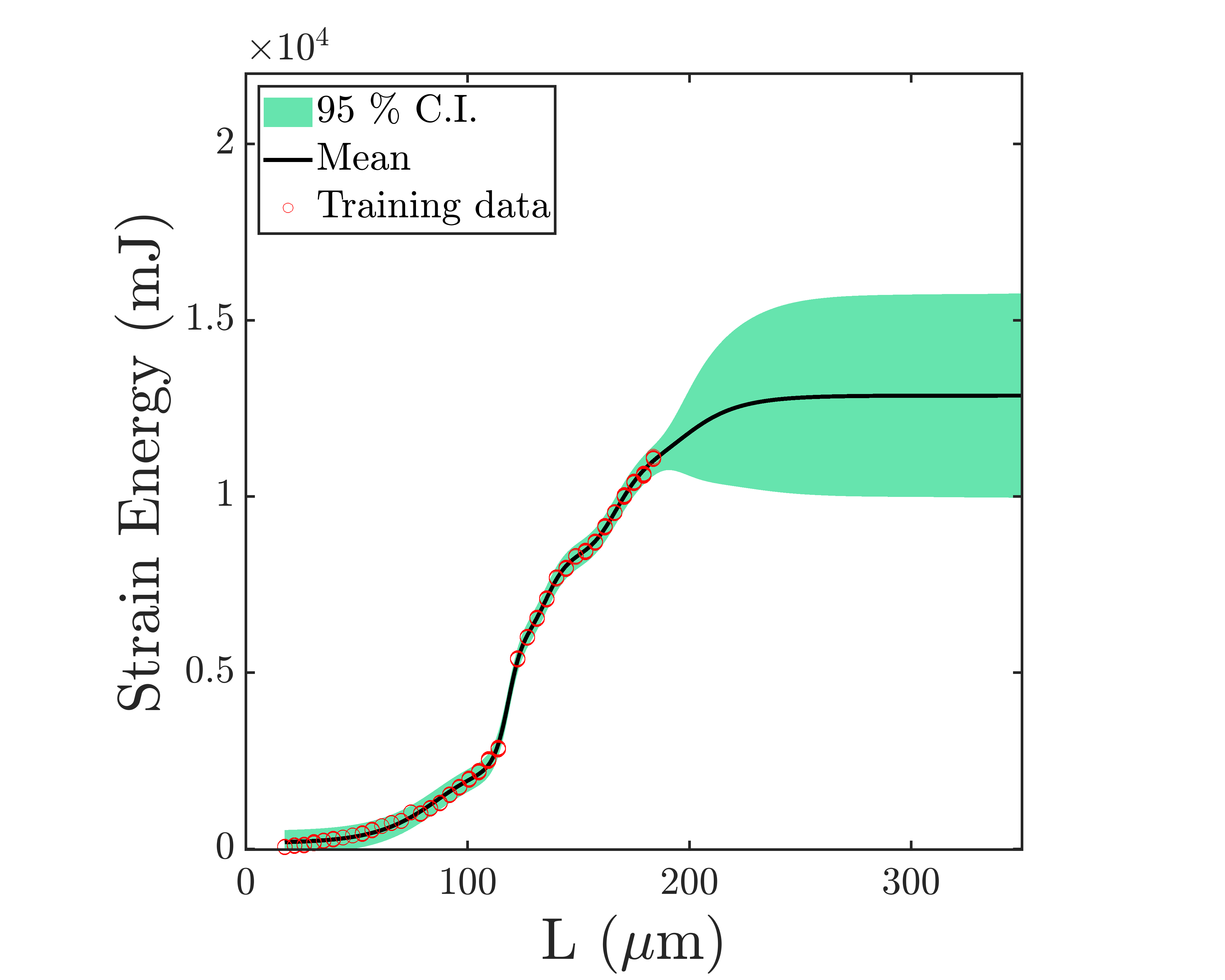}
    ~
    \includegraphics[trim=0.9in 0in 0.9in 0in, clip, width=0.48\textwidth]{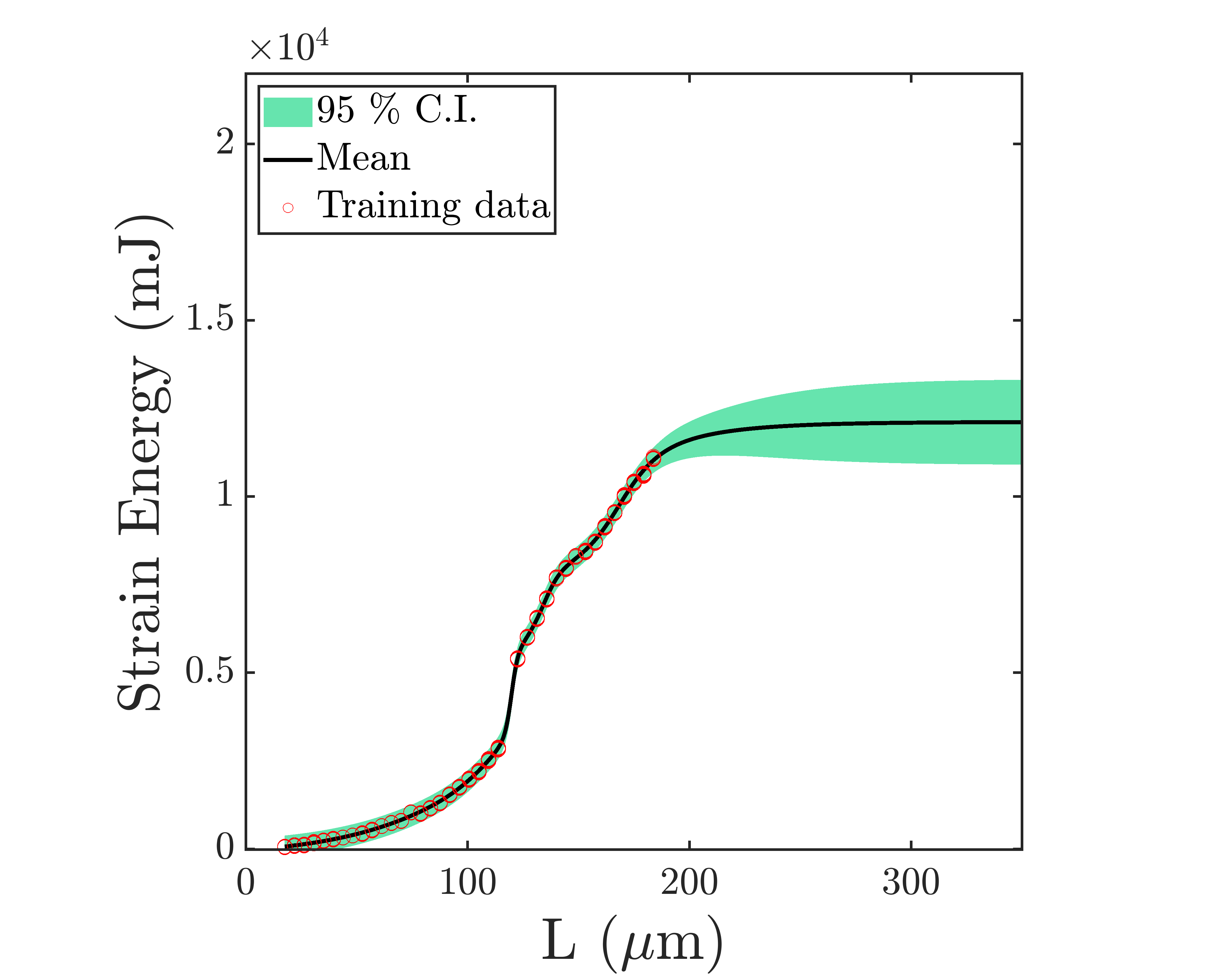}
    \\ (a) \hspace{2.2in} (b)\\
    % \includegraphics[trim=0in 0in 0in 0in, clip, width=0.48\textwidth]{draft_figs/1d_sprase_mean_D4.png}
    % ~
    % \includegraphics[trim=0in 0in 0in 0in, clip, width=0.48\textwidth]{draft_figs/1d_sprase_var_D4.png}
    % \\ (c) \hspace{2.2in} (d)\\
    \includegraphics[trim=0.9in 0in 0.9in 0in, clip, width=0.48\textwidth]{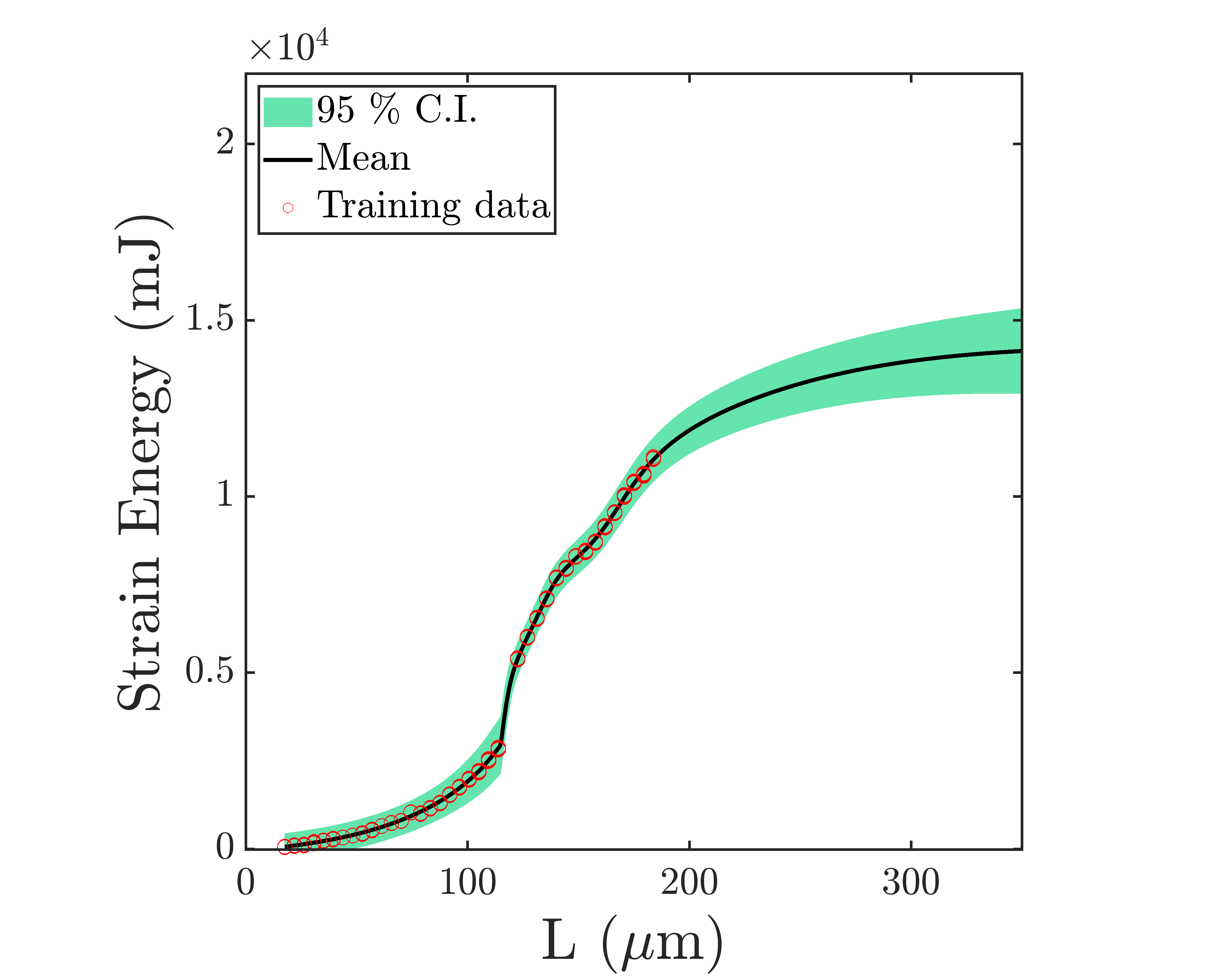}
    ~
    \includegraphics[trim=0.9in 0in 0.9in 0in, clip, width=0.48\textwidth]{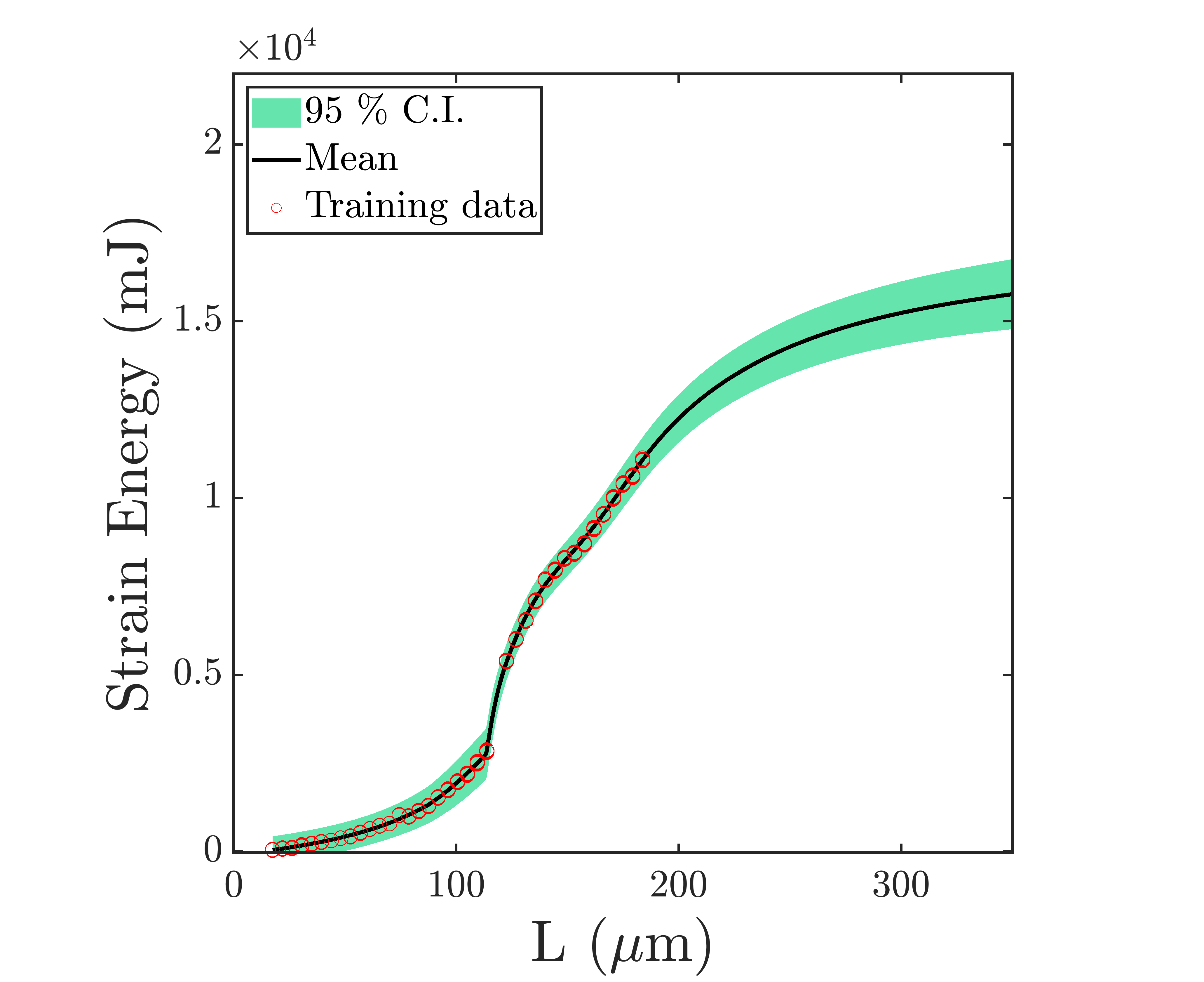}
    \\ (e) \hspace{2.2in} (f)  
    \vspace{-0.1in}
    \caption{
    Illustrative 1D example: 
    Network sparsification results for the BayesNN models.
    The mean and uncertainty predictions for
    (a) the fully connected network with $D=4$, $W=1600$, and $\ln \pi_{evid} (\bs D| \bs \xi) = -3821$, and 
    (b) the corresponding sparsified network using $TOL_{\bs\theta} = 0.025$, resulting in the elimination of 22\% of the parameters and yielding $\ln \pi_{evid} (\bs D| \bs \xi) = -3610$.
    (c) The fully connected network with $D=6$, $W=1700$, and $\ln \pi_{evid} (\bs D| \bs \xi) = -3874$, and 
    (d) the corresponding sparsified network using $TOL_{\bs\theta} = 0.05$, resulting in the elimination of 32\% of the parameters and yielding $\ln \pi_{evid} (\bs D| \bs \xi) = -3545$.
    }
    \label{fig:1d_sparse_D4}
\end{figure}

\end{paragraph}

%\clearpage
%+++++++++++++++++++++++++++
\subsubsection{OPAL-surrogate demonstration}\label{sec:porous_opal}
\noindent
This section presents the results of identifying the credible surrogate model for predicting the unobservable QoI, the strain energy of the prediction scenario $\mathcal{S}_p$, in the elasticity problem, as illustrated in Figure \ref{fig:aerogel}.
For training the surrogate models and computing the evidence and plausibility, we first infer network parameters using pre-training data $\bs D_c$ based on \eqref{eq:lev1} and \eqref{eq:lev2}. The resulting posterior serves as the prior for network parameters in hierarchical Bayesian inferences using training data $\bs D$ to evaluate posterior model plausibility.

\begin{paragraph}{Determining the initial model set $\mathcal{M}$ and categorization}
In defining a large model space, we use the model evidence (model plausibility) of BayesNNs with a single layer ($D=1$), varying widths ($W = [1,1200]$), and four activation functions, as shown in Figure \ref{fig:opal_grid}. We set the upper width limit at $W=600$ in the initial model set $\mathcal{M}$, representing a 10\% increase over the peak average log-evidences for different functions.
Numerical experiments indicate that selecting the upper bound within the range of $W=500$ and $W=800$ minimally impacts the validity and performance of the model identified by OPAL-surrogate, indicating the effectiveness of the proposed incremental search for initializing OPAL-surrogate.
Accordingly, we categorize the BayesNN models with $W = [1, 600]$ based on the number of layers as a measure of model complexity, such that Category 1 encompassing networks with $D = \{1,2\}$, Category 2 including those with $D = \{3,4\}$, and so forth.
\begin{figure}[htpb]
    \centering
    \includegraphics[trim=0.5in 0in 1.3in 0in, clip, width=.45\textwidth]{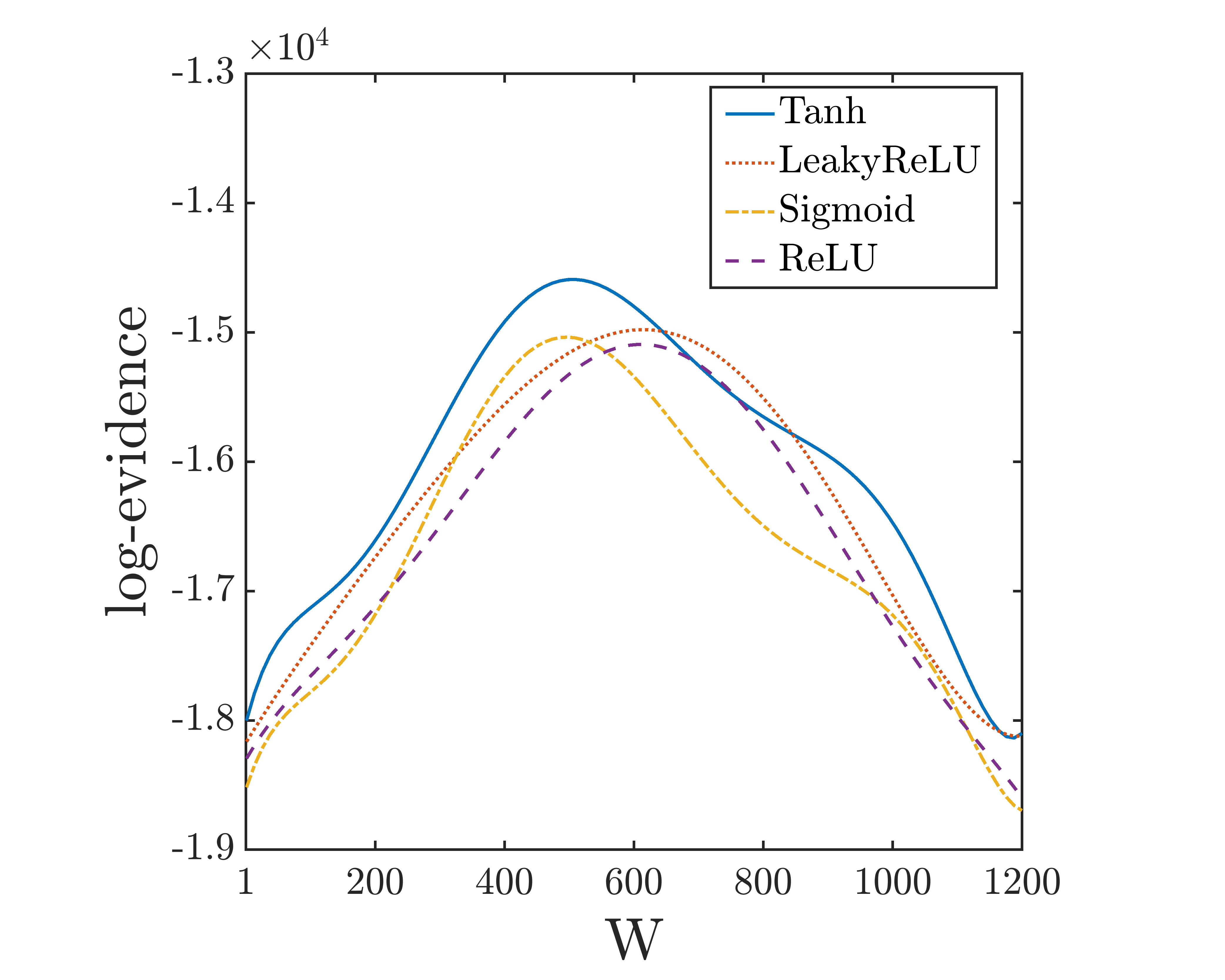}
    \vspace{-0.1in}
    \caption{
    Elasticity problem:
    The log-evidence (corresponding to model plausibility $\rho$) for BayesNN models with single layer ($D=1$) with different widths and activation functions.
    }
    \label{fig:opal_grid}
\end{figure}

\end{paragraph}

\vspace{-0.15in}
\begin{table}[h!]
\centering
\caption{Elasticity problem: Identifying the layerwise activation functions based on log-evidence $\pi_{evid} (\bs D |$ $\bs \xi_{(\cdot)}^l, \bs D_c, \hat{\mathcal{M}}^l, \mathcal{S}, \mathcal{S}_c)$. All models are fully connected networks $M^l_{F(D, W=600)}$ with depth $D$ and width $W=600$. }
\vspace{0.1in}
\begin{tabular}{cccccc}
\hline
Occam                & BayesNN      & \multicolumn{4}{c}{Log-evidence}       \\ \cline{3-6} 
categories           & model     & ReLU   & Leaky ReLU & Sigmoid & Tanh   \\ \hline
\multirow{2}{*}{$l=1$} & $D=1$   & -15732    & -15045        & -15317     & \textbf{-14846} \\
                       & $D=2$   & -15320 & -15246     & -15314  & \textbf{-14924} \\ \hline
\multirow{2}{*}{$l=2$} & $D=3$   & -15835 & \textbf{-14834}     & -15427  & -15340 \\
                        & $D=4$   & -15723 & -15256     & -15417  & \textbf{-15012} \\ \hline
% \multirow{2}{*}{$l=3$} & $D=5$   & -15946 & -15263     & -15529  & \textbf{-15036} \\
%                        & $D=6$   & -15630 & -15384     & -15435  & \textbf{-15133} \\ \hline
% \multirow{2}{*}{$l=4$} & $D=7$   & -15962 & -15187     & -15517  & \textbf{-15136} \\
%                        & $D=8$   & -15534 & -15215     & -15383  & \textbf{-15046} \\ \hline
\end{tabular}
\label{table:layer_act}
\end{table}

\begin{paragraph}{Occam Category 1}
Table \ref{table:layer_act} shows the values of model evidence $\pi_{evid}$$(\bs D |$ $\bs \xi_{(\cdot)}^l, \bs D_c, \hat{\mathcal{M}}^l, \mathcal{S}, \mathcal{S}_c)$ for fully connected networks $M^l_{F(D, W=600)}$ with different activation functions.
Accordingly, the \textit{Tanh} function is selected for the first and second layers.
Upon applying sparsification to $M^1_{F(D=1, W=600)}$, the plausible model in this category $M^1_I$ is identified that consists of 564 connections.
According to Table \ref{table:opal_elasticity}, the sparsification results in
elimination of 6\% of the parameters and
0.6\% improvement in log-evidence in $M^1_I$ compared to $M^1_{F(D=1, W=600)}$.
Figure \ref{fig:opal_cat1} illustrates the surrogate model predictions of both the fully connected and sparsified networks in comparison to the pre-training and training datasets. The observed increase in model prediction uncertainty at lower $E_s$ is attributed to higher uncertainty in high-fidelity simulation data at smaller elastic modulus values captured by the surrogate model.
\begin{figure}[htpb]
\vspace{-0.1in}
    \centering
    \includegraphics[trim=0.5in 0in 0.5in 0in, clip, width=.45\textwidth]{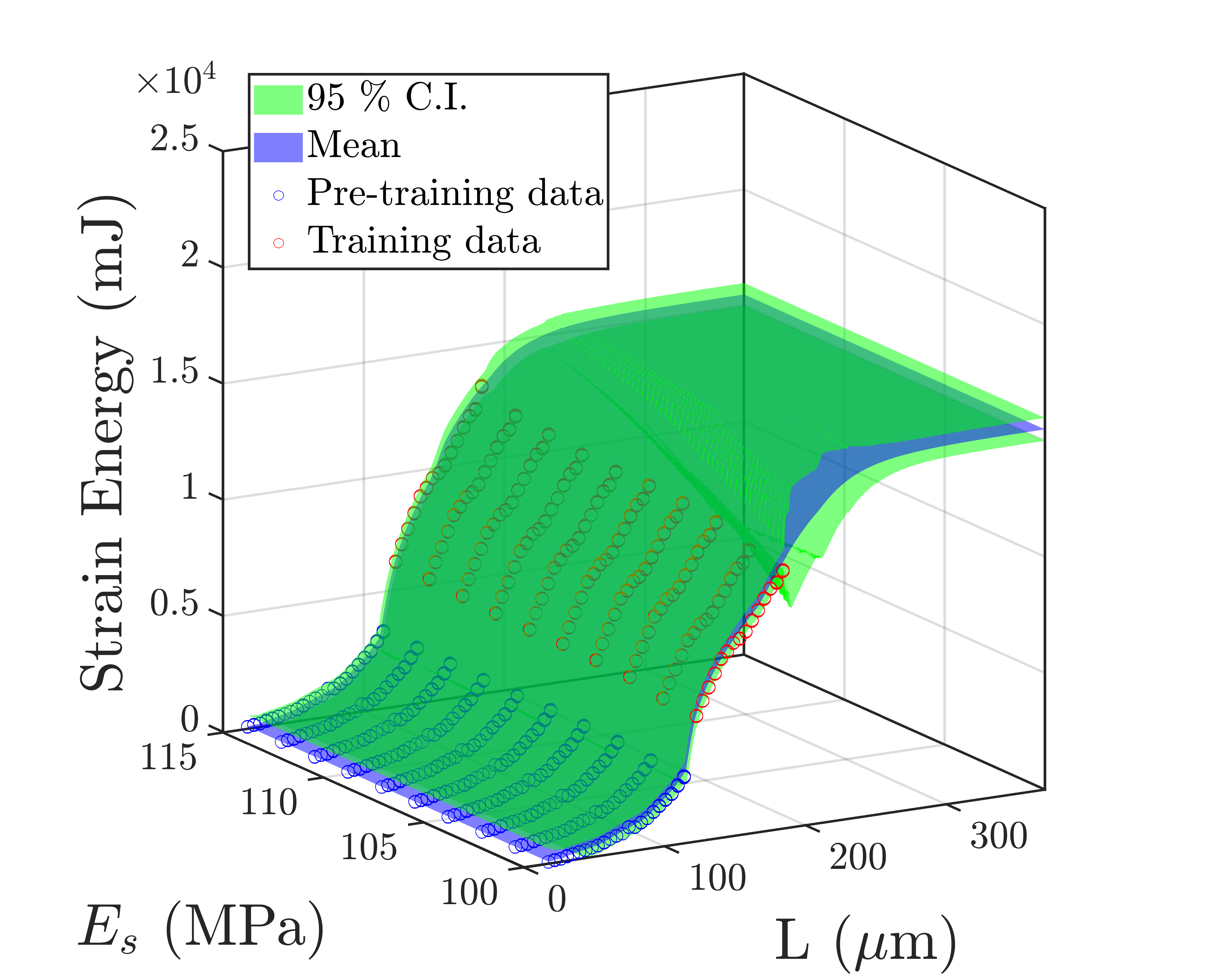}
    ~
    \includegraphics[trim=0.5in 0in 0.5in 0.0in, clip, width=.45\textwidth]{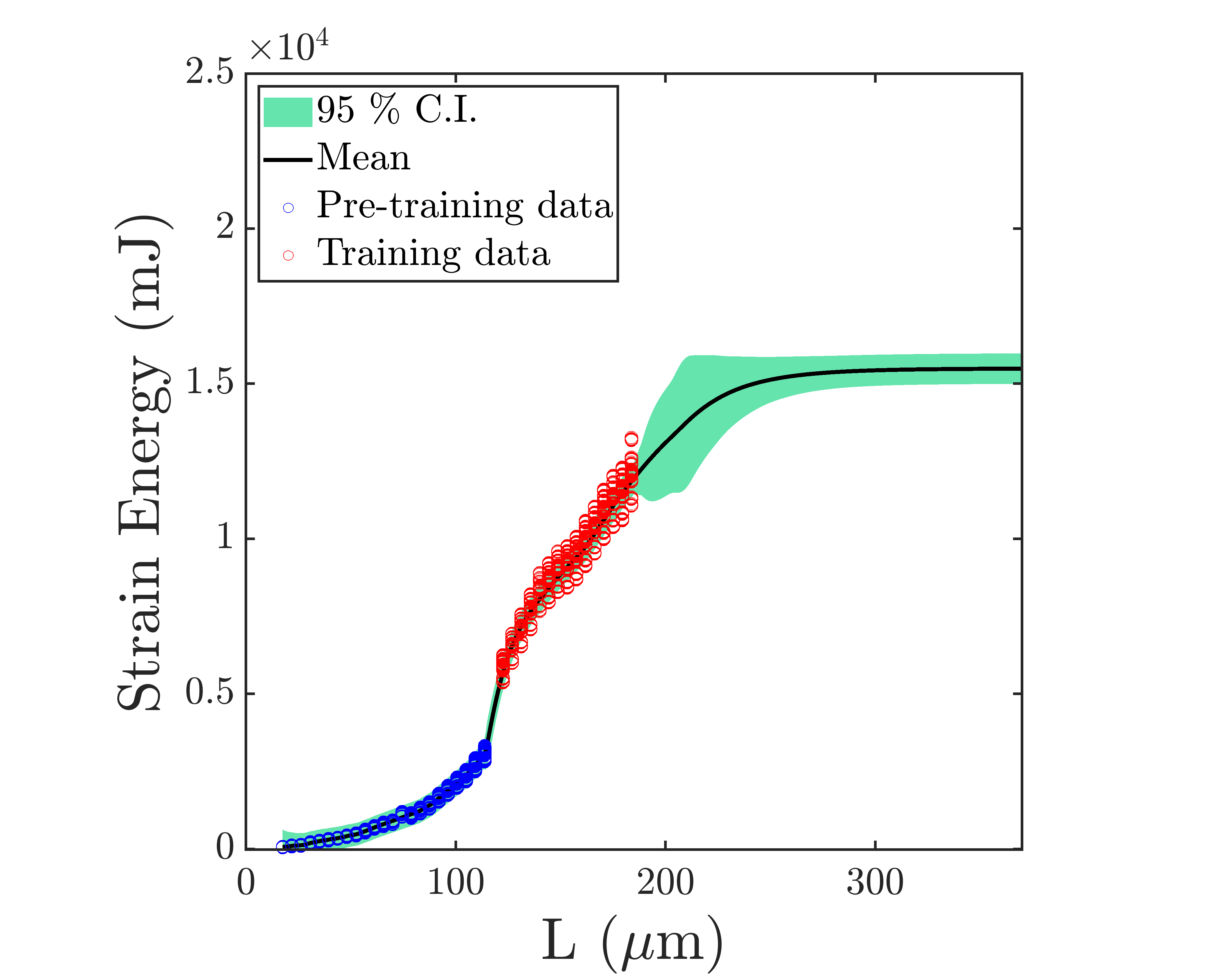}
    \\ (a) \\
\vspace{-0.0in}    
    \includegraphics[trim=0.5in 0in 0.5in 0in, clip, width=.45\textwidth]{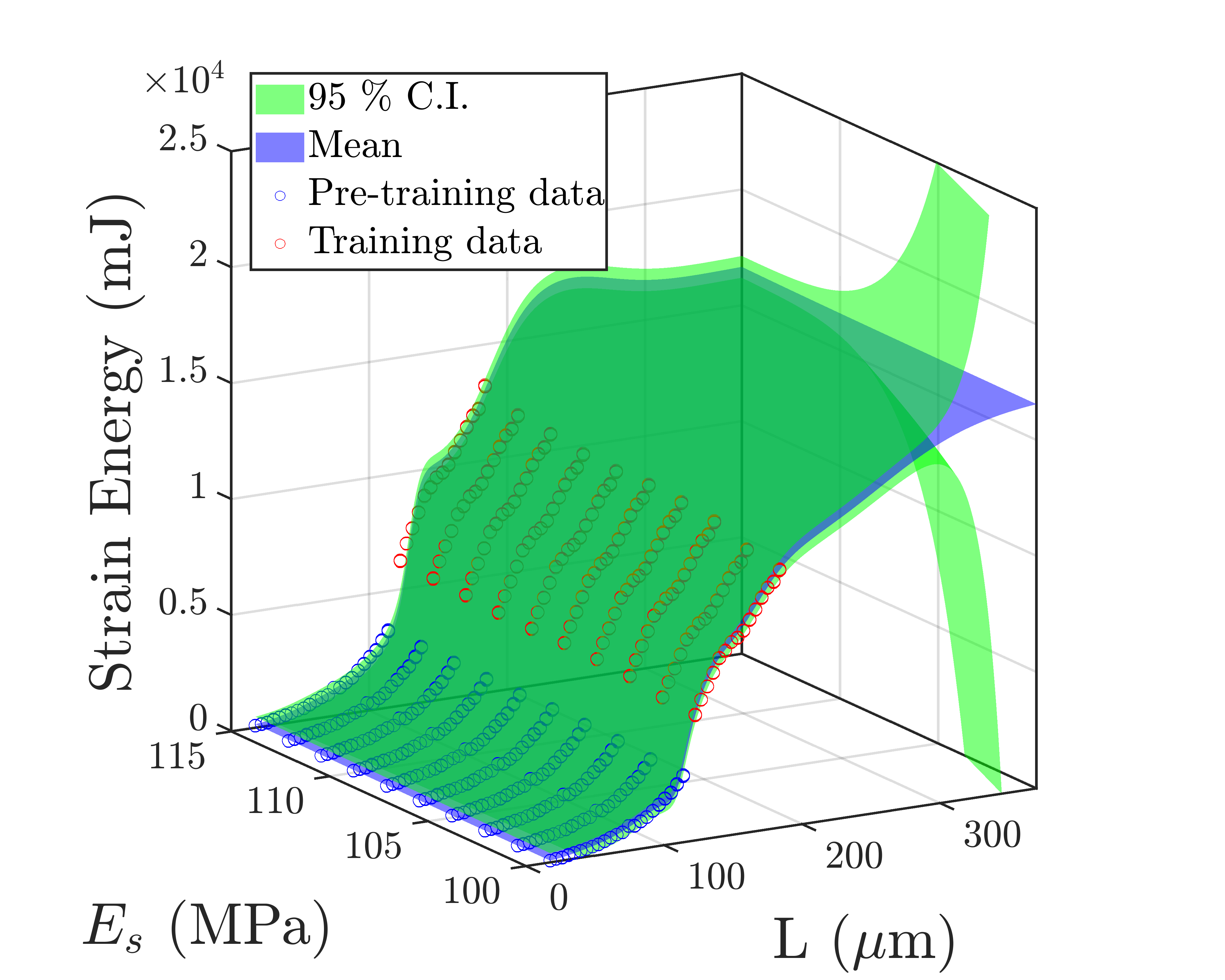}
    ~ 
    \includegraphics[trim=0.5in 0in 0.5in 0in, clip, width=.45\textwidth]{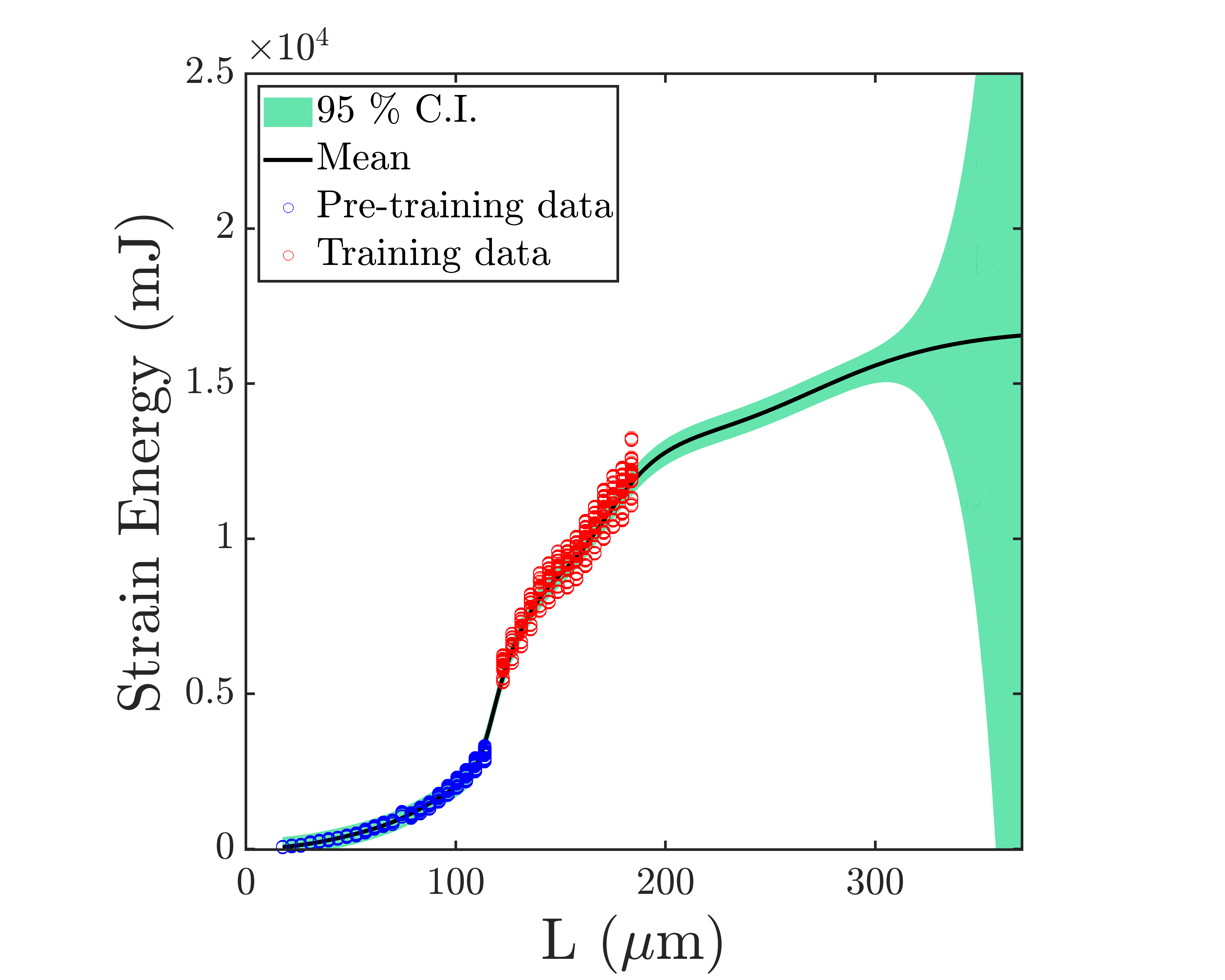} 
    \\ (b) \\ 
\vspace{-0.0in}
    \includegraphics[trim=0.5in 0in 0.5in 0in, clip, width=.45\textwidth]{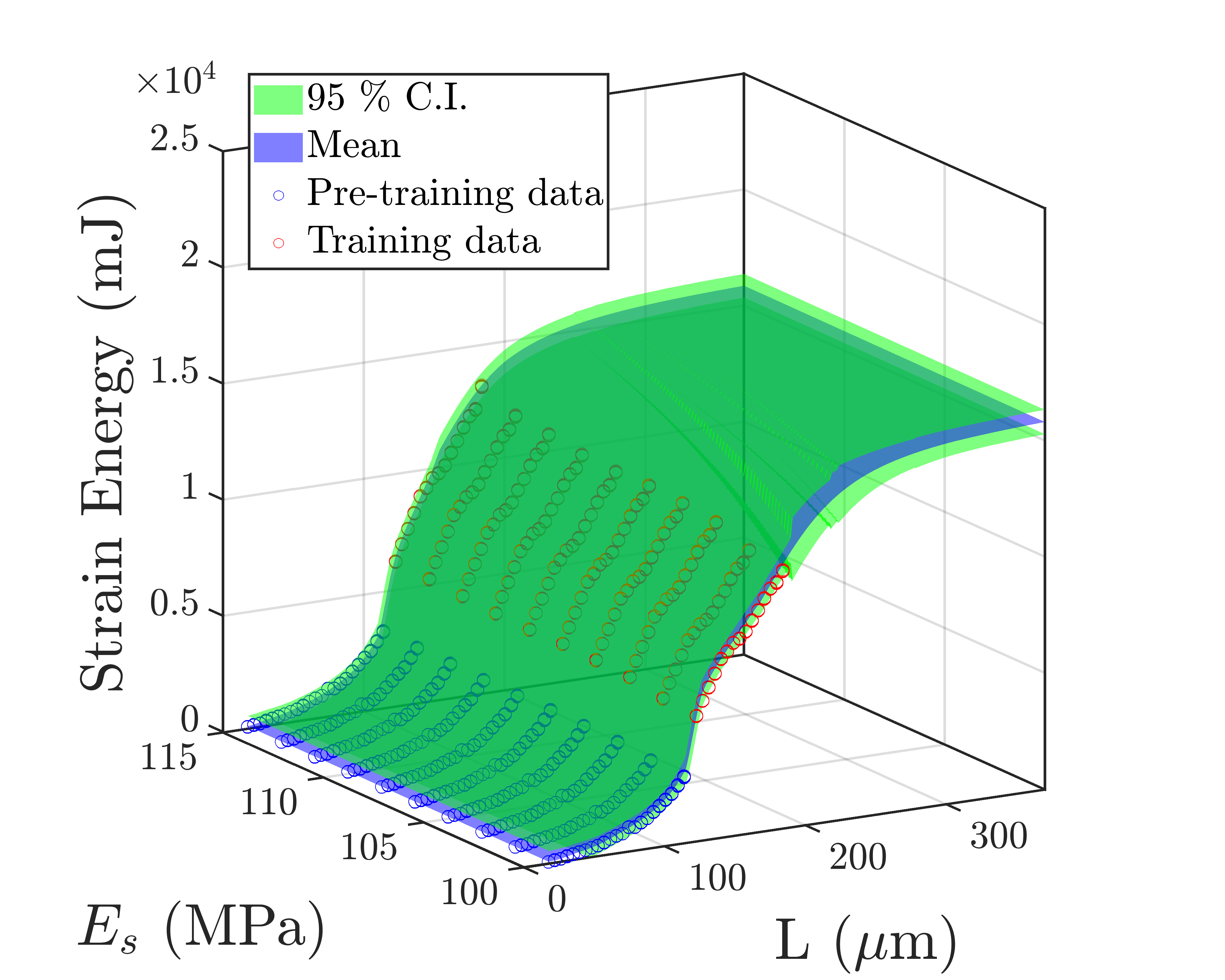}
    ~
    \includegraphics[trim=0.5in 0in 0.5in 0in, clip, width=.45\textwidth]{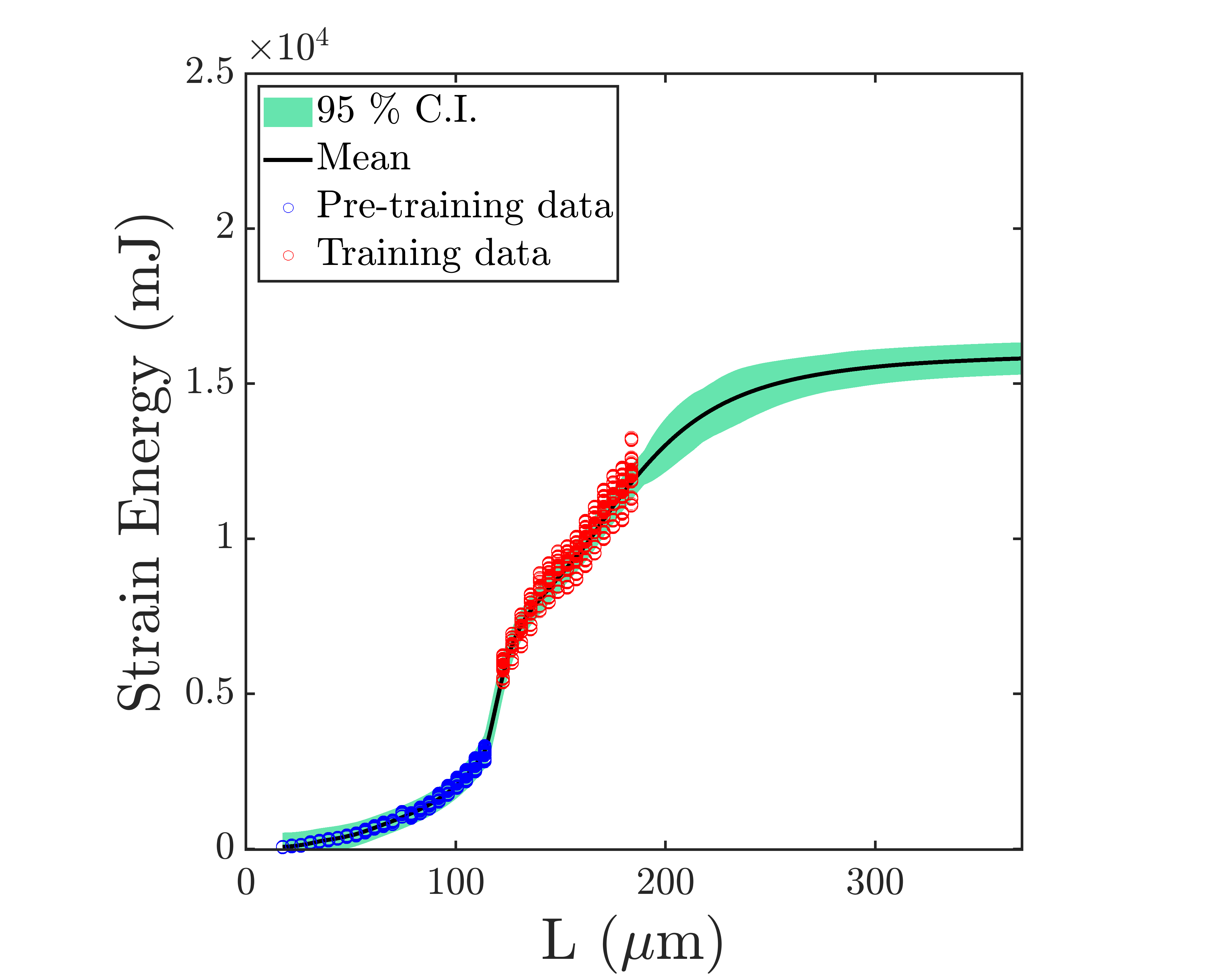}
    \\ (c)
\vspace{-0.12in}
    \caption{
    Elasticity problem:
    The mean and uncertainty predictions of surrogate models in Category 1 for
    (a) fully-connected network $M^1_{F(D=1, W=600)}$;
    (b) fully-connected network $M^1_{F(D=2, W=600)}$;
    (c) plausible sparsified network $M^1_I$ with $D=1$ and 564 connections.
    The left panels show 3D plots, while the right panels depict the marginalized plots over the $E_{s}$.
    }
    \label{fig:opal_cat1}
\end{figure}

Next, we assess the credibility of $M^1_I$ through a leave-out validation test. The comparison of validation measures for each leave-out data point with the corresponding validation tolerances reveals that $M^1_I$ fails the validation test and is deemed an invalid model,

$L_{LO}=179.4\mu m$:
\vspace{-0.1in}
\begin{eqnarray*}
    \mathbbm{d}_{DKL} = 0.0112 \nleqslant TOL_{DKL} = 0.008
    & , &
    \mathbbm{d}_{CDF} = 52.31 \nleqslant TOL_{CDF} = 45,    
\end{eqnarray*}

$L_{LO}=183.8 \mu m$:
\vspace{-0.1in}
\begin{eqnarray*}
    \mathbbm{d}_{DKL} = 0.0108 \nleqslant TOL_{DKL} = 0.008
    & , &
    \mathbbm{d}_{CDF} = 60.52 \nleqslant TOL_{CDF} = 45.   
\end{eqnarray*}
\end{paragraph}

\begin{paragraph}{Occam Category 2}
Following the same procedure leads to the selection of the \textit{Leaky ReLU} activation function for layer 3 and the \textit{Tanh} function for layer 4. Subsequent sparsification of $M^2_{F(D=3, W=600)}$ results in the plausible model within this category, $M^2_I$, comprising $D=3$ and 619245 connections (approximately 14\% parameters reduction upon sparsification). 
Figure \ref{fig:opal_cat2} presents the predictions of this surrogate model in comparison to high-fidelity datasets.
\begin{figure}[htpb]
\vspace{-0.1in}
    \centering
    \includegraphics[trim=0.5in 0in 0.5in 0in, clip, width=.48\textwidth]{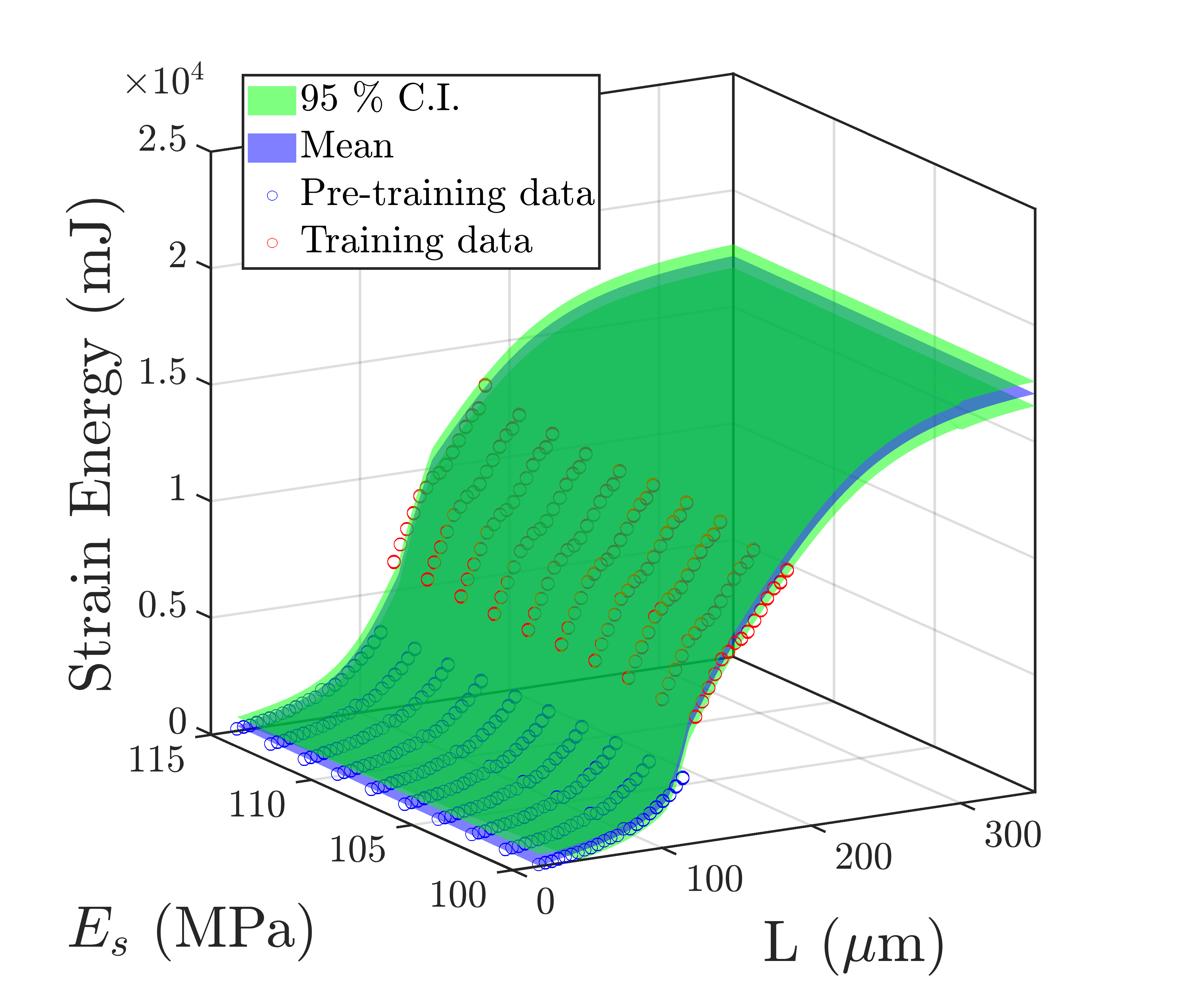}
    ~
    \includegraphics[trim=0.5in 0in 0.5in 0in, clip, width=.48\textwidth]{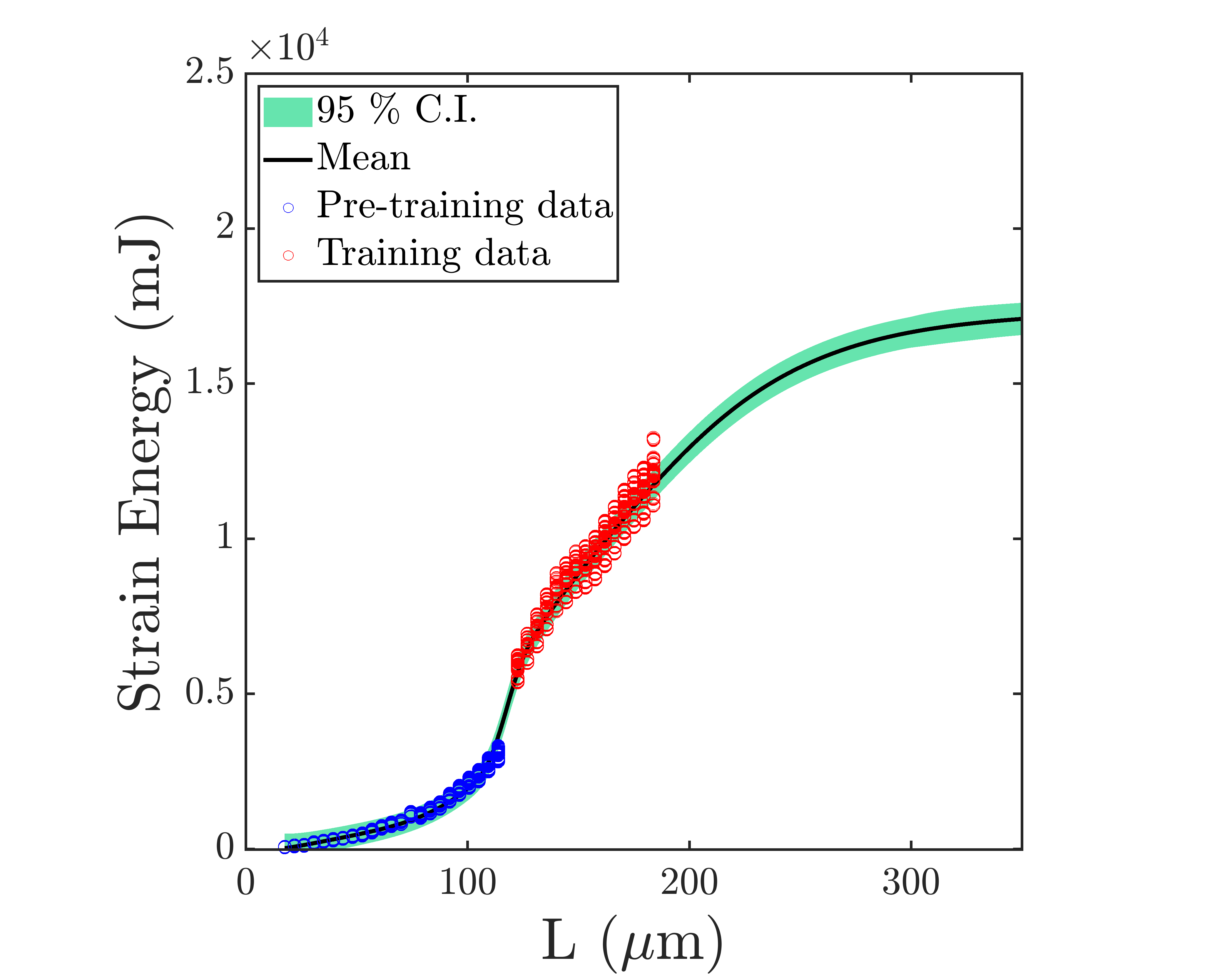}
    \vspace{-0.0in}
    \\ (a) \\
    % \vspace{-0.05in}
    % \includegraphics[trim=0.4in 0in 0.4in 0.2in, clip, width=.45\textwidth]{Figures_manuscript/mean_sparsified.png}
    % ~
    % \includegraphics[trim=0.4in 0in 0.4in 0.2in, clip, width=.45\textwidth]{Figures_manuscript/variance_sparsified.png}
    % \vspace{-0.05 in}
    % \\ (b) \\ 
    \includegraphics[trim = 0.6in 3.1in 0.6in 2.6in,clip,width =.85\textwidth]{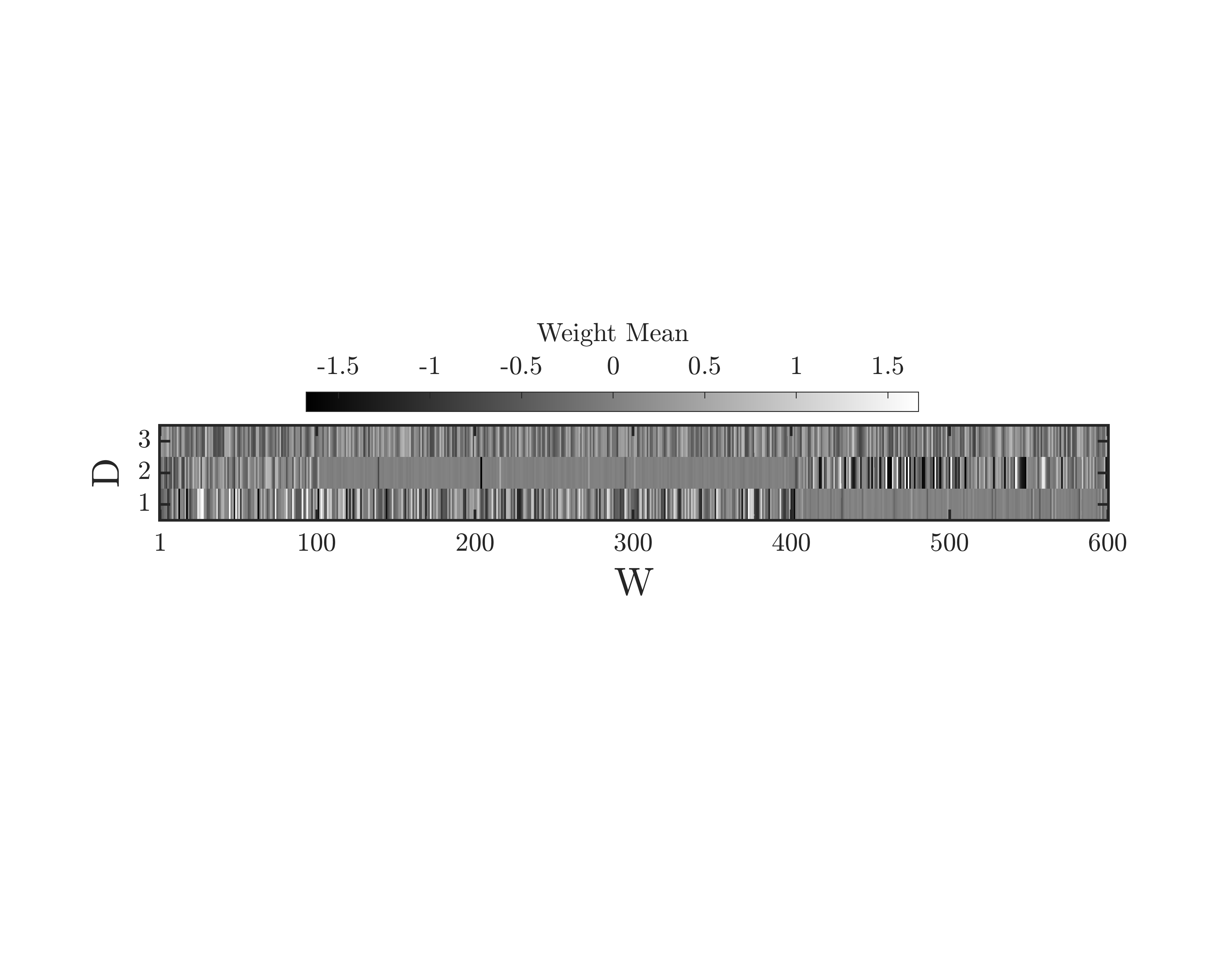}
    \\ \vspace{-0.1 in}
    \includegraphics[trim = 0.6in 3.1in 0.6in 3.2in,clip,width =.85\textwidth]{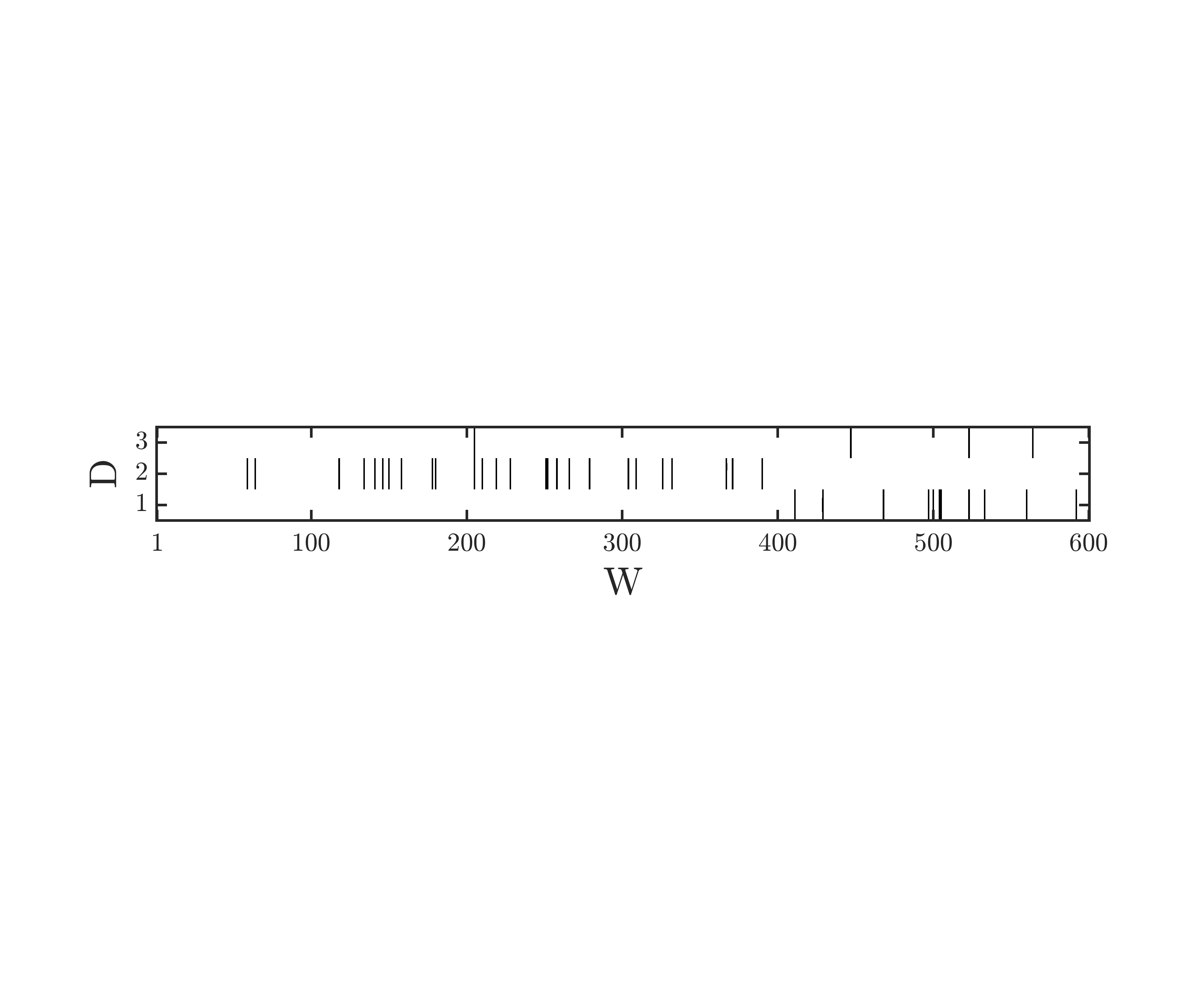}
    \vspace{-0.0 in}
    \\ (b) \\
    \vspace{-0.1 in}
    \caption{
    Elasticity problem:
    (a) The mean and uncertainty predictions of the plausible sparsified network $M^2_I$ in Category 2, identified as the ``best" surrogate model by OPAL-surrogate.The network consists of 3 layers, 619245 connections, and \textit{Tanh}, \textit{Tanh}, \textit{Leaky ReLU} activation functions for layers 1 to 3, respectively. 
    %The left panel shows 3D plot, while the right panel depict the marginalized plot over the $E_{s}$ axis.
    % (b) Probability distributions of the mean and variance of the network weights for the fully-connected network $M^2_{F(D=3, W=600)}$, utilizing Gaussian and Laplace priors. Network sparsification is performed to obtain $M^2_I$.
    (b) Visualization of weights pattern for $M^2_{F(D=3, W=600)}$ across network depth $D$ and width $W$ in the top panel. The bottom panel illustrates the sparsified pattern of $M^2_I$, with removed connections marked in black from each hidden layer using $TOL_{\bs \theta} = 0.05$. 
    }
    \label{fig:opal_cat2}
\end{figure}
Importantly, $M^2_I$ successfully passes the validation test based on both validation metrics, 

$L_{LO}=179.4 \mu m$:
\vspace{-0.1in}
\begin{eqnarray*}
    \mathbbm{d}_{DKL} = 0.0069 \leq TOL_{DKL} = 0.008
    & , &
    \mathbbm{d}_{CDF} = 40.42 \leq TOL_{CDF} = 45,    
\end{eqnarray*}

$L_{LO}=183.8 \mu m$:
\vspace{-0.1in}
\begin{eqnarray*}
    \mathbbm{d}_{DKL} = 0.0061 \leq TOL_{DKL} = 0.008
    & , &
    \mathbbm{d}_{CDF} = 34.55 \leq TOL_{CDF} = 45,  
\end{eqnarray*}
establishing it as the ``best" predictive surrogate model for the given pre-training and training data sets.
As depicted in Table \ref{table:opal_elasticity}, model $M^2_I$ demonstrates significantly improved predictive performance compared to $M^1_I$, as evident from the validation metrics (a 37\% reduction in $\mathbbm{d}_{DKL}$ and a 30\% decrease in $\mathbbm{d}_{CDF}$), and visually shown in Figure \ref{fig:val_obs}.

\begin{figure}[htpb]
\vspace{-0.1in}
    \centering
    \includegraphics[trim=1.0in 0in 1.4in 0in, clip, width=.45\textwidth]{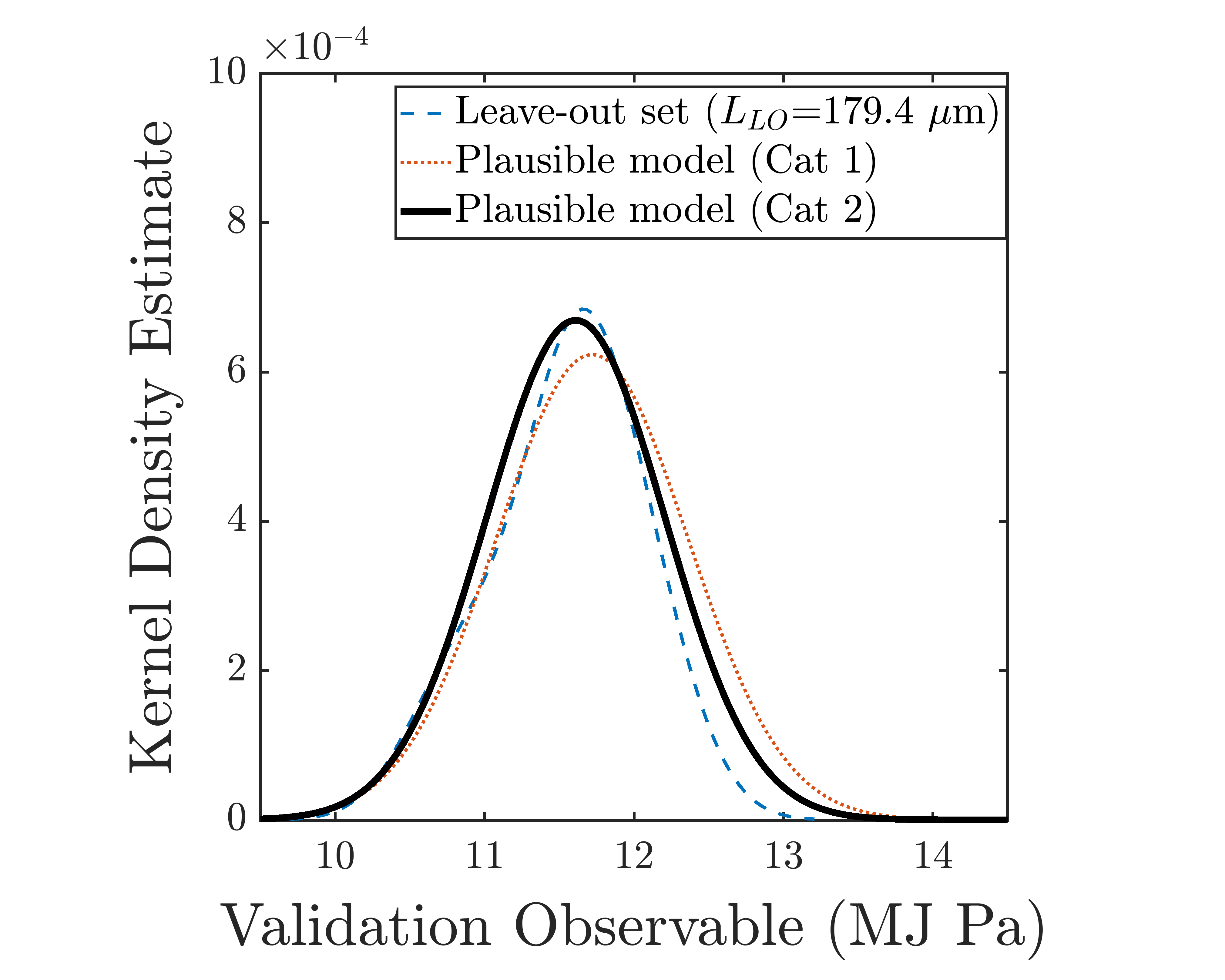}
    ~
    \includegraphics[trim=1.0in 0in 1.4in 0in, clip, width=.45\textwidth]{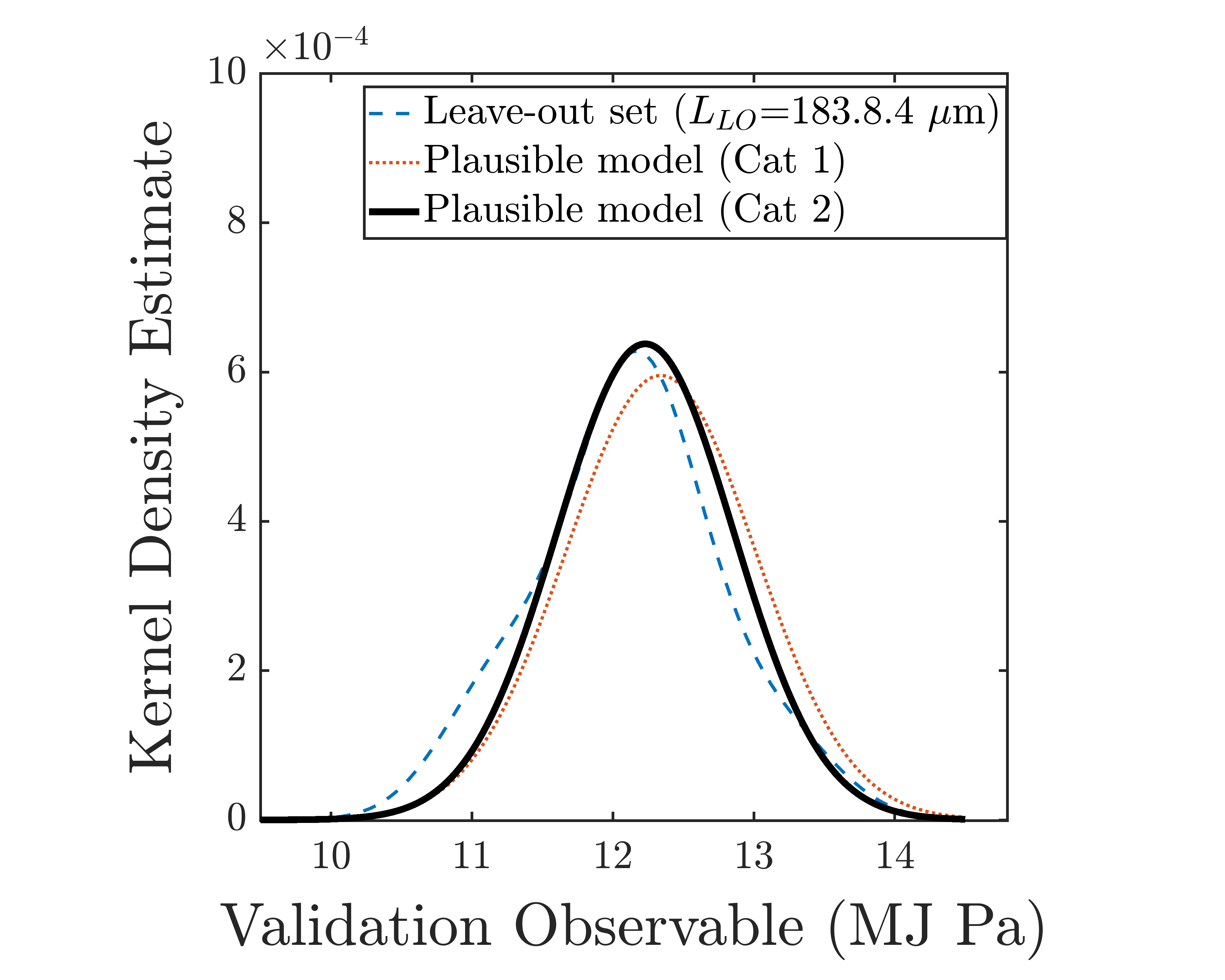}
    \\ (a) \hspace{2.2in} (b)
    \vspace{-0.1in}
    \caption{
    Elasticity problem:
    Comparison of the observables $\mathcal{Z}_D$ in \eqref{eq:observ} corresponding to leave-out validation sets and the predicted $\mathcal{Z}_M$ from plausible models $M^1_I$ and $M^2_I$ for
    (a) $L_{LO}=179.4 \mu m$;
    (b) $L_{LO}=183.8 \mu m$.
    }
    \label{fig:val_obs}
\end{figure}

\end{paragraph}

\vspace{-0.1in}
\begin{table}[h!]
\scriptsize
\centering
\caption{
Elasticity problem:
Results from the implementation of OPAL-surrogate for the identification of the predictive BayesNN surrogate models.
$M^l_{F(D,W)}$ denotes fully connected networks, and $M^l_I$ represents the plausible sparsified model in each Occam Category $l$.
The values of log-evidence $\pi_{evid} (\bs D | \bs \xi_{(\cdot)}^l, \bs D_c, \hat{\mathcal{M}}^l, \mathcal{S}, \mathcal{S}_c)$ and
validation metrics corresponding to leave-out sets $L_{LO}=\{179.4, 183.8\} \mu m$ are presented. OPAL-surrogate identifies $M^2_I$ at Category 2 as the ``best" BayesNN surrogate model.
}
\vspace{0.1in}
\begin{tabular}{ccccccccc}
\hline
Occam      & BayesNN & Activation & $M^l_{F(D,W)}$      & $M^l_I$      & \multicolumn{2}{c}{$\mathbbm{d}_{DKL}$} & \multicolumn{2}{c}{$\mathbbm{d}_{CDF}$} \\
categories & models   & function   & log-evid & log-evid & $179.4 \mu m$           & $183.8 \mu m$          & $179.4 \mu m$           & $183.8 \mu m$          \\ \hline
$l=1$ & $D=1$ & Tanh       & -14846 & -14758 & 0.0112 & 0.0108 & 52.31  & 60.52  \\
      & $D=2$ & Tanh       & -14924 &        &        &        &        &        \\ \hline
$l=2$ & $D=3$ & {Leaky ReLU} & {-14834} & {-14691} & \textbf{0.0073} & \textbf{0.0064} & \textbf{43.62}  & \textbf{36.71}  \\
      & $D=4$ & Tanh       & -15012 &        &        &        &        &        \\ \hline
$l=3$ & $D=5$ & Tanh       & -15036 & -14882 & 0.0124 & 0.0116 & 85.26  & 77.42  \\
      & $D=6$ & Tanh       & -15133 &        &        &        &        &        \\ \hline
$l=4$ & $D=7$ & LeakyReLU  & -15187 &        &        &        &        &        \\
      & $D=8$ & Tanh       & -15046 & -14962 & 0.0287 & 0.0258 & 153.48 & 143.21 \\ \hline
\end{tabular}
\label{table:opal_elasticity}
\end{table}

\begin{paragraph}{Exploring higher categories}
Although OPAL-surrogate concludes upon identifying the model $M^2_I$ in Category 2, as depicted in Table \ref{table:opal_elasticity} and Figure \ref{fig:opal_cat34}, an investigation into the performance of more complex models in higher categories reveals compromised performance. Specifically, models $M^3_I$ and $M^4_I$ are deemed invalid based on the specified accuracy tolerance. This highlights the observation that larger networks with increased free parameters do not necessarily translate to improved predictive capabilities.

\begin{figure}[htpb]
\vspace{-0.1in}
    \centering
    \includegraphics[trim=0.9in 0in 0.9in 0in, clip, width=.48\textwidth]{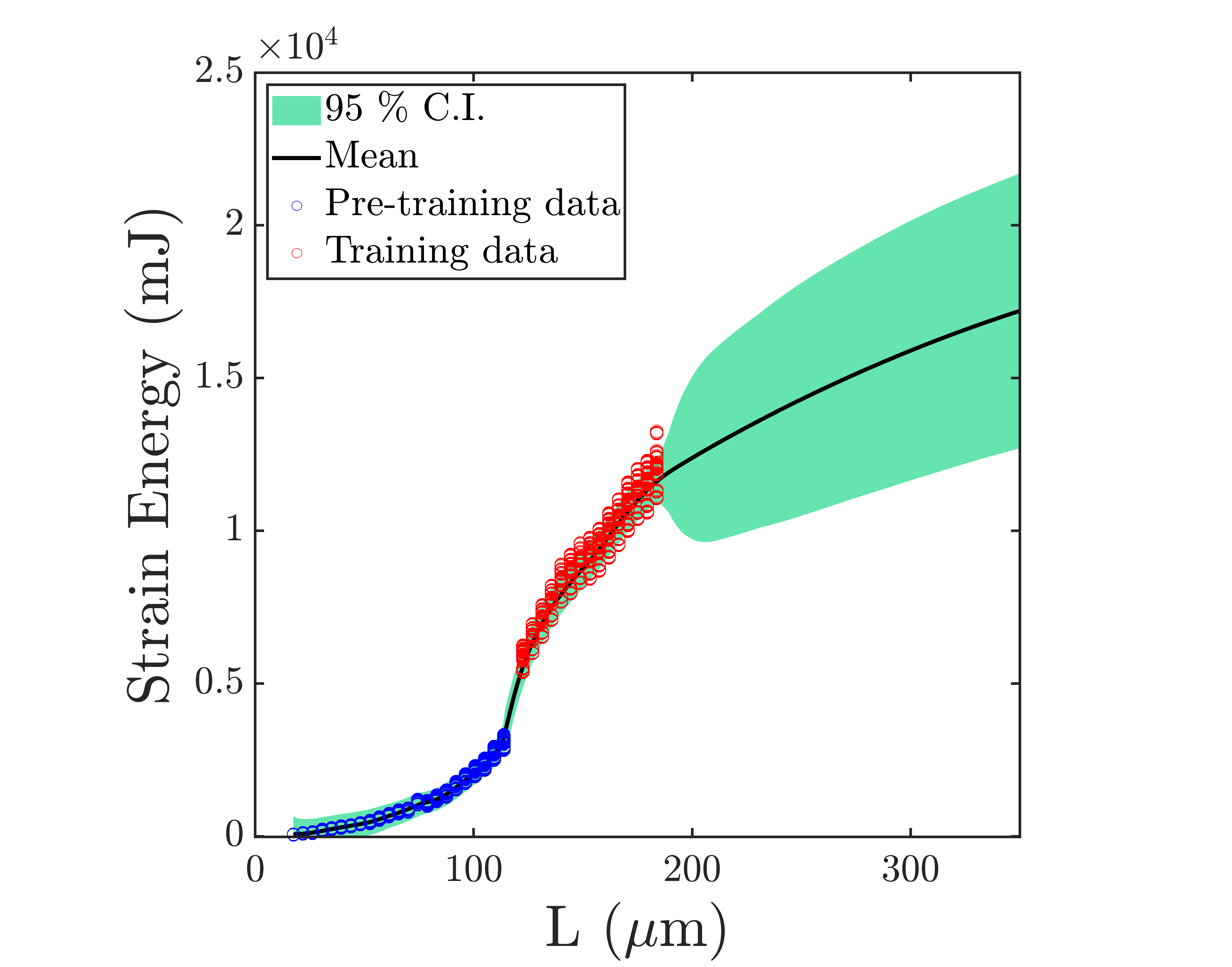}
    ~
    \includegraphics[trim=0.9in 0in 0.9in 0in, clip, width=.48\textwidth]{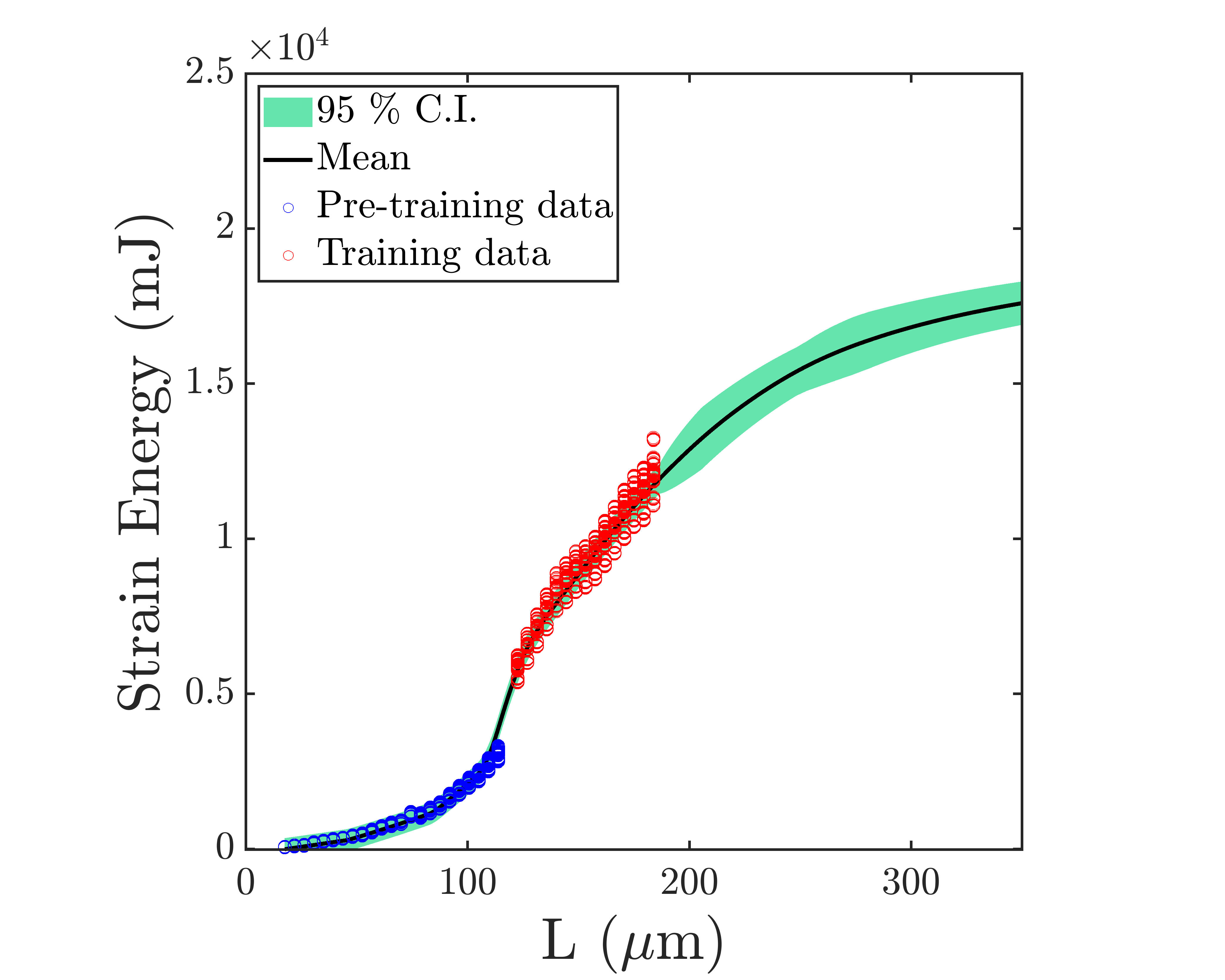}
    \\ (a) \hspace{2.2in} (b)
    \vspace{-0.1in}
    \caption{
    Elasticity problem:
    The mean and uncertainty predictions, marginalized over the $E_{s}$, of surrogate models in Categories 3 and 4:
    (a) plausible sparsified network $M^3_I$ with $D=5$ and 1264240 connections;
    (b) fully-connected network $M^4_{F(D=7, W=600)}$.
    Both panels depict the marginalized plots over the $E_{s}$.
    %The activation functions for each layer, and the associated model evidence are detailed in Table \ref{table:opal_elasticity}.
    }
    \label{fig:opal_cat34}
\end{figure}

\end{paragraph}

\begin{paragraph}{Prediction and forward UQ}
Finally, we utilize the surrogate model, $M^2_I$, to predict and quantify uncertainty in the target QoI (strain energy) in the prediction scenario $\mathcal{S}_P$.
In Figure \ref{fig:qoi}(a), the surrogate model prediction at $E_s = 110$ MPa and domain size $L=297.5 \mu m$ is compared with one realization of the microstructure in $\mathcal{S}_P$, as modeled by high-fidelity simulation.
Figure \ref{fig:qoi}(b) presents the probability distributions of the QoI for uncertain elastic modulus of the silica aerogel $E_{s} \sim \mathcal{N}(107.5, 112.5)$, as obtained from previous studies \cite{tan2022predictive}.
The estimated mean and 95\% CI of the QoI for the extrapolation prediction by the BayesNN surrogate model are 
$ \mathbb{E}(QoI) = 16615, \quad CI(QoI) = [15565, 17665] mJ$.
However, considering only the mean of the surrogate model prediction (corresponding to deterministic neural network training), the estimated uncertainty becomes 
$\mathbb{E}(QoI) = 16660, \quad CI(QoI) = [16350,16970] mJ$. 
These results suggest that while the mean prediction of the surrogate model demonstrates relatively acceptable performance, 
an evaluation of model prediction uncertainty using BayesNN highlights the intrinsic limitation of neural networks in extrapolation.

\begin{figure}[htpb]
    \centering
    \includegraphics[trim=0.9in 0in 0.9in 0in, clip, width=.45\textwidth]{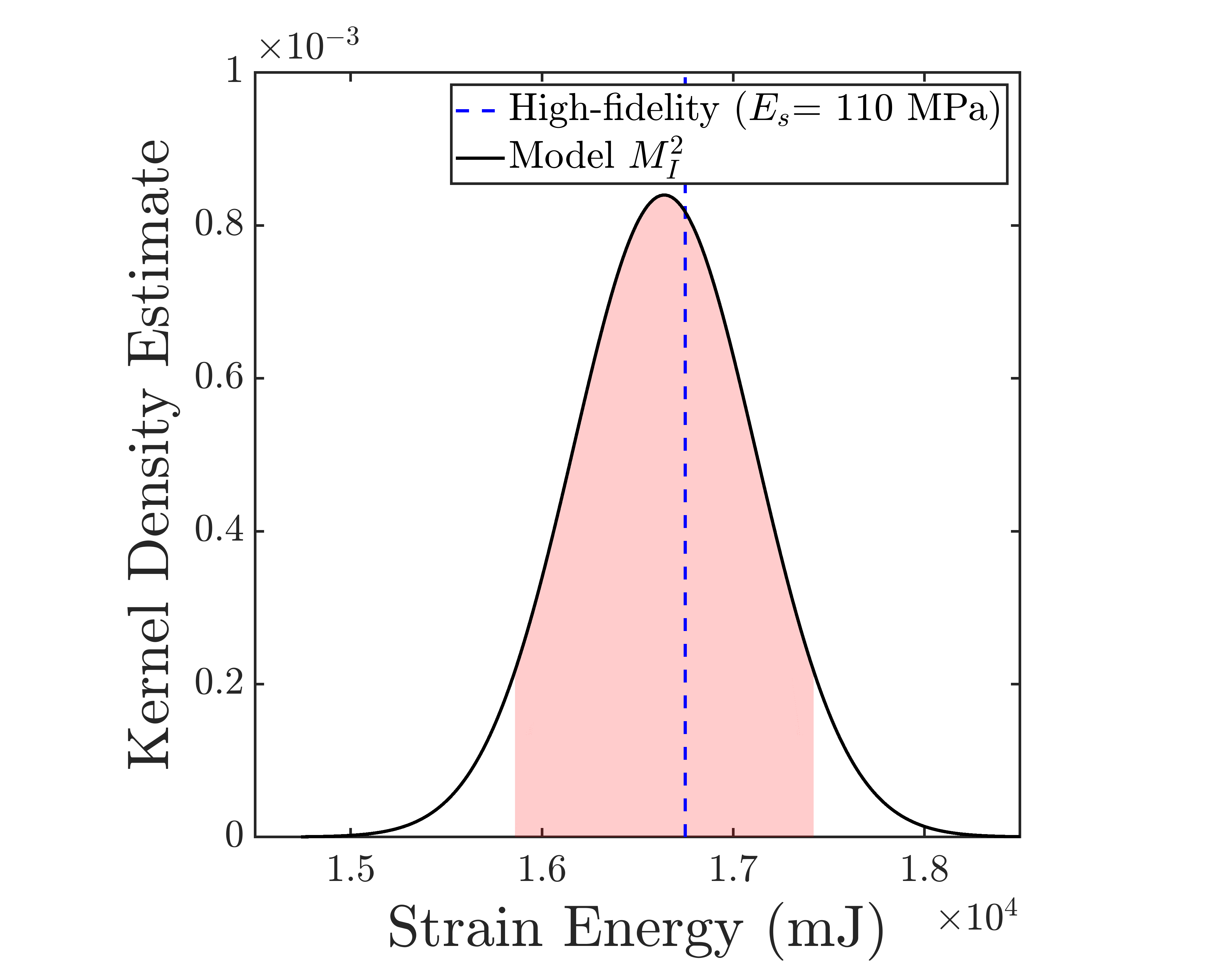}~
    \includegraphics[trim=0.9in 0in 0.9in 0in, clip, width=.45\textwidth]{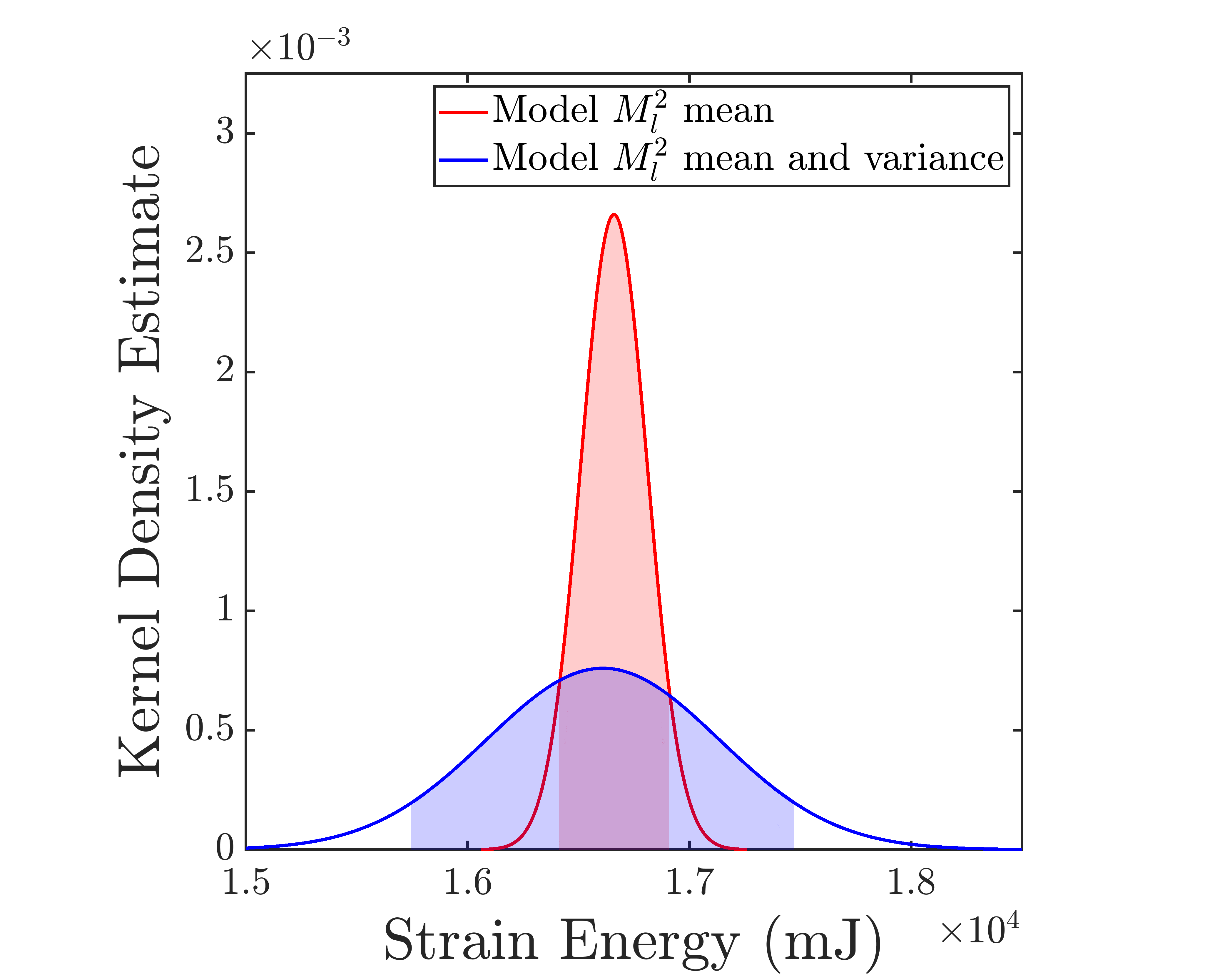}
    \\ (a) \hspace{2.2in} (b)
    \vspace{-0.1in}
    \caption{
    Elasticity problem:
    Prediction of the QoI, strain energy, in the prediction scenario $\mathcal{S}_P$, utilizing the BayesNN surrogate model $M^2_I$, considering:
    (a) elastic modulus $E_s = 110$ MPa and in comparison with high-fidelity simulation for one microstructure realization;
    (b) uncertain elastic modulus $E_s \sim \mathcal{N}(107.5, 112.5)$ MPa and with and without the inclusion of uncertainty bounds in the model predictions.
    }
    \label{fig:qoi}
\end{figure}

\end{paragraph}

%++++++++++++++++++++++++++++++++++++++++++++++++++++++++++++++++++++++++
\subsection{Direct numerical simulations of turbulent combustion}\label{sec:ablate}
\noindent
Our second application focuses on the combustion dynamics of shear-induced ablation in solid fuels used in hybrid rocket motors. Specifically, we leverage the Ablative Boundary Layers at the Exascale (ABLATE) software framework \cite{ABLATE}, which integrates direct numerical simulations of turbulent reacting flows with thermochemical species equations. The focal point of our investigation centers on modeling slab burner experiments according to the setup in \cite{georgalis2023combined, surina2022measurement}, designed to study the reacting boundary layer combustion in hybrid rockets and measuring fuel regression rates.
\blue{Turbulent combustion involves complex interactions among chemical reactions, flame structures, radiation, and soot, occurring on scales much smaller than flow transport. ABLATE's Navier-Stokes solver comprehensively models these phenomena by simulating fully compressible and reactive gas phases.}

\begin{figure}[htpb]
    \centering
    \includegraphics[trim=0in 0in 0in 0in, clip, width=.95\textwidth]{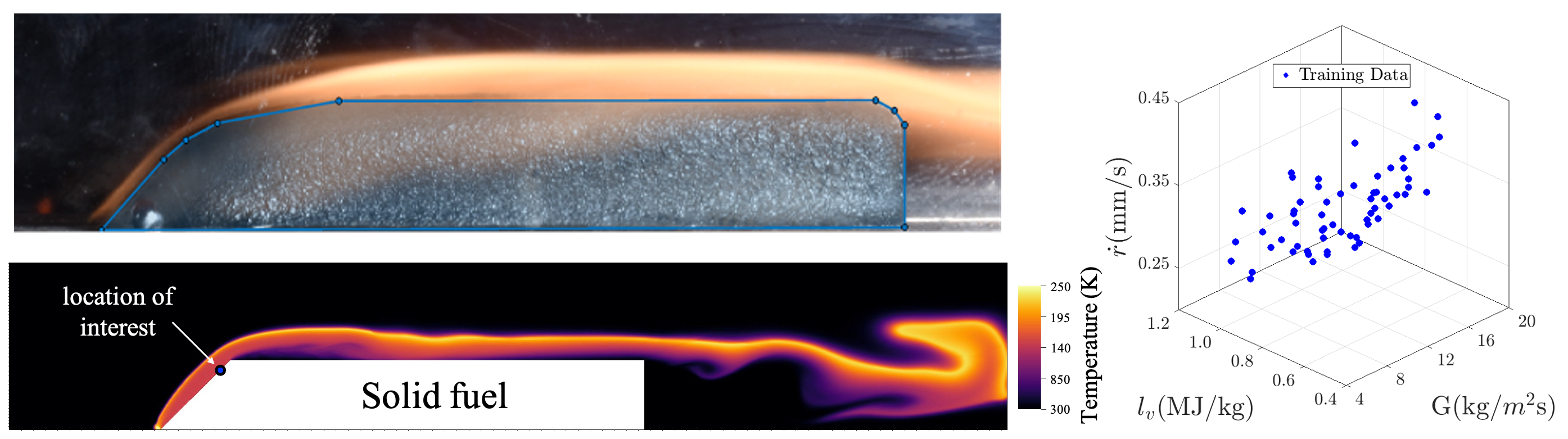} 
    \vspace{-0.1in}
    \caption{
    Turbulent combustion problem:
    The top-left panel illustrates the side profile of the solid fuel in the 2D slab burner experiment \cite{georgalis2023combined, surina2022measurement}.
    The bottom-left panel displays a representative 2D ABLATE simulation result of the slab burner, highlighting the location on the fuel surface for obtaining fuel regression rate ($\dot{r}$) training data.
    The right panel shows the training data used for surrogate modeling, comprising an ensemble of $\dot{r}$ for various values of the oxidizer flux ($G$) and latent heat of vaporization ($l_v$).
    }
    \label{fig:ablate}
\end{figure}

Figure \ref{fig:ablate} depicts a representative simulation result from a 2D slab burner employed in this study, with Polymethyl methacrylate serving as the solid fuel and $O_2$ as the oxidizer in ABLATE simulations. The key output of interest is the fuel regression rate, denoted as $\dot{r}$, representing the rate of fuel recession during combustion and a critical parameter influencing motor thrust and hybrid rocket motor geometry.
Given the high computational costs associated with ABLATE simulations, the ultimate aim of the surrogate model is to facilitate Bayesian calibration and validation of ABLATE by utilizing regression rate measurements from slab burner experiments. To this end, we develop a BayesNN surrogate for the fuel regression rate, incorporating two input parameters of ABLATE. The first input is the oxidizer flux $G=[5, 20] Kg/m^2s$, corresponding to the range of inlet velocities in the slab burner experiment \cite{georgalis2023combined, surina2022measurement}, and the second input parameter is the latent heat of vaporization $l_v=[6 \times 10^5, 11 \times 10^5] J/Kg$.
The training dataset $\bs D$, comprising $N_D = 64$ simulation ensembles, captures the fuel regression rate $\dot{r}$ at a critical location on the slab boundary (illustrated in Figure \ref{fig:ablate}), with the inputs $G$ and $l_v$ sampled using Latin hypercube sampling.
% validation
For credibility assessment, we consider the leave-out set $\bs D_{LO}$ containing data points within $G_{LO} = [9.89, 14.63] Kg/m^2s$.
The validation observables are,
\begin{equation}\label{eq:ablate_observ}
    \mathcal{Z}_D = \int_{G_{LO}^{low}}^{G_{LO}^{up}} \int_{l_v^{low}}^{l_v^{up}} u_{\bs D} \; d l_v \; d G, 
    \quad\quad
    \mathcal{Z}_M = \int_{G_{LO}^{low}}^{G_{LO}^{up}} \int_{l_v^{low}}^{l_v^{up}}
    u_{\bs \theta} \; d l_v \; d G,
\end{equation}
where $u_{\bs D}$ and $u_{\bs \theta}$ are the fuel regression rate $\dot{r}$, evaluated using the high-fidelity simulations and surrogate models, respectively, and
$G_{LO}^{low} = 9.89 Kg/m^2s$,
$G_{LO}^{up} = 14.63 Kg/m^2s$,
$l_v^{low} = 6 \times 10^5 J/Kg$, and
$l_v^{up} = 11 \times 10^5 J/Kg$.

\begin{paragraph}{OPAL-surrogate demonstration}
Following the approach outlined in Section \ref{sec:porous_opal}, we set the upper limit for the width at $W=300$ in the initial BayesNN surrogate model set $\mathcal{M}$, guided by Figure \ref{fig:combustion_opal_grid}. Given the smaller dataset, a more strict categorization strategy is employed for models with $W = [1, 300]$, with each category corresponding to a distinct width range.
\begin{figure}[htpb]
\vspace{-0.2in}
    \centering
    \includegraphics[trim=0.5in 0in 1.4in 0in, clip, width=.45\textwidth]{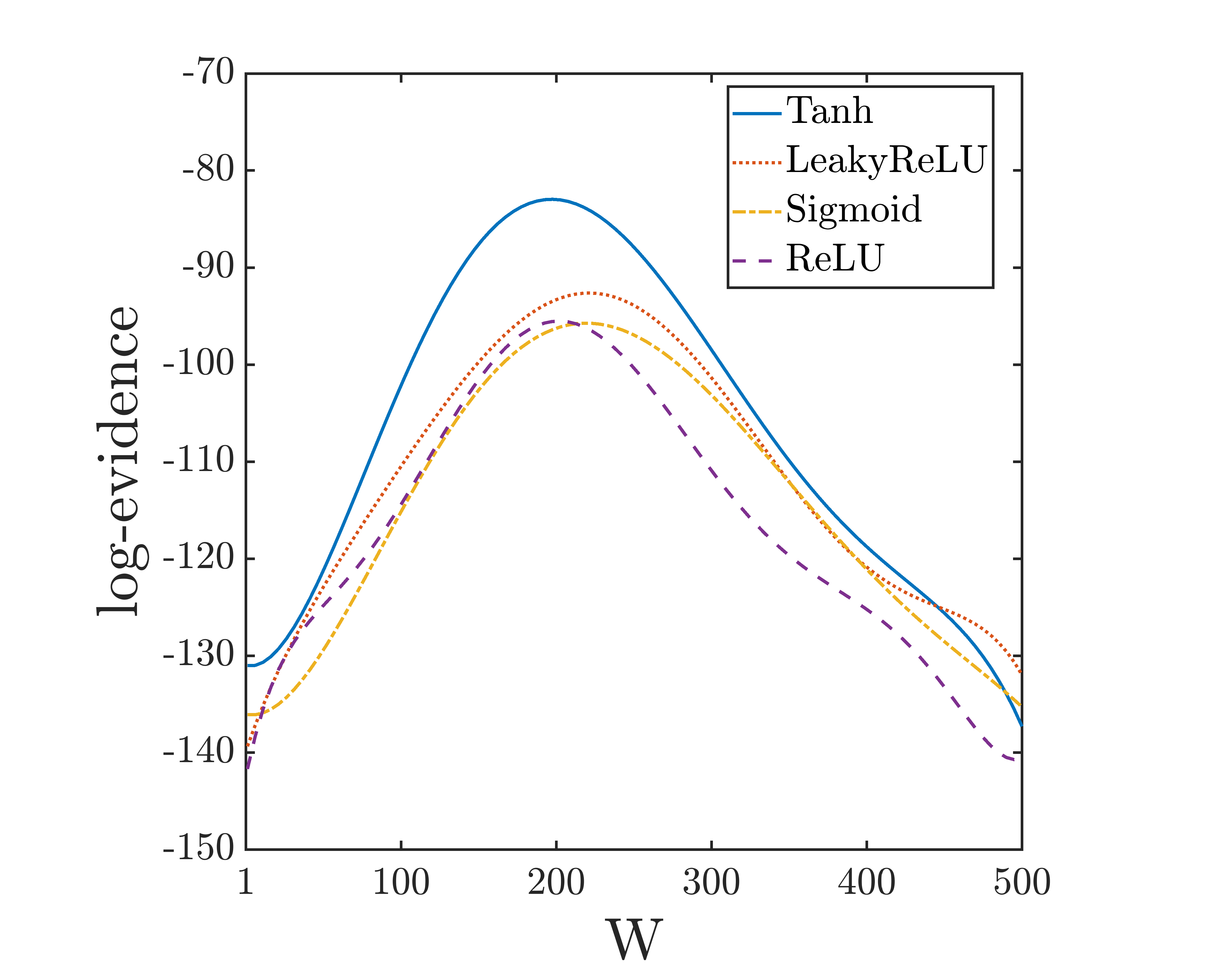}
    \vspace{-0.15in}
    \caption{
    Turbulent combustion problem:
    The log-evidence for BayesNN models with a single layer ($D=1$) with different widths and activation functions.
    }
    \label{fig:combustion_opal_grid}
\end{figure}
\vspace{-0.15in}
\begin{table}[h!]%-------------------------------------------------------
\footnotesize
\centering
\caption{
Turbulent combustion problem:
Results of OPAL-surrogate implementation for the identification of $M^2_I$ at Category 2 as the ``best" BayesNN surrogate model.
}
\begin{tabular}{cccccc}
\hline
Occam      & BayesNN & Activation & $M^l_{F(D,W)}$      & $M^l_I$      & $\mathbbm{d}_{DKL}$ \\
categories & models   & function   & log-evid & log-evid &          \\ \hline
$l=1$ & $D=1$ & Tanh       & -99.5 & -95.5 & 0.0127   \\
$l=2$ & $D=2$ & {Tanh} & {-94.7} & {-90.6} & \textbf{0.0084}   \\ 
$l=3$ & $D=3$ & LeakyReLU & -103.6 & -98.7 & 0.0156   \\ \hline
\end{tabular}
\label{table:opal_combustion}
\end{table}%-------------------------------------------------------

Table \ref{table:opal_combustion} presents the results of OPAL-surrogate for the first three categories. 
Figures \ref{fig:combustion_opal_cat1_fc} and \ref{fig:combustion_opal_cat1_sparse}(a) illustrate the predictions of both fully connected and sparsified networks in Category 1, comparing them to the training datasets. The plausible model $M^1_I$ comprises 245 connections, approximately 19\% parameter reduction from $M^1_{F(D=1, W=300)}$ following sparsification with $TOL_{\bs \theta} = 0.075$. However, considering $TOL_{DKL} = 0.01$, this model is deemed invalid. Figure \ref{fig:combustion_opal_cat1_sparse}(a) illustrates the predictions of $M^1_I$ compared to the leave-out data for credibility assessment.
\begin{figure}[htpb]
\vspace{-0.1in}
    \centering
    \includegraphics[trim=0.0in 0in 1.0in 0.0in, clip, width=.5\textwidth]{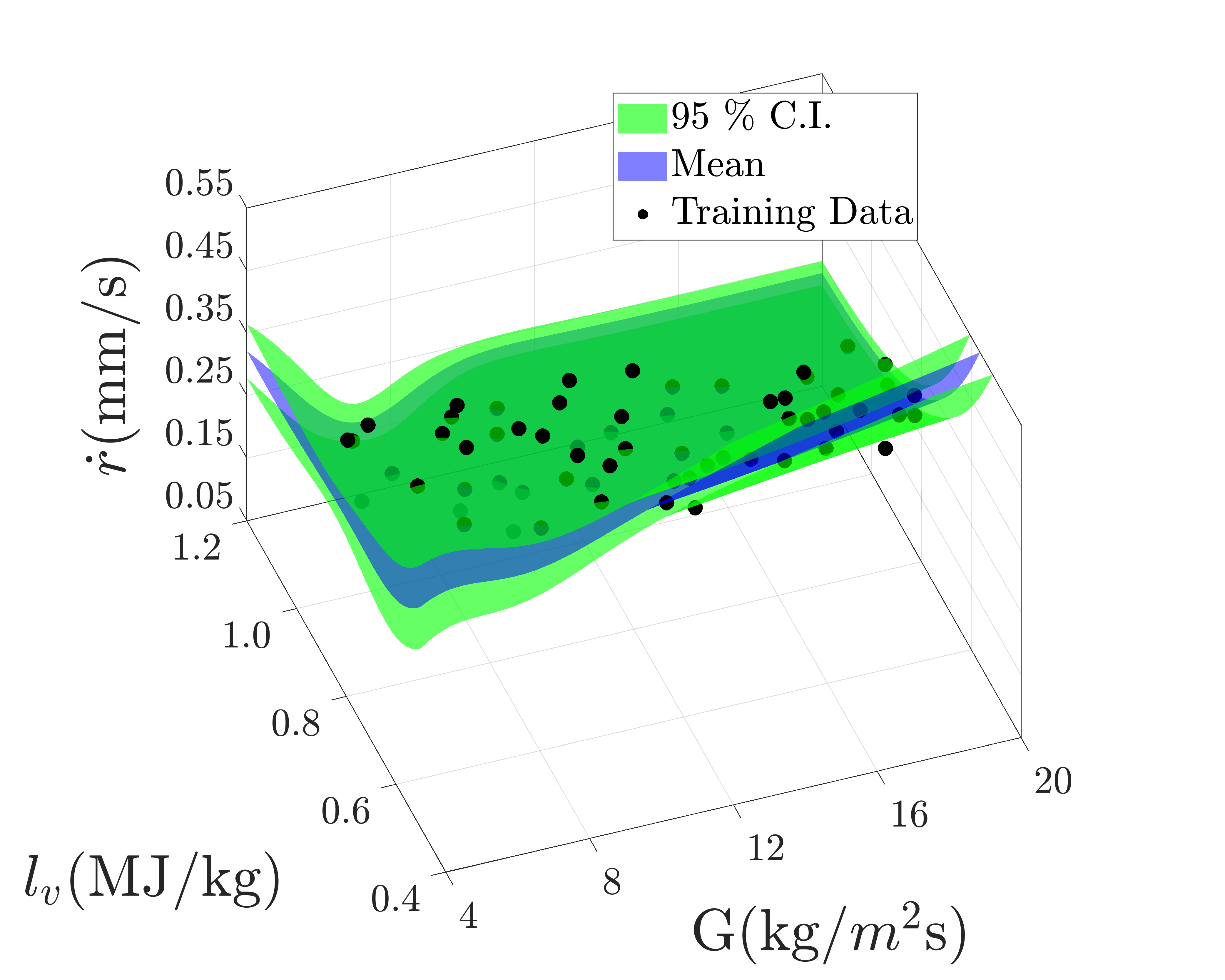}
    \\
    \includegraphics[trim=0.5in 0in 0.5in 0.0in, clip, width=.48\textwidth]{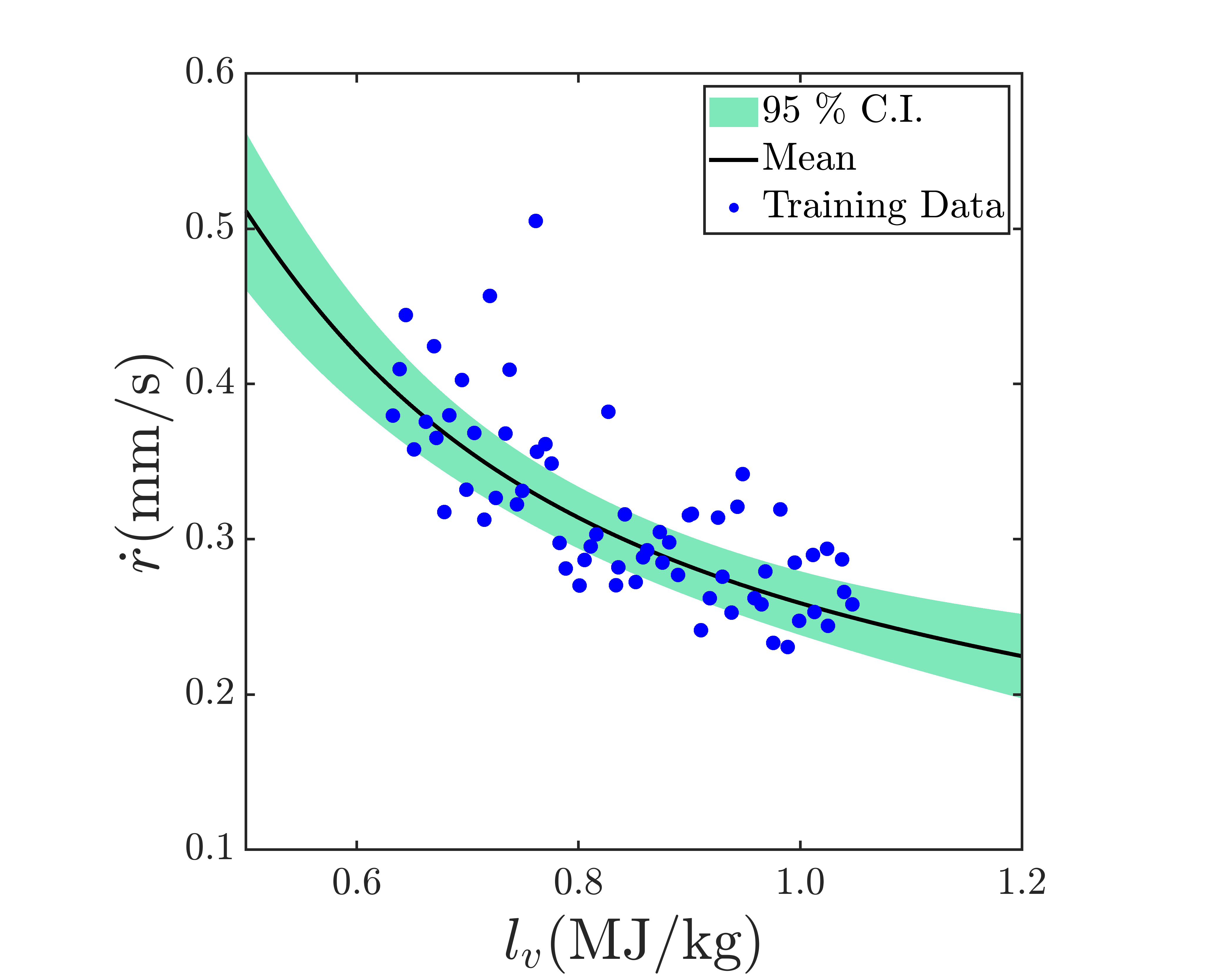}
    ~   
    \includegraphics[trim=0.5in 0in 0.5in 0in, clip, width=.48\textwidth]{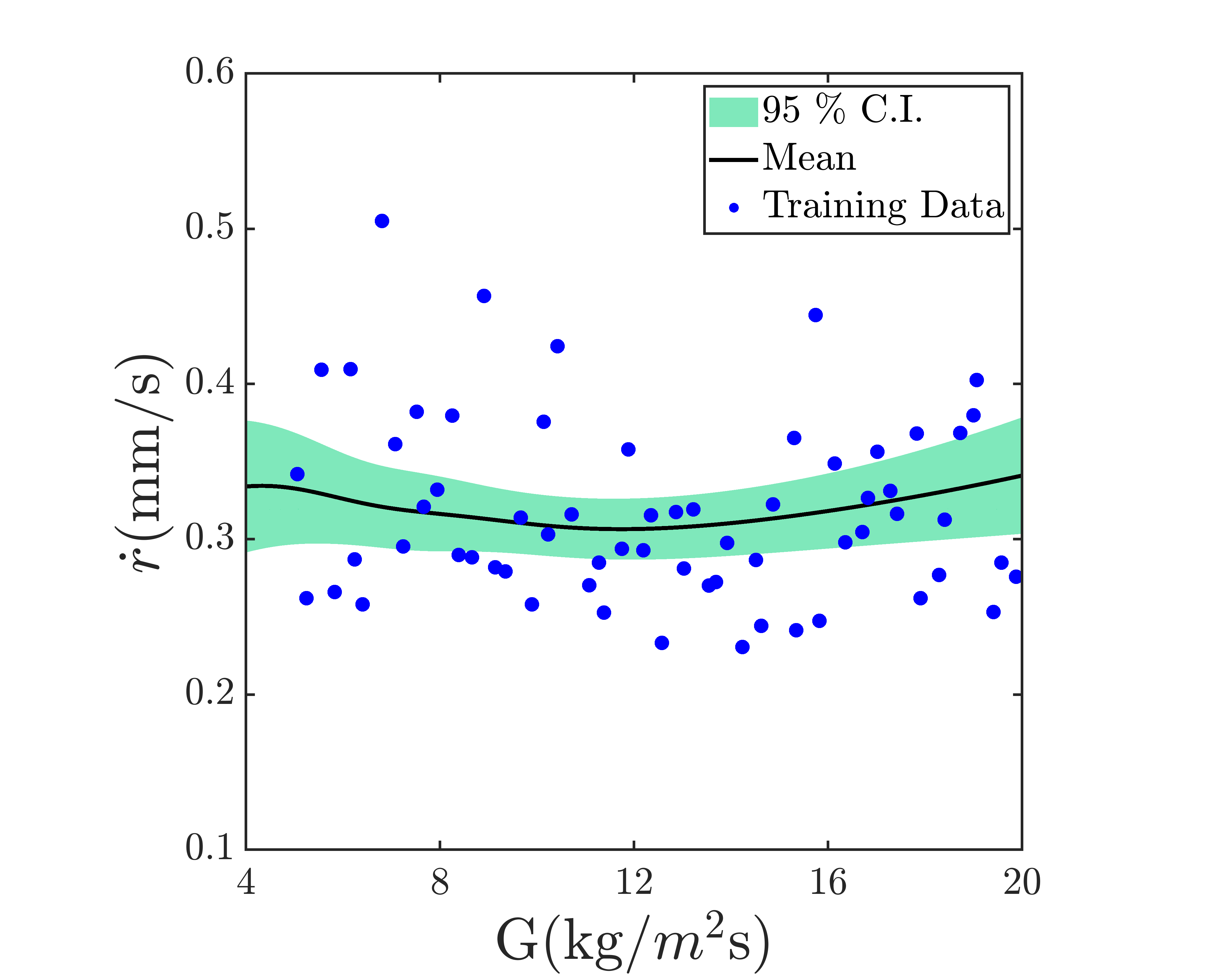}
    \vspace{-0.1in}
    \caption{
    Turbulent combustion problem:
    The mean and uncertainty predictions of the fully-connected network $M^1_{F(D=1, W=300)}$ in Category 1.
    The top panel shows the 3D plot and the bottom panels display the marginalized plots over the $l_{v}$ and $G$.
    }
    \label{fig:combustion_opal_cat1_fc}
\end{figure}

\begin{figure}[htpb]
\vspace{-1.0in}
    \centering
    \includegraphics[trim=0.0in 0in 1.0in 0.0in, clip, width=.45\textwidth]{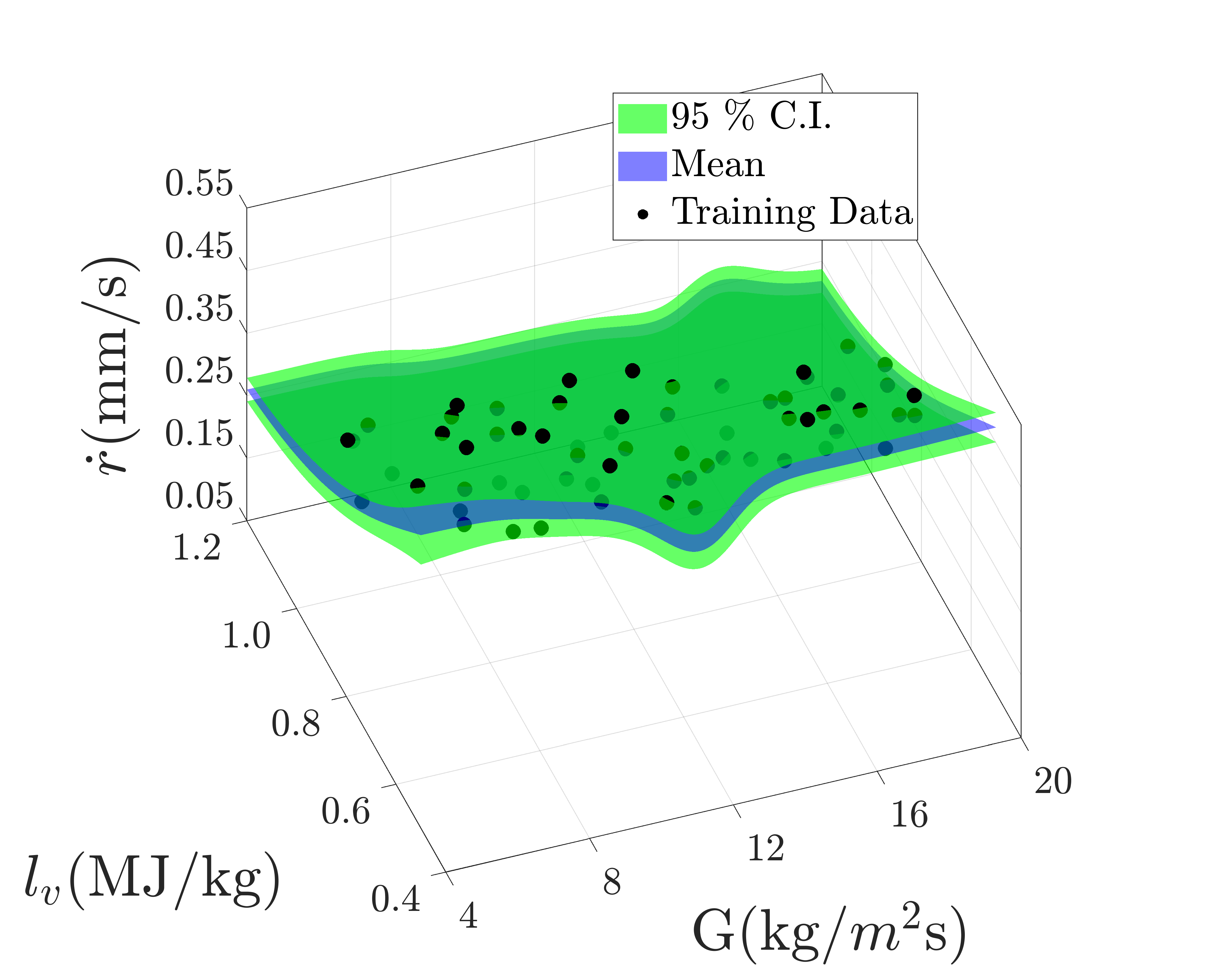}
    \\
    \includegraphics[trim=0.5in 0in 0.5in 0.0in, clip, width=.4\textwidth]{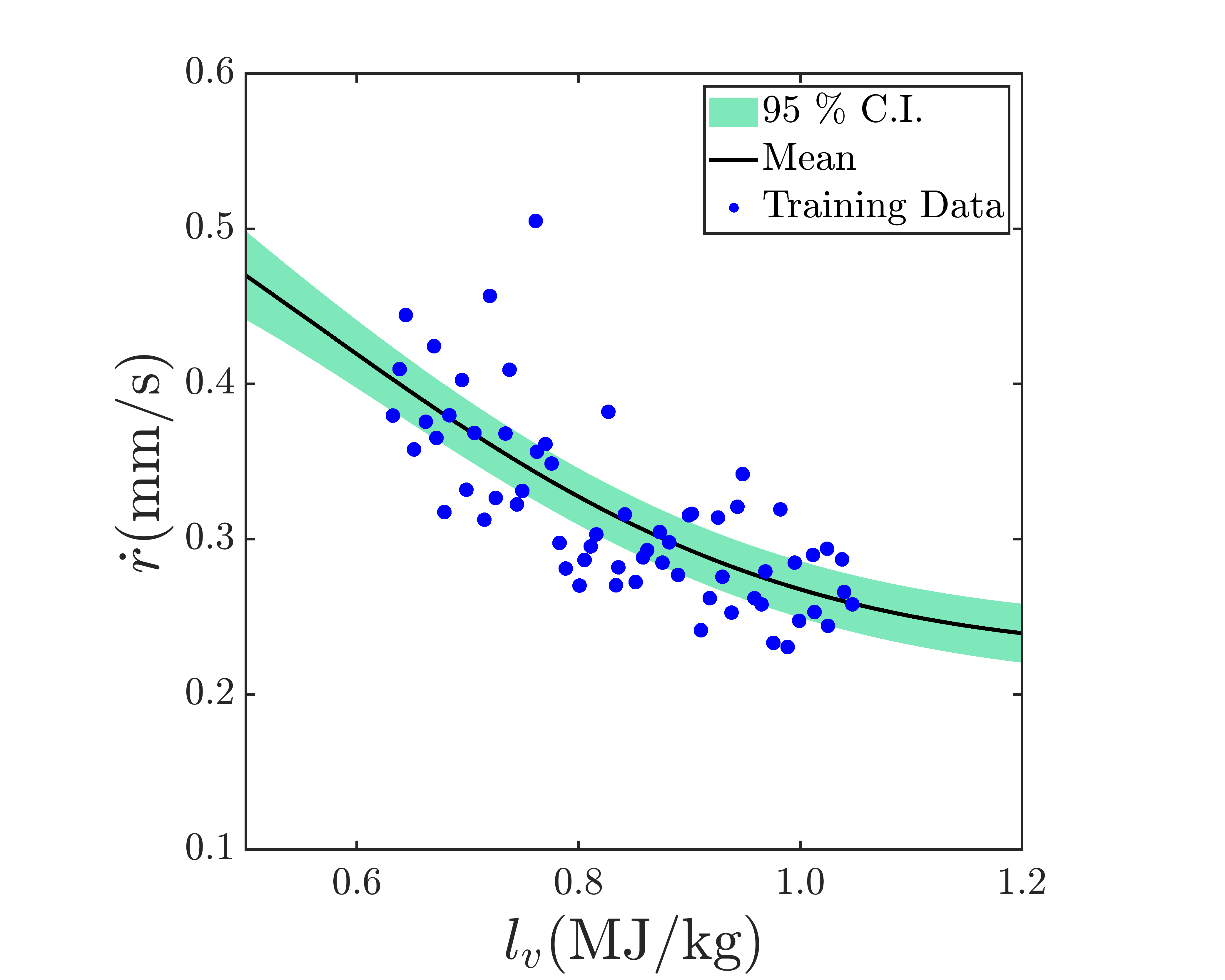}
    ~   
    \includegraphics[trim=0.5in 0in 0.5in 0in, clip, width=.4\textwidth]{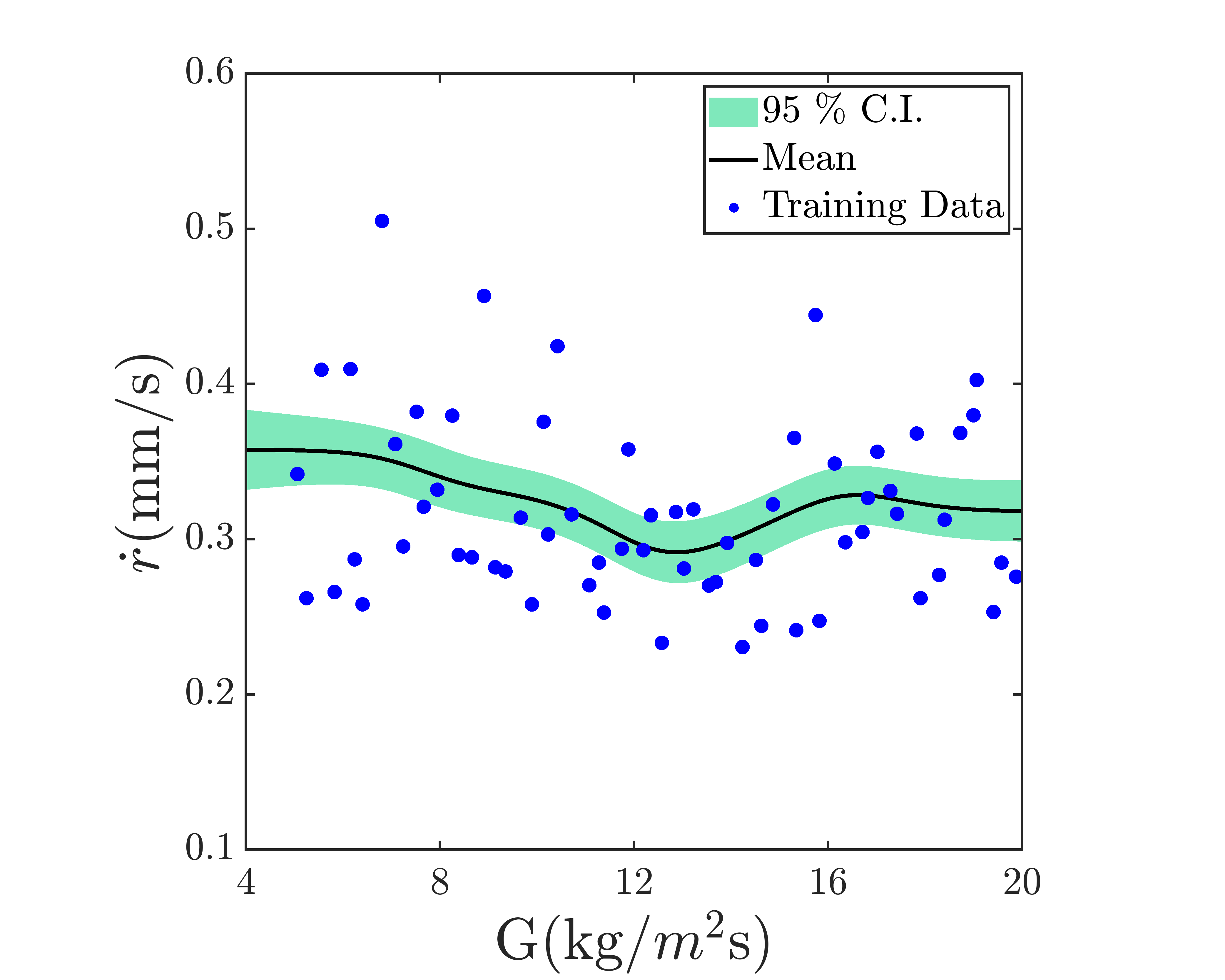}
    \\ (a) \\
    \includegraphics[trim=0.0in 0in 1.0in 0.0in, clip, width=.45\textwidth]{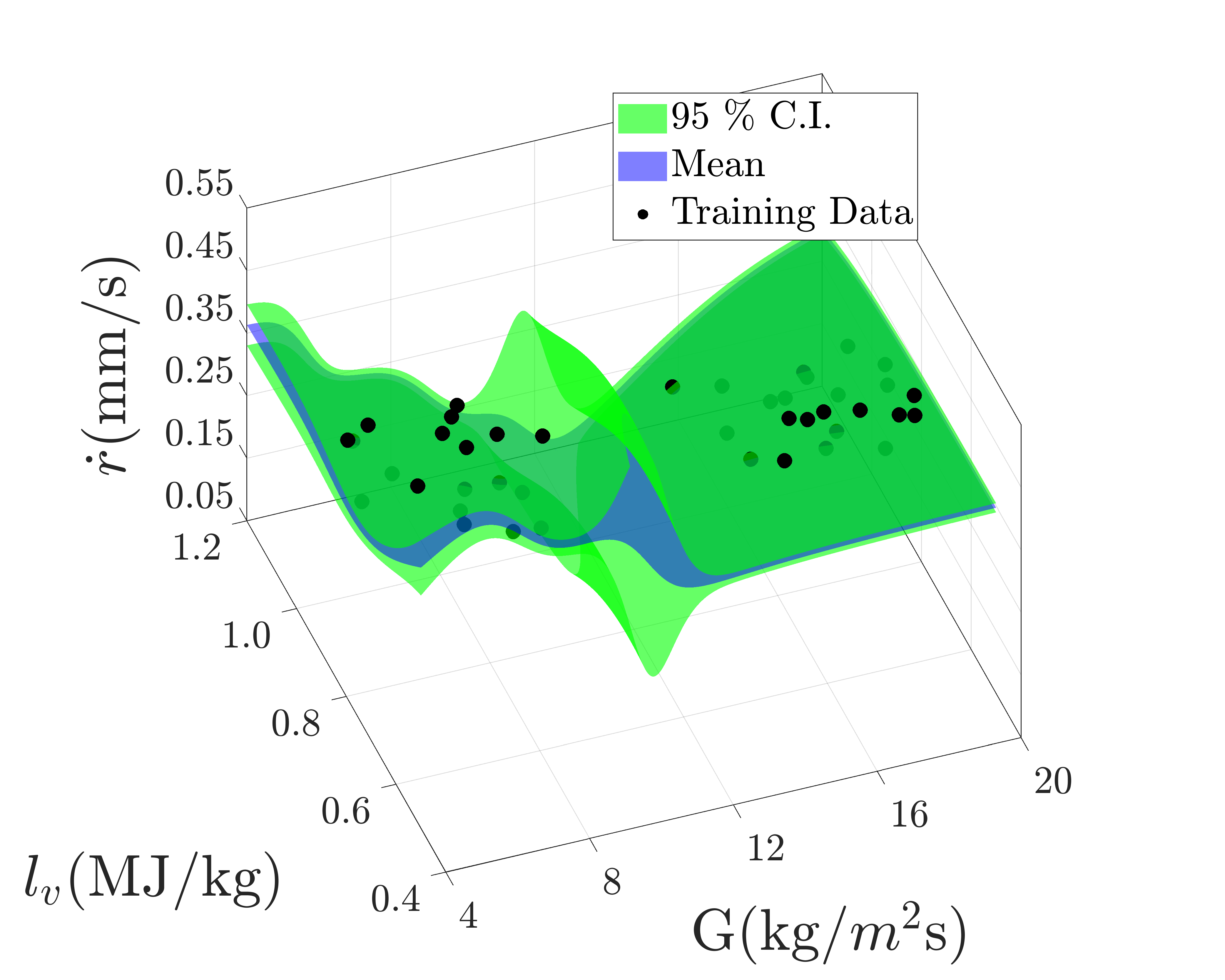}
    \\
    \includegraphics[trim=0.5in 0in 0.5in 0.0in, clip, width=.4\textwidth]{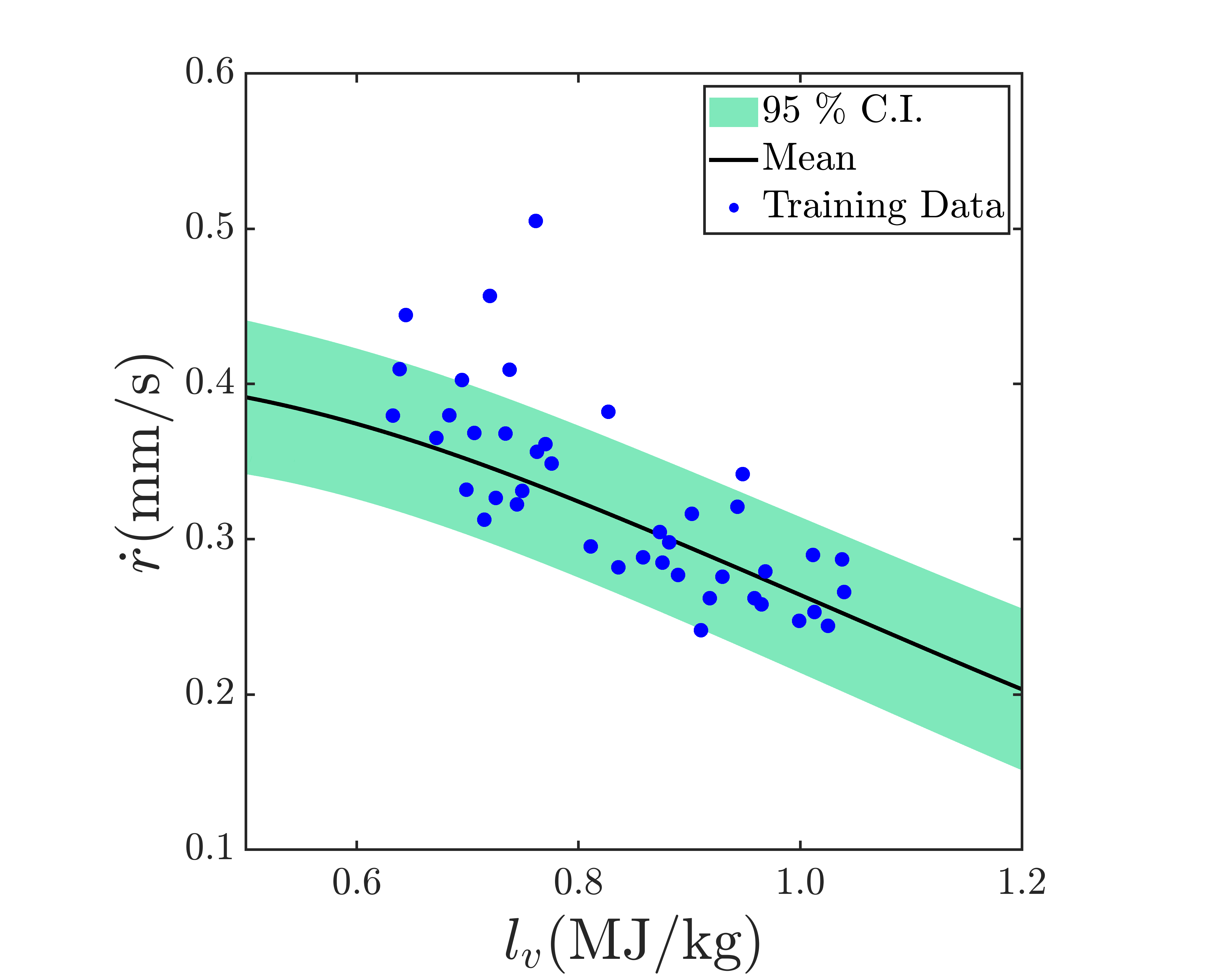}
    ~   
    \includegraphics[trim=0.5in 0in 0.5in 0in, clip, width=.4\textwidth]{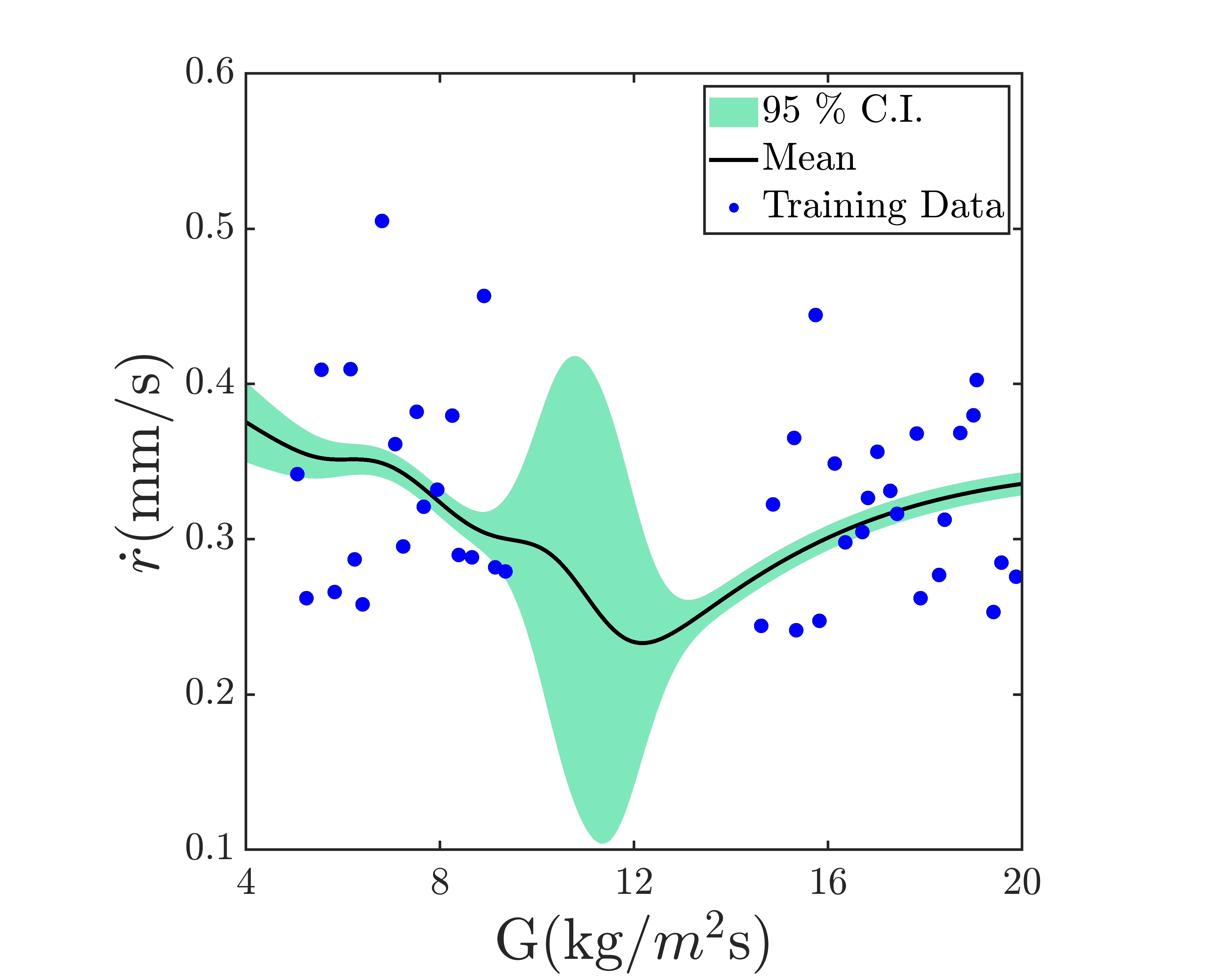}
    \\ (b)
    \vspace{-0.1in}
    \caption{
    Turbulent combustion problem:
    The mean and uncertainty predictions of the plausible sparsified network $M^1_I$ in Category 1 with $D=1$ and 245 connections in comparison with 
    (a) training data set; (b) leave-out data set.
    The top panel shows the 3D plot and the bottom panels display the marginalized plots over the $l_{v}$ and $G$.
    }
    \label{fig:combustion_opal_cat1_sparse}
\end{figure}

%===== Cat2 ======
%
\begin{figure}[htpb]
    \centering
    \vspace{-1.5 in}
    \includegraphics[trim=0.0in 0in 1.0in 0.0in, clip, width=.45\textwidth]{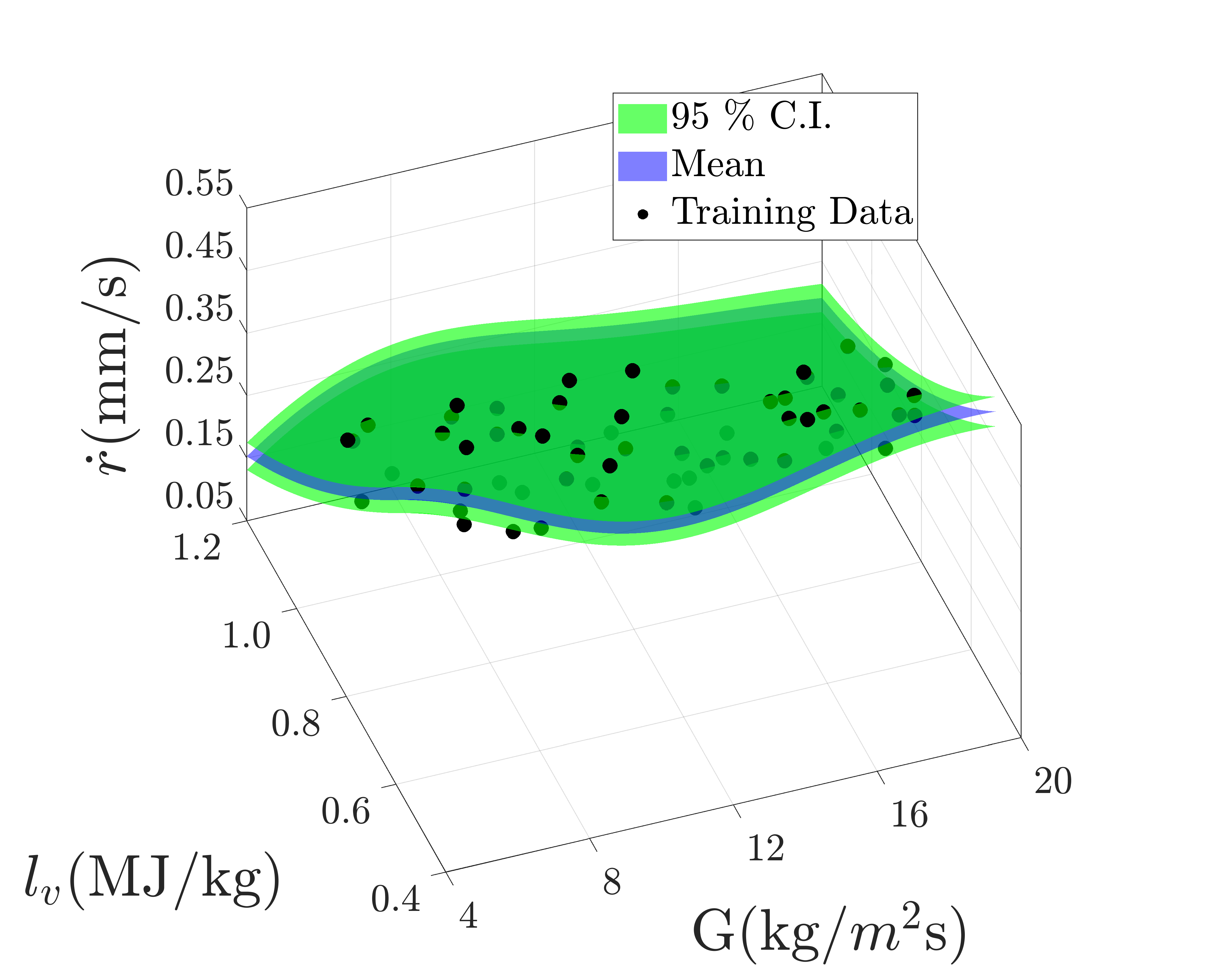}
    \\
    \includegraphics[trim=0.5in 0in 0.5in 0.0in, clip, width=.4\textwidth]{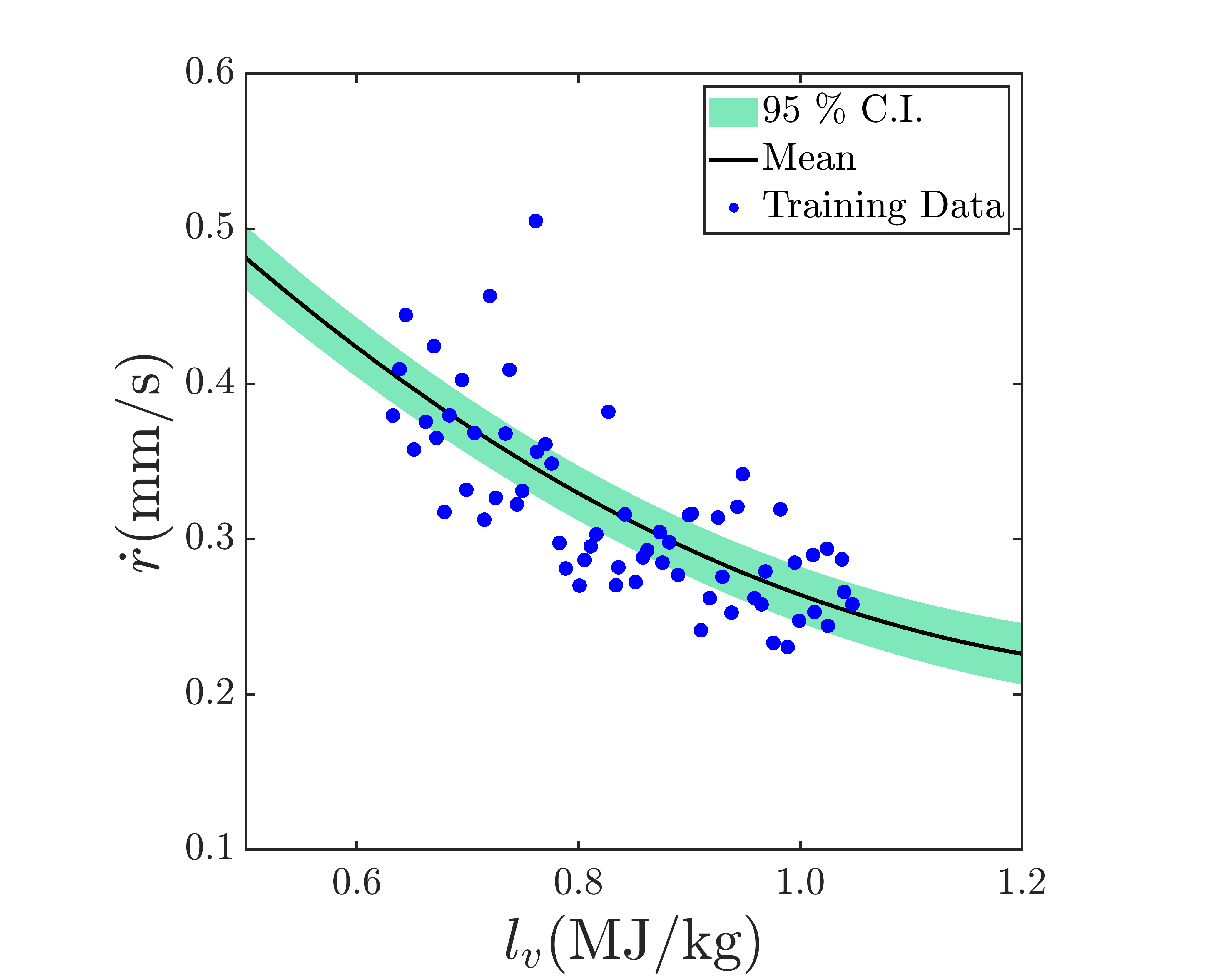}
    ~   
    \includegraphics[trim=0.5in 0in 0.5in 0in, clip, width=.4\textwidth]{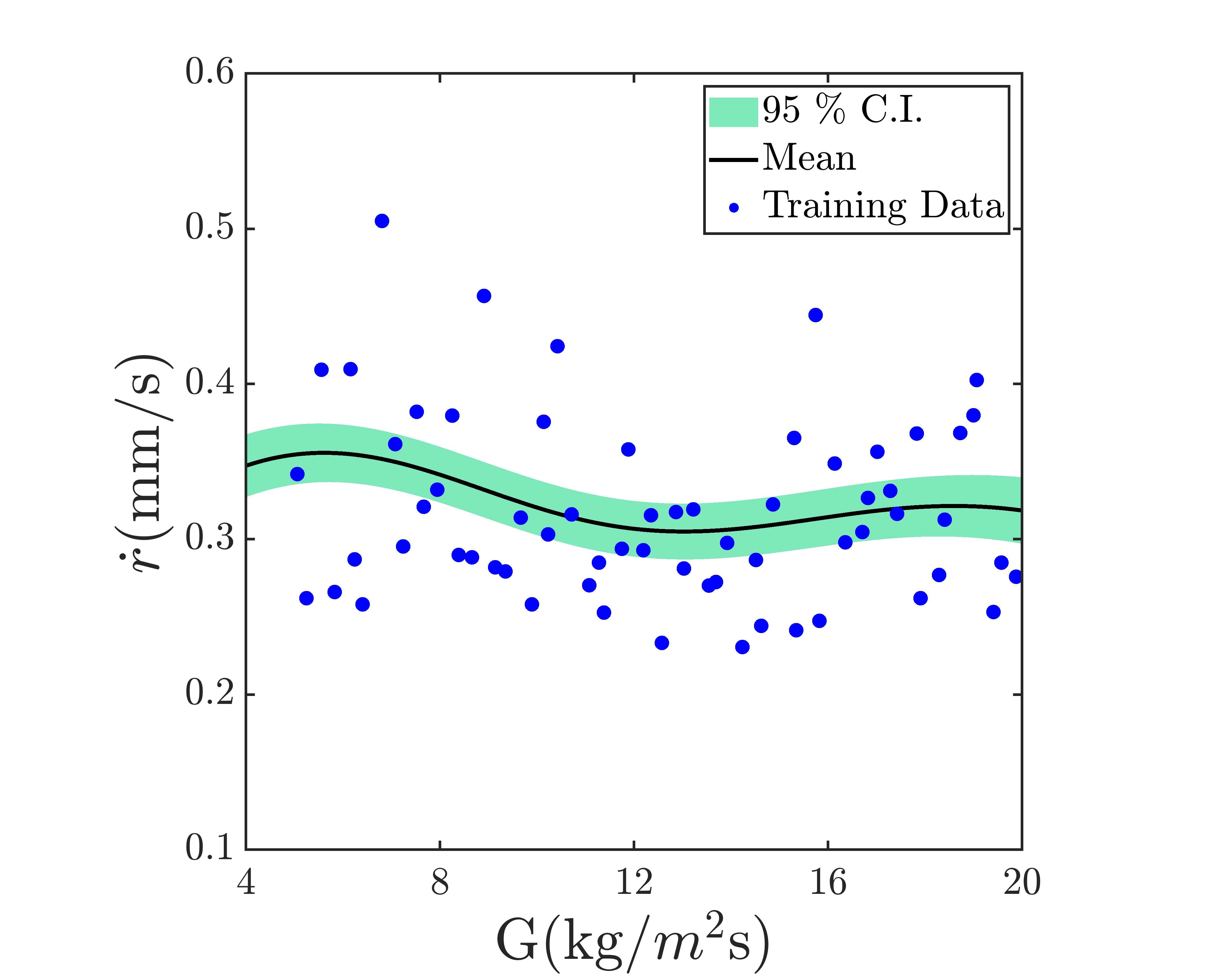}
    \\ (a) \\
    \includegraphics[trim=0.5in 0in 0.5in 0.0in, clip, width=.4\textwidth]{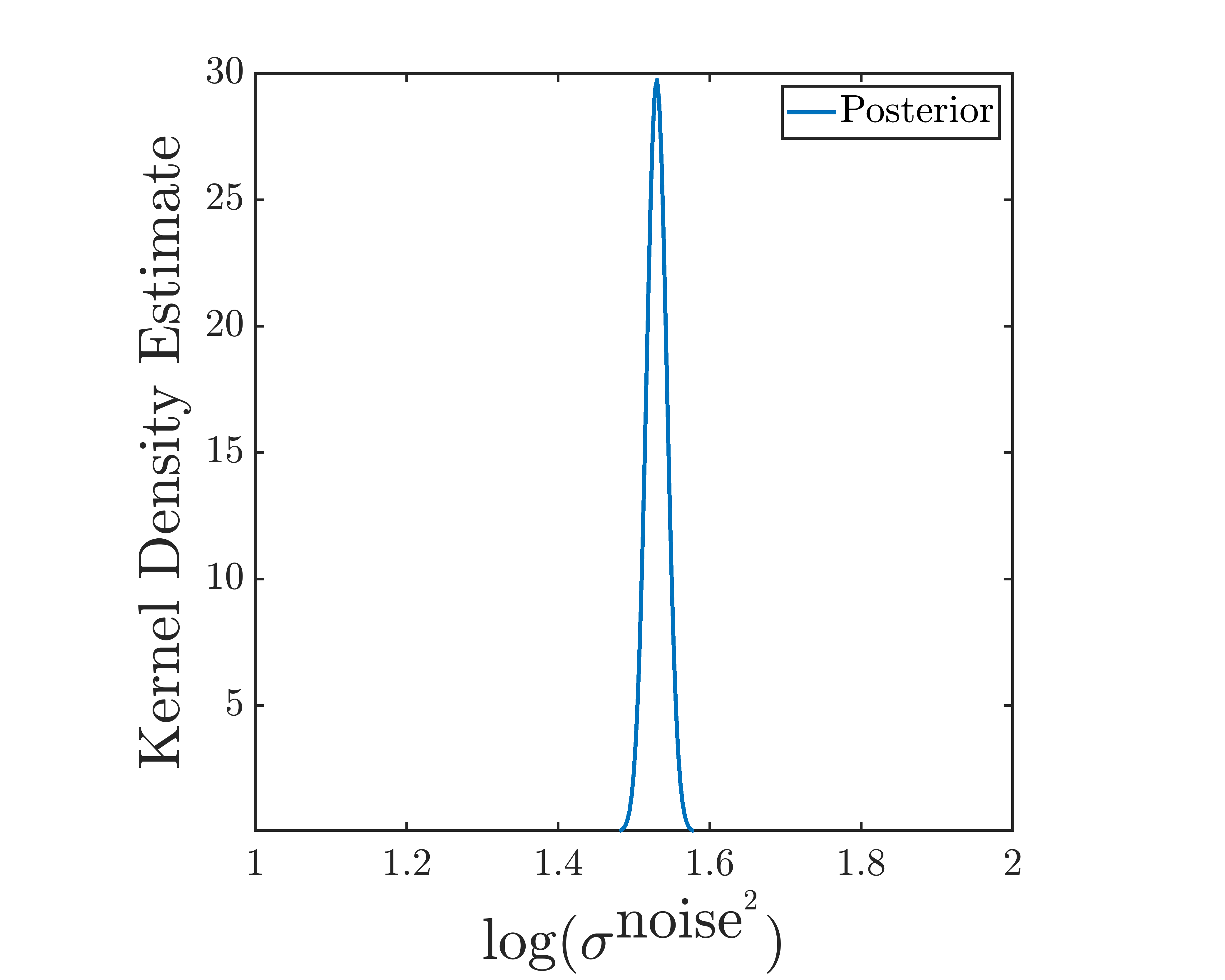}
    ~   
    \includegraphics[trim=0.5in 0in 0.5in 0in, clip, width=.4\textwidth]{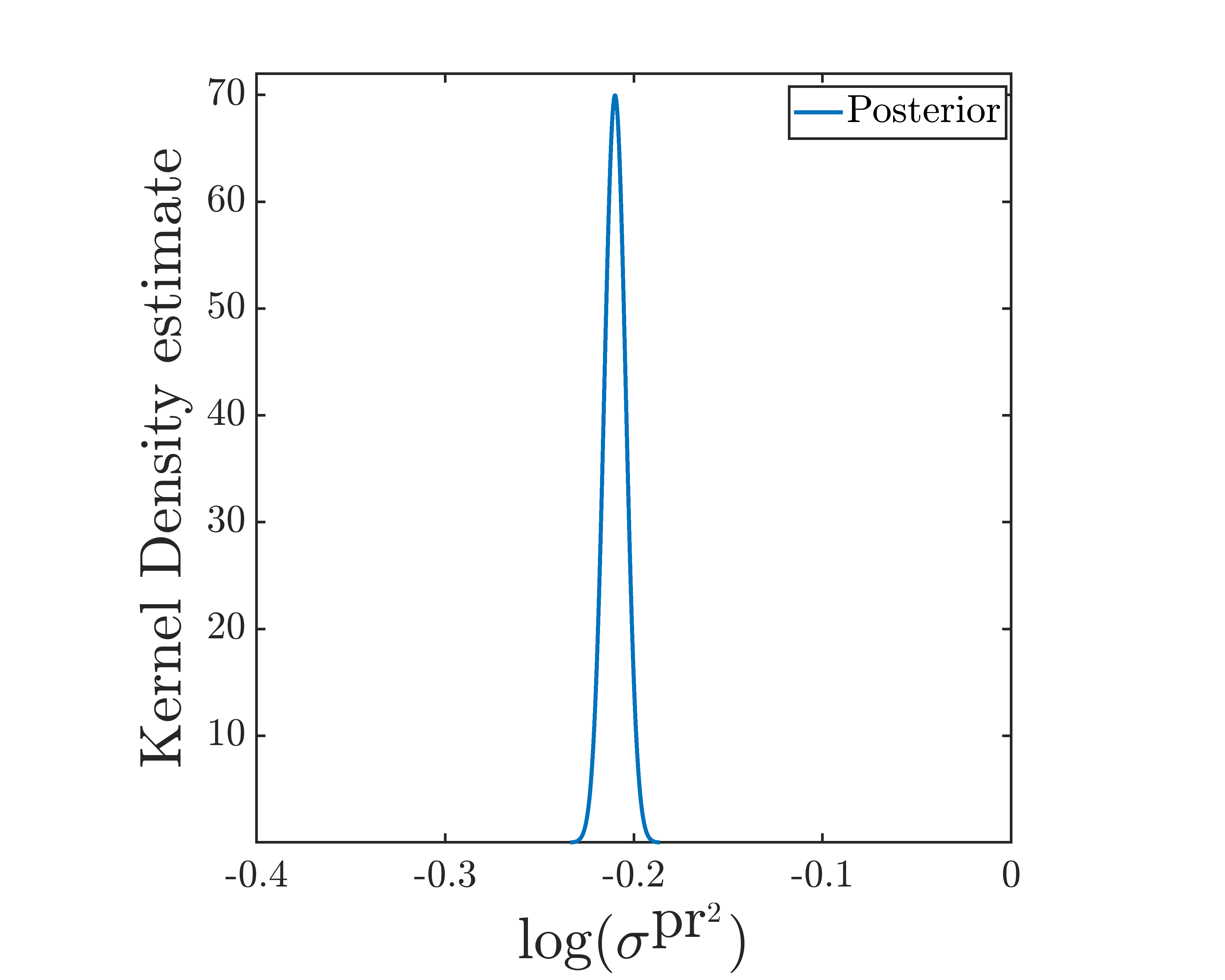}
    \\ (b) \\
    \includegraphics[trim=0.6in 3.1in 0.6in 2.5in, clip, width=.75\textwidth]{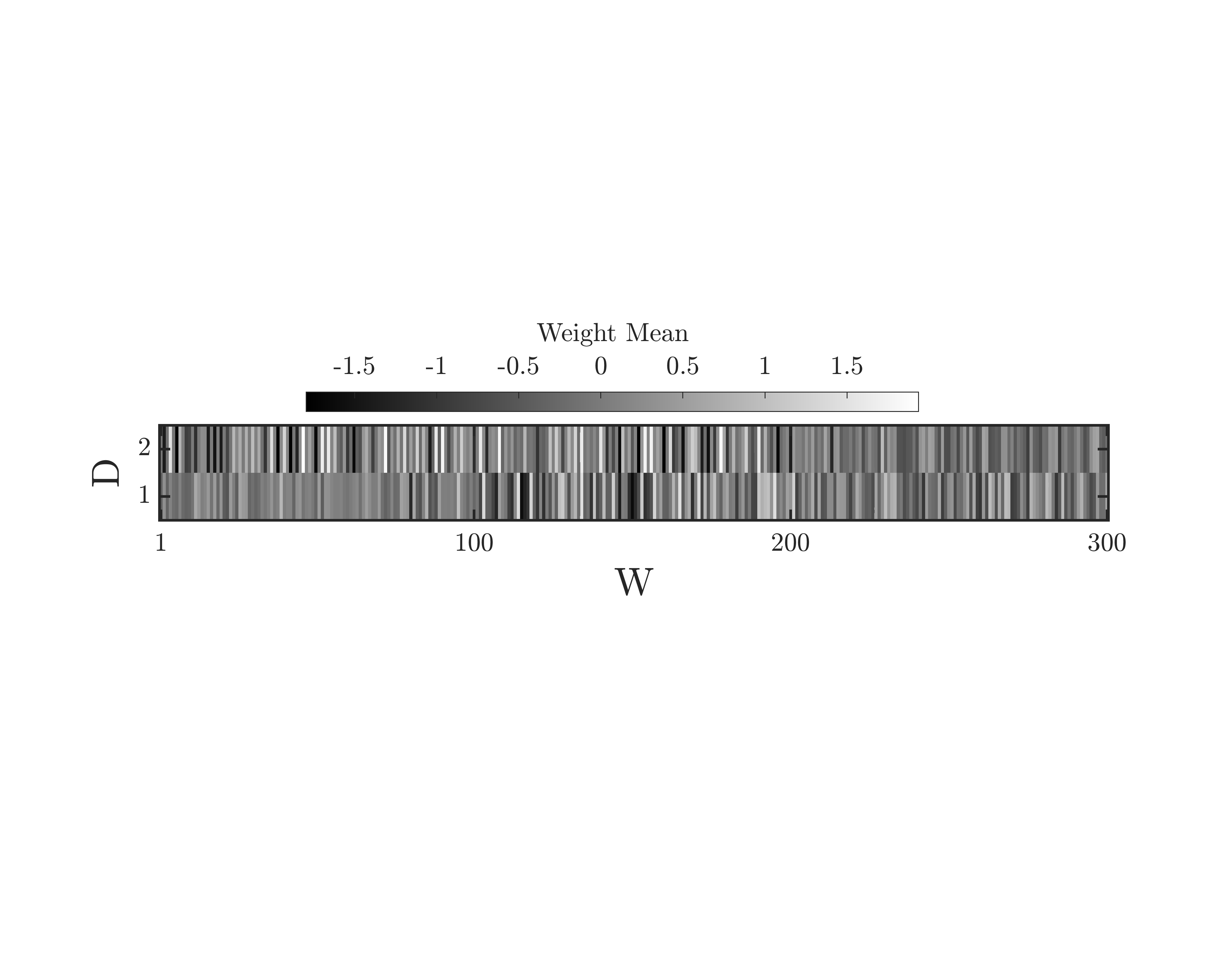} \\
    \includegraphics[trim=0.6in 3.1in 0.6in 3.3in, clip, width=.75\textwidth]{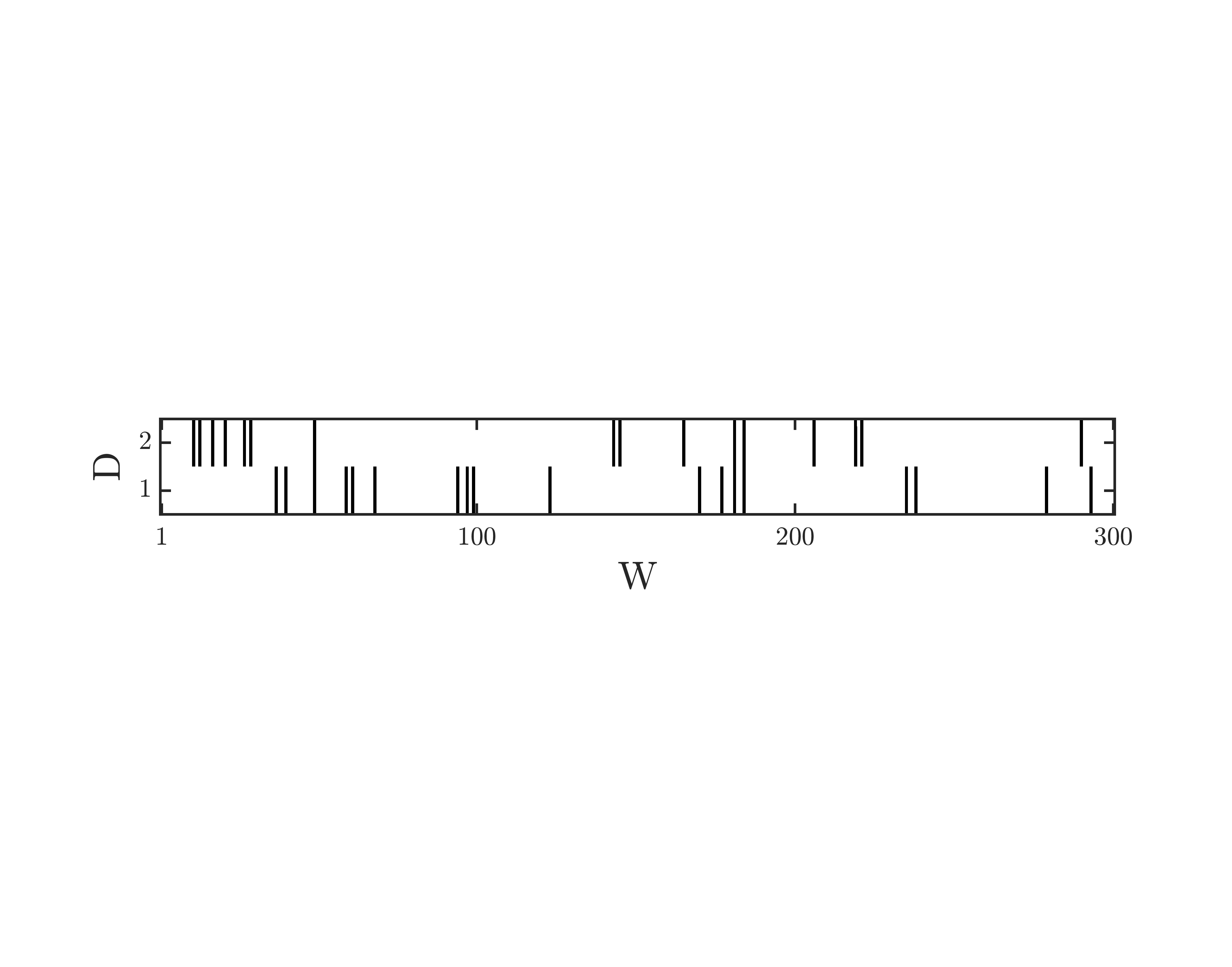}
    \\ (c) \\
    \vspace{-0.15in}
    \caption{
    Turbulent combustion problem:
    (a) The mean and uncertainty predictions of the plausible sparsified network $M^2_I$ in Category 2, identified as the ``best" surrogate model by OPAL-surrogate.
    The network consists of 2 layers, 65025 connections, and \textit{Tanh} activation functions for both layers. 
    (b) Posterior of the inference hyper-parameters \blue{considering the priors $\pi_{pr}(\log(\sigma^{{\rm pr^2}}))=\mathcal{U}(-10, 10)$ and $\pi_{pr}(\log(\sigma^{{\rm noise^2}}))=\mathcal{U}(-6, 6)$}.   
    (c) Visualization of weights pattern for $M^2_{F(D=2, W=300)}$ across network depth $D$ and width $W$ in the top panel. The bottom panel illustrates the sparsified pattern of $M^2_I$, with removed connections marked in black from each hidden layer using $TOL_{\bs \theta} = 0.12$. 
    }
    \label{fig:combustion_opal_cat2_sparse}
\end{figure}

Moving to Category 2,
 the \textit{Tanh} activation function is chosen for layer 2. Sparsification of $M^2_{F(D=2, W=300)}$ with 90000 connections, results in the plausible model $M^2_I$ for this category, comprising $D=2$ and 65025 connections, representing approximately 28\% reduction in parameters after sparsification with $TOL_{\bs \theta} = 0.12$. 
 As shown in Tabel \ref{table:opal_combustion}, $M^2_I$ successfully passes the validation test establishing it as the ``best'' predictive surrogate model.
Figure \ref{fig:combustion_opal_cat2_sparse} illustrates the predictions of $M^2_I$ in comparison to the data. The means of the posteriors of the inference hyperparameters shown in this figure are obtained by maximizing the evidence in \eqref{eq:evid_lev2_la}, and the posterior variances are approximated using \eqref{eq:var_hyper}.
As indicated in Table \ref{table:opal_combustion}, model $M^2_I$ demonstrates significantly improved predictive performance compared to $M^1_I$ and $M^3_I$ (see Figure \ref{fig:combustion_opal_cat3_sparse}) in both lower and higher categories. The comparison of validation observables is presented in Figure \eqref{fig:combustion_val_obs}.
\begin{figure}[htpb]
\vspace{-0.1in}
    \centering
    \includegraphics[trim=0.0in 0in 1.0in 0in, clip, width=.47\textwidth]{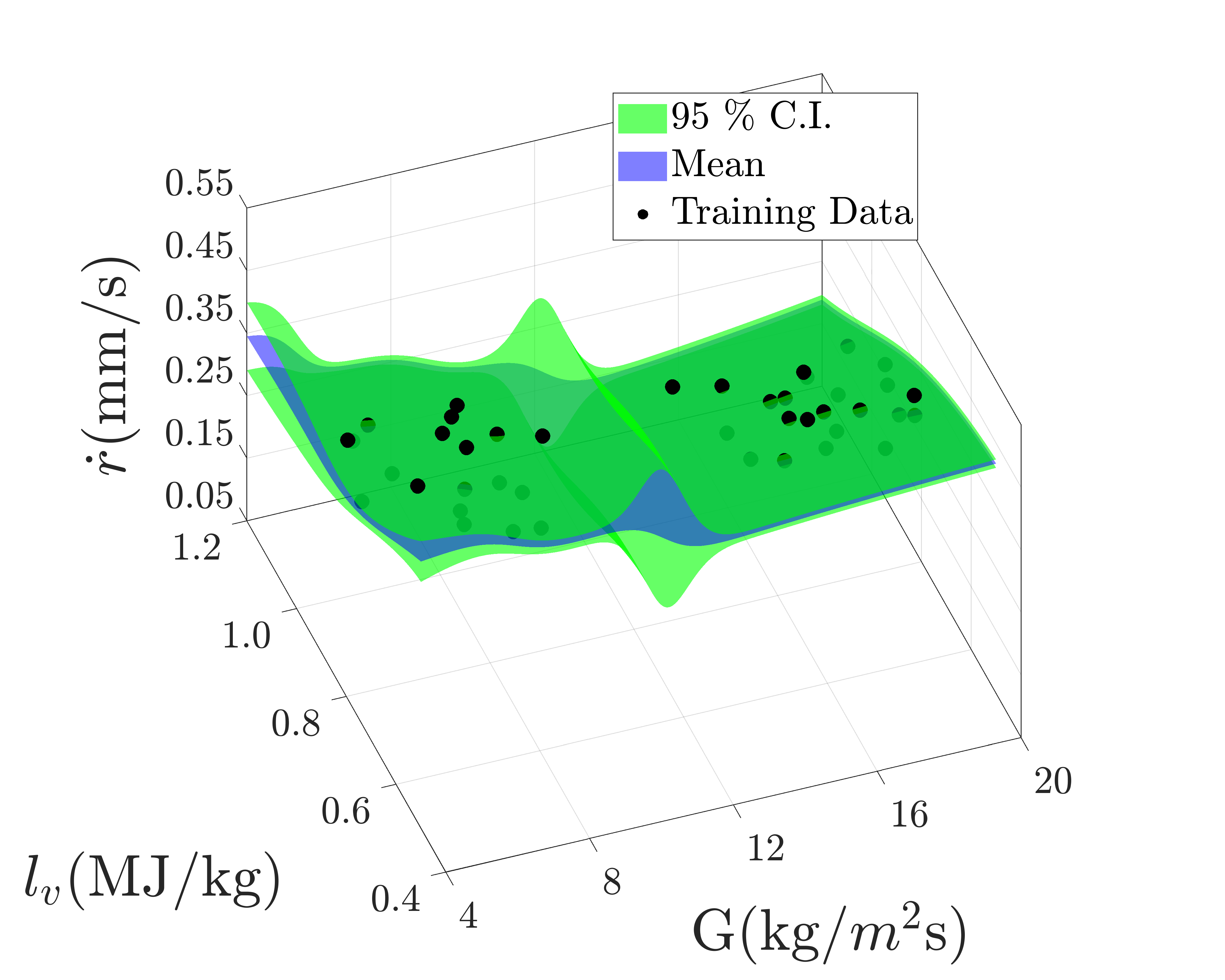}
    \\
    \includegraphics[trim=0.5in 0in 0.5in 0.0in, clip, width=.42\textwidth]{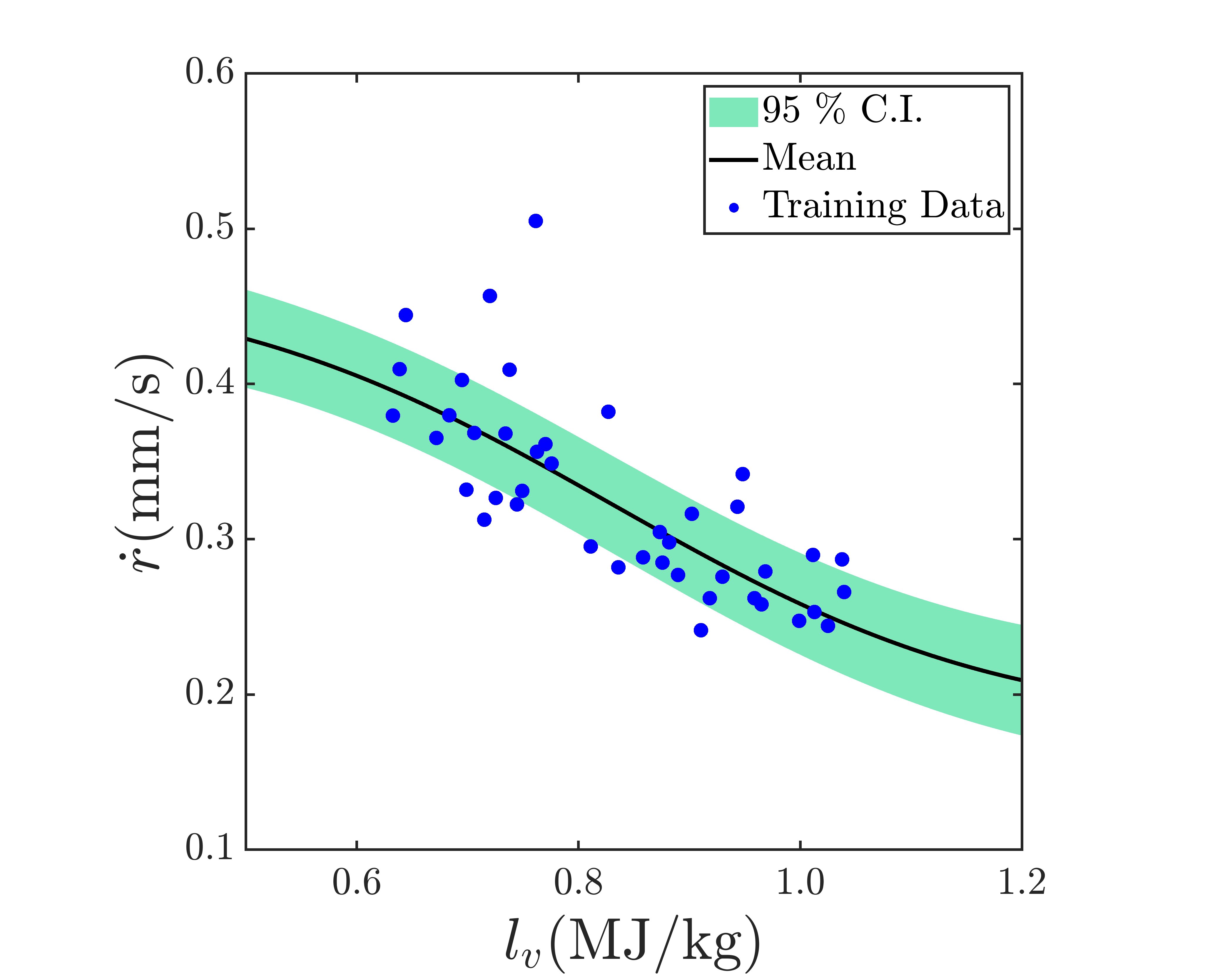}
    ~   
    \includegraphics[trim=0.5in 0in 0.5in 0in, clip, width=.42\textwidth]{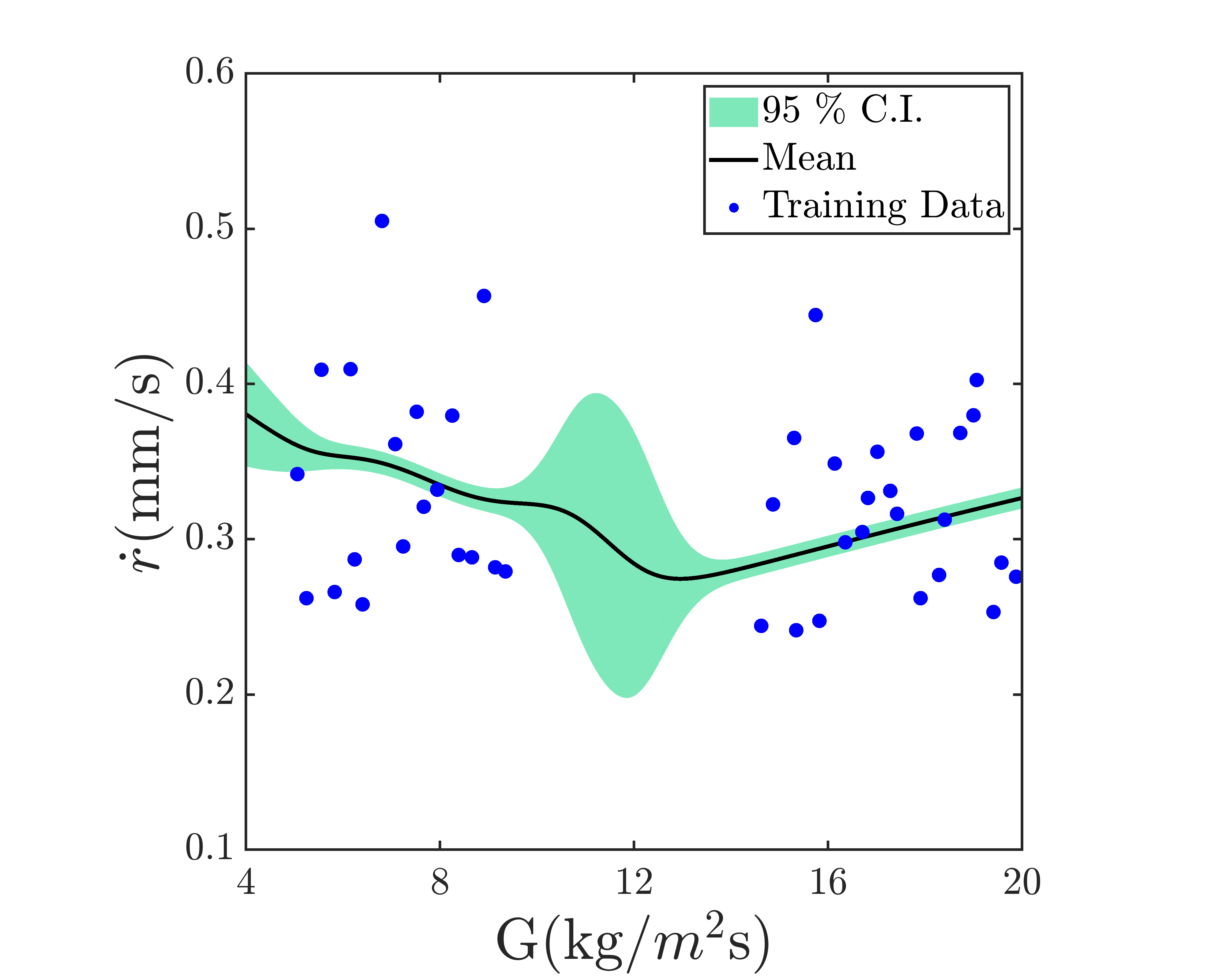}
    \vspace{-0.1 in}
    \caption{
    Turbulent combustion problem:
    The mean and uncertainty predictions of the plausible sparsified network $M^2_I$ in Category 2 in comparison with the leave-out data.
    The top panel shows the 3D plot and the bottom panels display the marginalized plots over the $l_{v}$ and $G$.
    }
    \label{fig:combustion_opal_cat2_lo}
\end{figure}
\begin{figure}[htpb]
\vspace{-0.1in}
    \centering
    \includegraphics[trim=1.0in 0in 1.6in 0in, clip, width=.35\textwidth]{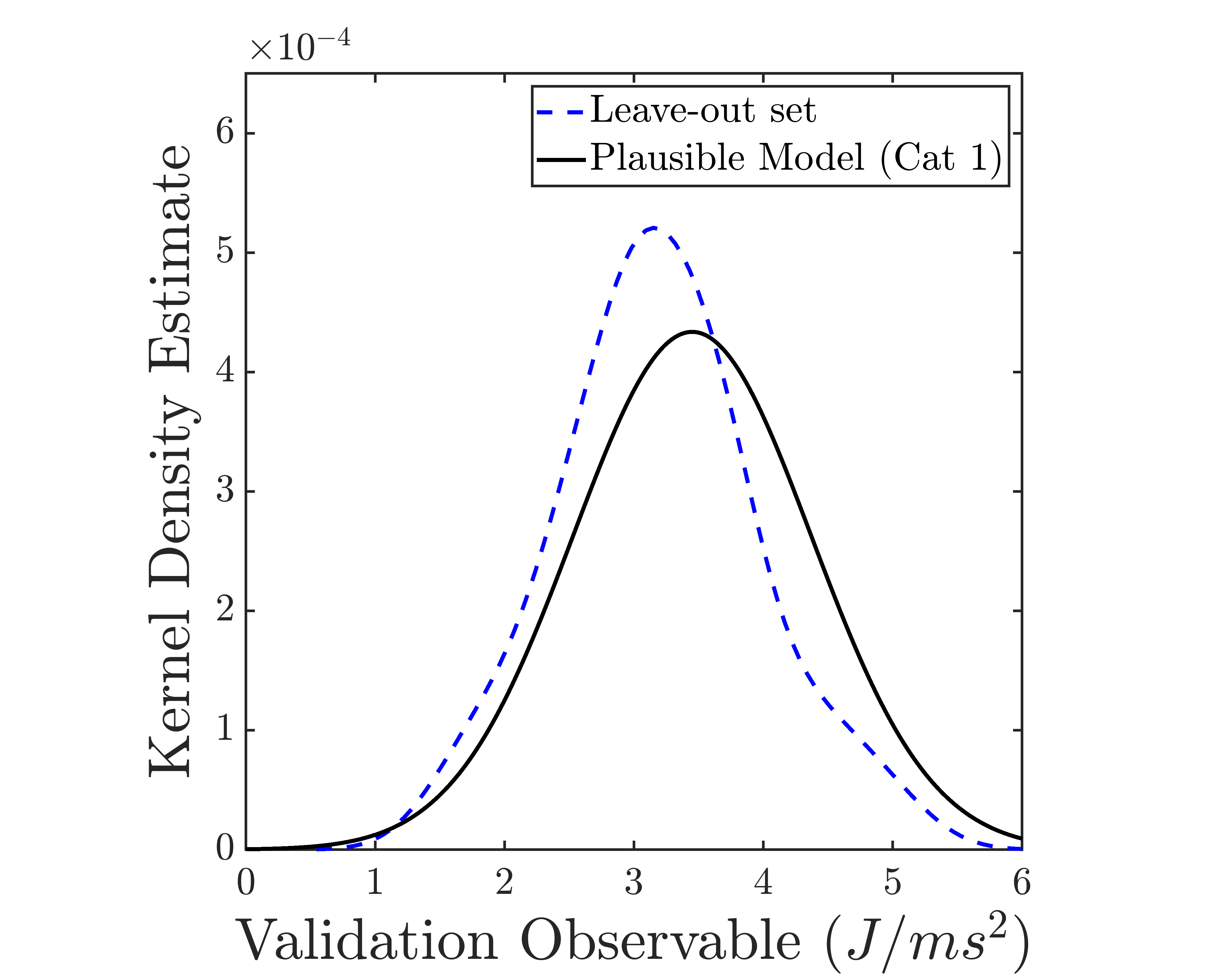}
     ~
    \includegraphics[trim=1.0in 0in 1.6in 0in, clip, width=.35\textwidth]{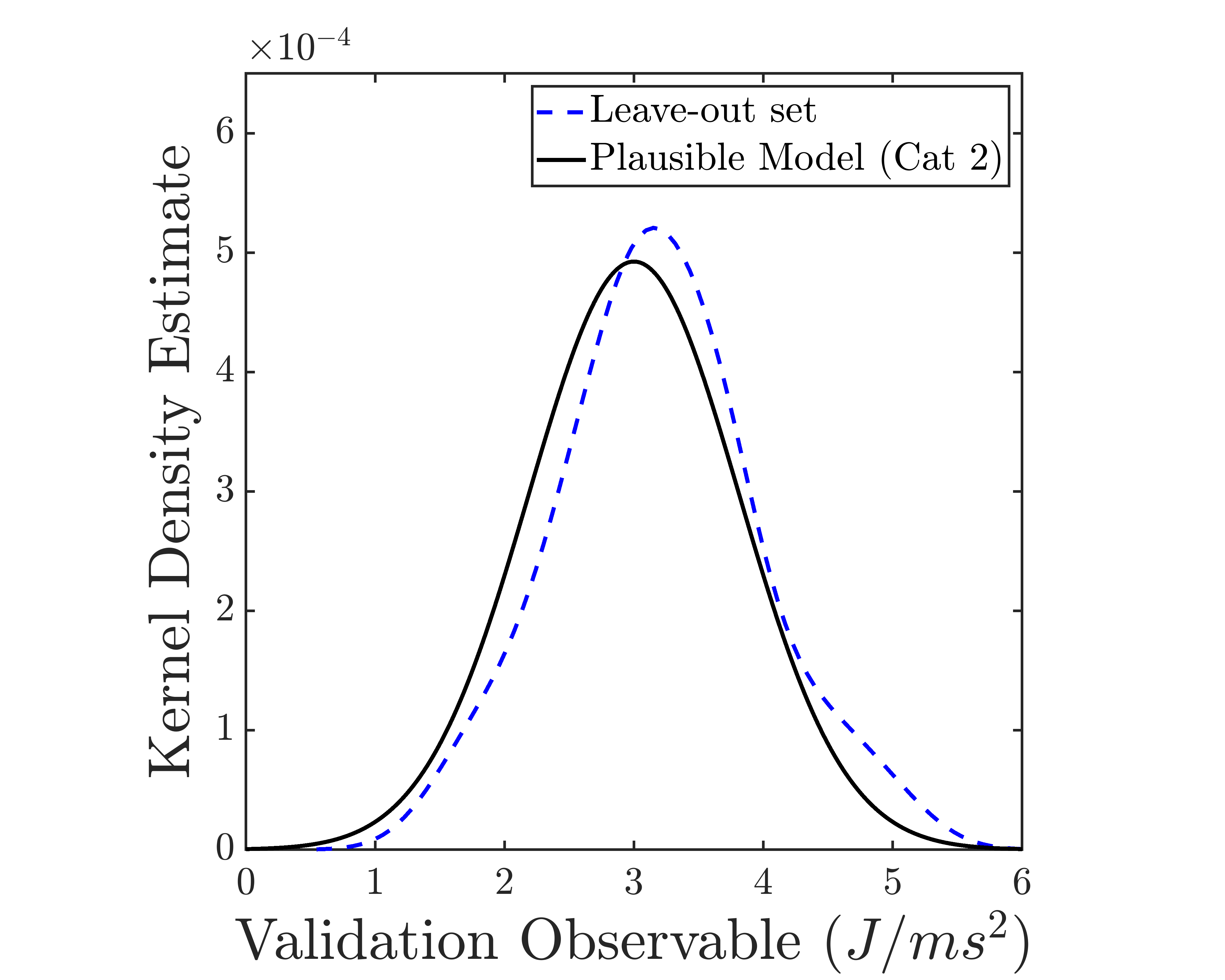}
    \\ (a) \hspace{2.2in} (b)
    \vspace{-0.1in}
    \caption{
    Turbulent combustion problem:
    Comparison of the observables $\mathcal{Z}_D$ in \eqref{eq:ablate_observ} corresponding to leave-out validation sets and the predicted $\mathcal{Z}_M$ from plausible models $M^1_I$ and $M^2_I$.
    }
    \label{fig:combustion_val_obs}
\end{figure}

\end{paragraph}

%++++++++++++++++++++++++++++++++++++++++++++++++++++++++++++++++++++++++
\section{Conclusions}\label{sec:conclusions}

%-------------------------------------------------------
% 1. Summarize the key methods and results (two paragraphs)
%-------------------------------------------------------
\noindent
This study introduces OPAL-surrogate, a systematic framework for identifying predictive BayesNN surrogate models within the expansive space of potential models characterized by diverse architectures and hyperparameters. Leveraging hierarchical Bayesian inferences and the concept of model plausibility, OPAL-surrogate efficiently and adaptively adjusts model complexity until satisfying validation criteria. We also stress the significance of well-organized neural network parameters, empowering the surrogate model to effectively capture multiscale interactions encoded in the training data for physics-based simulations. To achieve this, we propose a method involving the sequential addition of fully connected layers with large widths and the elimination of irrelevant weights through an effective network sparsification guided by model evidence.

\begin{figure}[htpb]
\vspace{-0.5in}
    \centering
    \includegraphics[trim=0.0in 0in 1.0in 0in, clip, width=.45\textwidth]{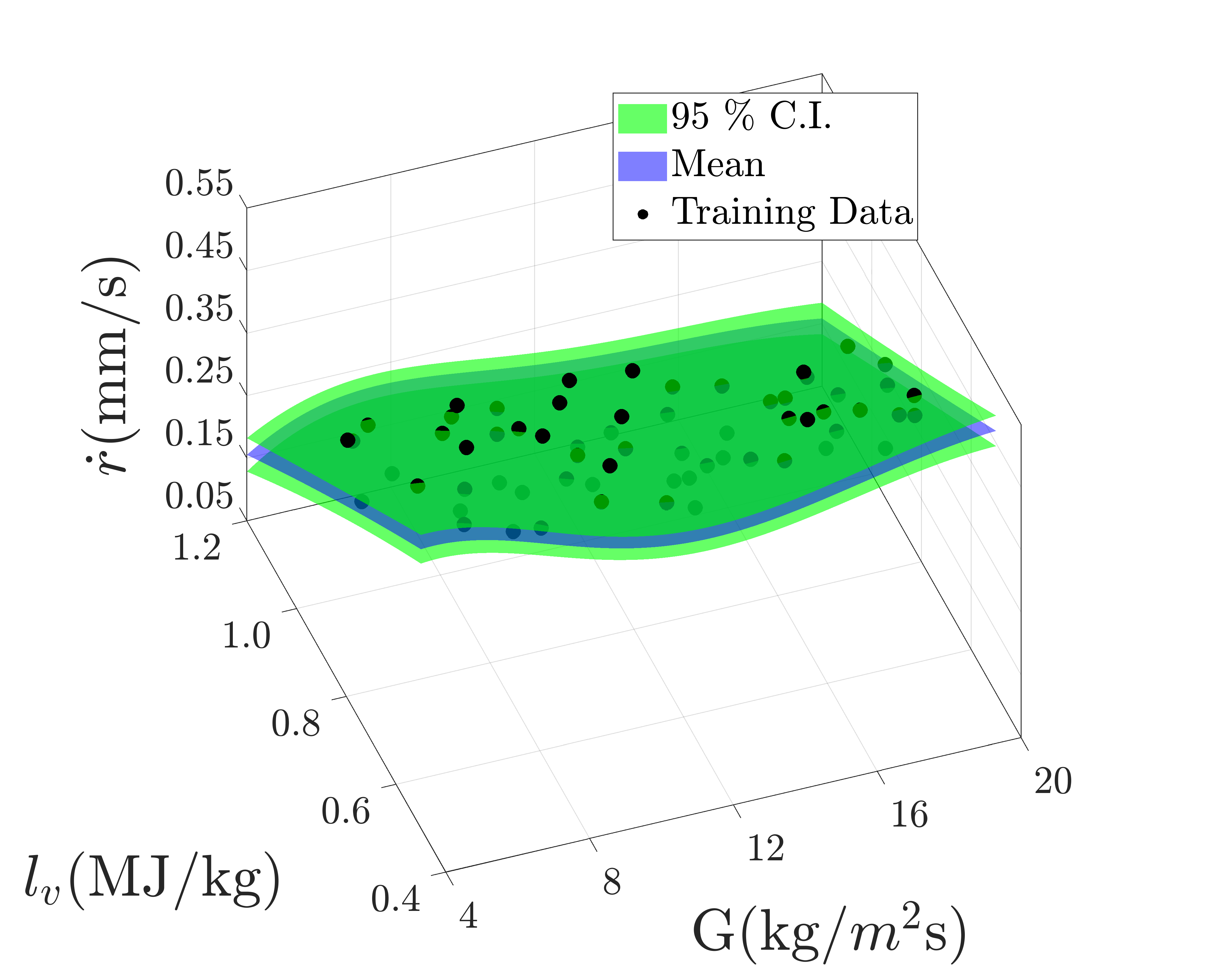}
    \\
    \includegraphics[trim=0.5in 0in 0.5in 0.0in, clip, width=.4\textwidth]{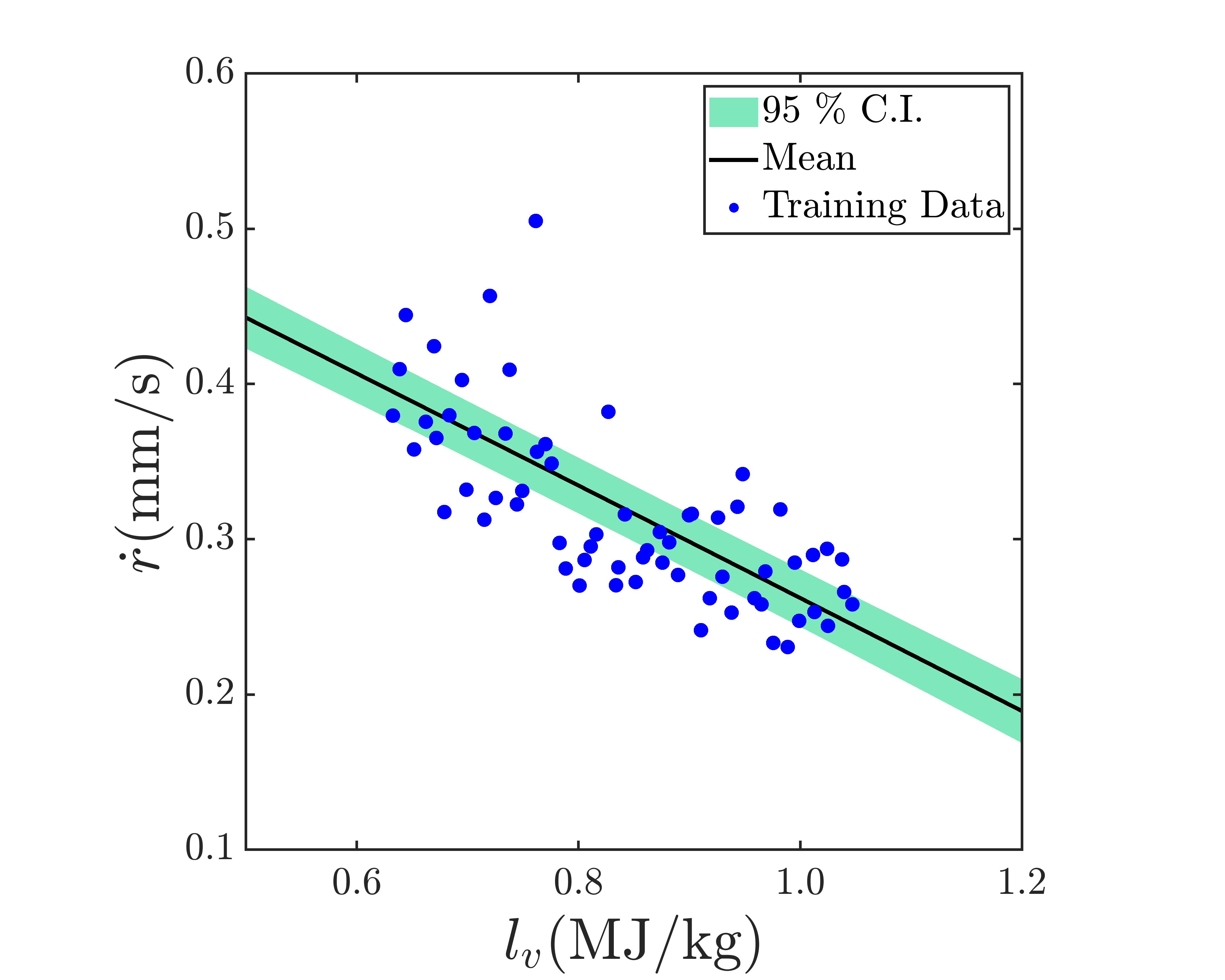}
    ~   
    \includegraphics[trim=0.5in 0in 0.5in 0in, clip, width=.4\textwidth]{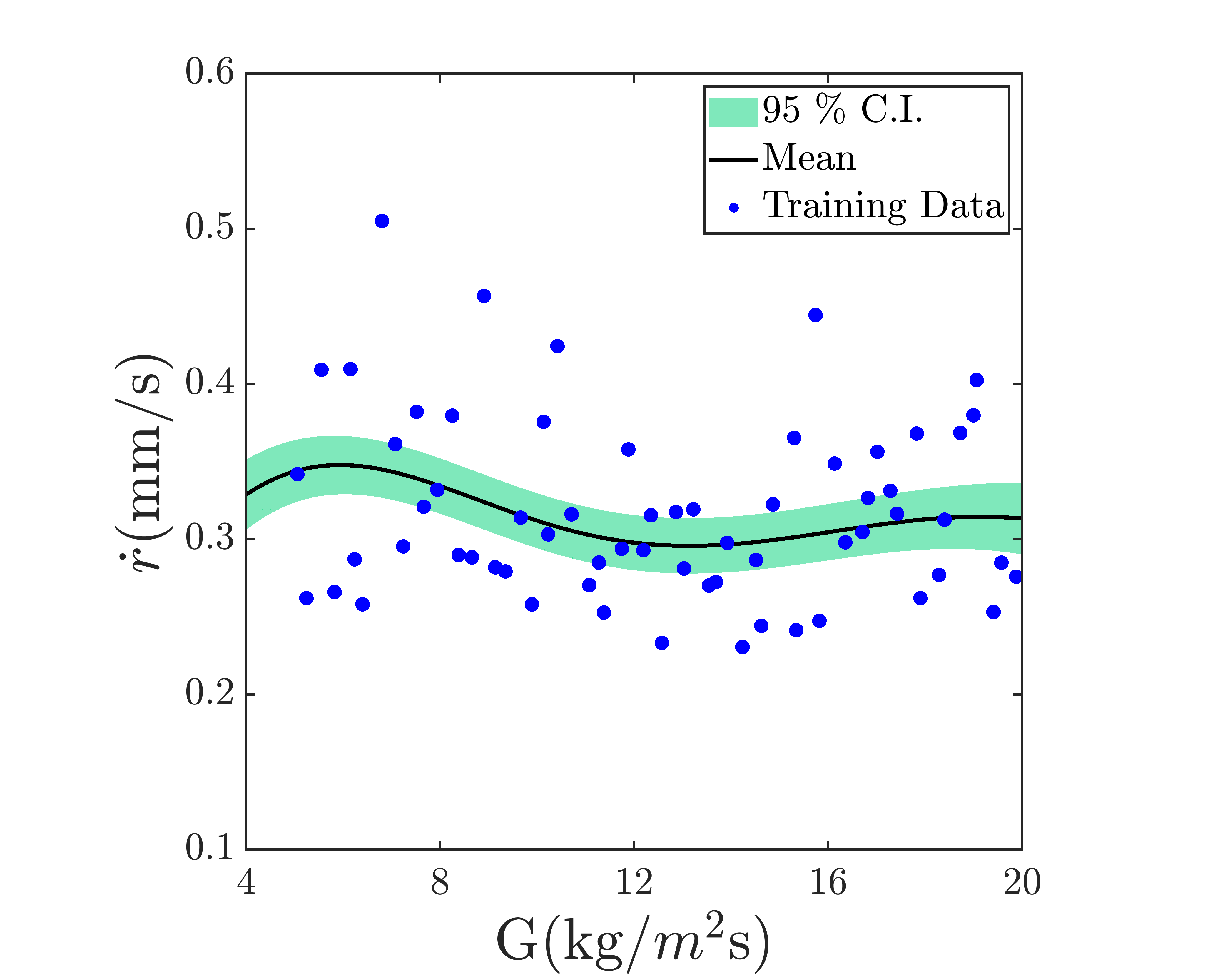}
    \vspace{-0.1in}
    \caption{
    Turbulent combustion problem:
    The mean and uncertainty predictions of the plausible sparsified network $M^3_I$ in Category 3 with $D=3$ and 135000 connections in comparison with training data.
    The top panel shows the 3D plot and the bottom panels display the marginalized plots over the $l_{v}$ and $G$.
    }
    \label{fig:combustion_opal_cat3_sparse}
    \vspace{-0.2in}
\end{figure}

Two applications of OPAL-surrogate demonstrate that the identified architecture and hyperparameters of the BayesNN model, achieved through a balanced trade-off between model complexity and validity, result in enhanced accuracy and reliability in predictions. The first example involves surrogate modeling of elastic deformation in porous materials, aiming to facilitate quantifying uncertainty in unobservable QoI for domain sizes where high-fidelity simulation is computationally prohibitive.
The results suggest that, despite a substantial training dataset comprising 1500 data points and constructing the prior for network parameters from the pre-training dataset, the extrapolation prediction capability of the identified BayesNN model remains credible only up to 1.5 times the size of the training domain. Beyond this point, the accuracy and reliability of the prediction rapidly decline, highlighting the necessity of imposing physical constraints to enhance the extrapolation abilities of neural networks, e.g., \cite{ raissi2019physics, jha2024, li2023isogeometric}.
The second numerical experiment involves a turbulent combustion flow model of a slab burner, where OPAL-surrogate successfully identifies the BayesNN surrogate model for the fuel regression rate based on 64 training data points.

We highlight a fundamental misconception contributing to overconfident predictions -- the notion that the complexity of a predictive model must always be limited when training data is scarce. In contrast, the OPAL-surrogate framework relies on Bayesian inference without the constraint of modifying the model and prior based on the data volume. In fact, we argue that there is never enough high-fidelity data from physical simulations for surrogate model construction. Thus, OPAL-surrogate embodies an open-ended inference approach, continuously refining the predictive surrogate model, as that additional data and information could potentially \textit{falsify} a model initially presumed to be valid. 
In constructing the initial set of possible models, \cite{mackay1995probable, box1979, box2011bayesian} propose incorporating models we genuinely believe in, along with every conceivable sub-model, ensuring that the model selection strategy can identify the sub-model that best explains the data.

% 2. Limitations of the study and possible extensions (one paragraph)
%-------------------------------------------------------

In advancing the surrogate modeling using the proposed framework in this work, several avenues can be explored in future studies.
Firstly, more accurate Bayesian solutions may be adopted, recognizing that the posterior distribution is only asymptotically normally distributed as the number of data points approaches infinity, and in neural networks, they might be multi-modal. Despite this, given the Laplace approximation's high efficiency and scalability compared to other existing algorithms, it may be worthwhile to expand the definition of the BayesNN model, incorporating various solutions with differing complexities into the initial model set. Subsequently, OPAL-surrogate can discern the one that provides valid predictions. This approach is justified by the primary goal of the surrogate model, which is to faithfully approximate the predictive distribution rather than the accuracy of parameter inference.
Secondly, there is a need for comprehensive investigations into effective methods for revealing neural network sparsity patterns associated with multiscale physical phenomena. This can involve exploring techniques like probabilistic sparse masks, e.g., \cite{zhou2021effective, williams1995bayesian} and leveraging architecture search algorithms, e.g., \cite{optuna_2019}, with the potential to enhance sparsification efficiency after replacing the performance measure with model evidence.
%
%Thirdly, expanding upon \cite{paquette2023optimal, tan2022toward}, a \textit{goal-oriented design} of training scenarios within Step 5 of OPAL-surrogate could be investigated. This involves employing active subspace methods to generate high-fidelity data that accurately reflects the structure of the prediction QoI.
%
%Fourthly, addressing the unresolved challenge of surrogate model-form uncertainty is imperative for achieving improved accuracy and reliability in extrapolation. One potential path involves exploiting the underlying structure of high-fidelity solutions through a-\textit{posteriori error estimation} and its associated computable uncertainty bounds, e.g., \cite{oden2002estimation, vemaganti2001estimation, oden2015amses, jha2024, cao2023residual, jha2022goal}.
%
Lastly, in the future, we aim to enhance the versatility of OPAL-surrogate to accommodate a broader range of possible models, e.g., different classes of neural operators \cite{oLeary2022learning, oLeary2022derivative, paris2021learning, Karniadakis2019deeponet, li2020fourier, Karniadakis2023physics}. This expansion could significantly reinforce its capacity to identify the appropriate surrogate model for a given problem.

%-------------------------------------------------------
% 3. Summarize the main take-home points from the study
%-------------------------------------------------------

In conclusion, this study highlights the substantial challenges associated with the discovery and assessing the credibility of neural network-based surrogate models for complex multiscale and multiphysics simulations.
The introduced framework aims to tackle some of these challenges by highlighting the crucial interplay between model complexity and rigorous validation. It provides a foundation for future research to enhance its efficacy and extend its applicability across diverse classes of surrogate models.

%++++++++++++++++++++++++++++++++++++++++++++++++++++++++++++++++++++++++
\section*{Acknowledgments}

DF and PKS extend their sincere gratitude for the financial support received from the U.S. National Science Foundation (NSF) CAREER Award CMMI-2143662. 
DF also appreciates the partial support from the U.S. Department of Energy’s (DoE) National Nuclear Security Administration (NNSA) under the Predictive Science Academic Alliance Program III (PSAAP III) DE-NA0003961.
Additionally, the authors would like to acknowledge the support provided by the Center for Computational Research at the University at Buffalo.

Sandia National Laboratories is a multimission laboratory managed and operated by National Technology \& Engineering Solutions of Sandia, LLC, a wholly owned subsidiary of Honeywell International Inc., for the U.S. Department of Energy’s National Nuclear Security Administration under contract DE-NA0003525. This paper describes objective technical results and analysis. Any subjective views or opinions that might be expressed in the paper do not necessarily represent the views of the U.S. Department of Energy or the United States Government.

%========================================================================
% Appendix
% \appendix

% \section{Sample Appendix Section}
% \label{sec:sample:appendix}
% \blue{
% Lorem ipsum dolor sit amet, consectetur adipiscing elit, sed do eiusmod tempor section \ref{sec:sample1} incididunt ut labore et dolore magna aliqua. Ut enim ad minim veniam, quis nostrud exercitation ullamco laboris nisi ut aliquip ex ea commodo consequat. Duis aute irure dolor in reprehenderit in voluptate velit esse cillum dolore eu fugiat nulla pariatur. Excepteur sint occaecat cupidatat non proident, sunt in culpa qui officia deserunt mollit anim id est laborum.
% }

%========================================================================
% Bibliography
%========================================================================


 \bibliographystyle{elsarticle-num} 
 \bibliography{refs}

\begin{thebibliography}{10}
\expandafter\ifx\csname url\endcsname\relax
  \def\url#1{\texttt{#1}}\fi
\expandafter\ifx\csname urlprefix\endcsname\relax\def\urlprefix{URL }\fi
\expandafter\ifx\csname href\endcsname\relax
  \def\href#1#2{#2} \def\path#1{#1}\fi

\bibitem{TRIPATHY2018565}
R.~K. Tripathy, I.~Bilionis, Deep {UQ}: Learning deep neural network surrogate models for high dimensional uncertainty quantification, Journal of Computational Physics 375 (2018) 565--588.

\bibitem{ZHU2018415}
Y.~Zhu, N.~Zabaras, Bayesian deep convolutional encoder–decoder networks for surrogate modeling and uncertainty quantification, Journal of Computational Physics 366 (2018) 415--447.

\bibitem{Georgalis23}
G.~Georgalis, K.~Retfalvi, P.~E. Desjardin, A.~Patra, Combined data and deep learning model uncertainties: An application to the measurement of solid fuel regression rate, International Journal for Uncertainty Quantification 13~(5) (1 2023).
\newblock \href {https://doi.org/10.1615/int.j.uncertaintyquantification.2023046610} {\path{doi:10.1615/int.j.uncertaintyquantification.2023046610}}.

\bibitem{scarabosio2019goal}
L.~Scarabosio, B.~Wohlmuth, J.~T. Oden, D.~Faghihi, Goal-oriented adaptive modeling of random heterogeneous media and model-based multilevel monte carlo methods, Computers \& Mathematics with Applications 78~(8) (2019) 2700--2718.

\bibitem{li2023surrogate}
Y.~Li, Y.~Wang, L.~Yan, Surrogate modeling for bayesian inverse problems based on physics-informed neural networks, Journal of Computational Physics 475 (2023) 111841.

\bibitem{cao2023bayesian}
L.~Cao, K.~Wu, J.~T. Oden, P.~Chen, O.~Ghattas, Bayesian model calibration for diblock copolymer thin film self-assembly using power spectrum of microscopy data and machine learning surrogate, Computer Methods in Applied Mechanics and Engineering 417 (2023) 116349.

\bibitem{luo2023efficient}
D.~Luo, T.~O'Leary-Roseberry, P.~Chen, O.~Ghattas, Efficient pde-constrained optimization under high-dimensional uncertainty using derivative-informed neural operators, arXiv preprint arXiv:2305.20053 (2023).

\bibitem{tan2024scalable}
J.~Tan, D.~Faghihi, A scalable framework for multi-objective pde-constrained design of building insulation under uncertainty, Computer Methods in Applied Mechanics and Engineering 419 (2024) 116628.

\bibitem{chattopadhyay2023oceannet}
A.~Chattopadhyay, M.~Gray, T.~Wu, A.~B. Lowe, R.~He, Oceannet: A principled neural operator-based digital twin for regional oceans, arXiv preprint arXiv:2310.00813 (2023).

\bibitem{he2023hybrid}
Q.~He, M.~Perego, A.~A. Howard, G.~E. Karniadakis, P.~Stinis, A hybrid deep neural operator/finite element method for ice-sheet modeling, arXiv preprint arXiv:2301.11402 (2023).

\bibitem{kapteyn2021probabilistic}
M.~G. Kapteyn, J.~V. Pretorius, K.~E. Willcox, A probabilistic graphical model foundation for enabling predictive digital twins at scale, Nature Computational Science 1~(5) (2021) 337--347.

\bibitem{mowlavi2023optimal}
S.~Mowlavi, S.~Nabi, Optimal control of pdes using physics-informed neural networks, Journal of Computational Physics 473 (2023) 111731.

\bibitem{zohdi2022digital}
T.~Zohdi, A digital-twin and machine-learning framework for precise heat and energy management of data-centers, Computational Mechanics 69~(6) (2022) 1501--1516.

\bibitem{wu2023large}
K.~Wu, T.~O’Leary-Roseberry, P.~Chen, O.~Ghattas, Large-scale {B}ayesian optimal experimental design with derivative-informed projected neural network, Journal of Scientific Computing 95~(1) (2023) 30.

\bibitem{stuckner2021optimal}
J.~Stuckner, M.~Piekenbrock, S.~M. Arnold, T.~M. Ricks, Optimal experimental design with fast neural network surrogate models, Computational Materials Science 200 (2021) 110747.

\bibitem{arzani2023interpreting}
A.~Arzani, L.~Yuan, P.~Newell, B.~Wang, Interpreting and generalizing deep learning in physics-based problems with functional linear models, arXiv preprint arXiv:2307.04569 (2023).

\bibitem{yuan2022towards}
L.~Yuan, H.~S. Park, E.~Lejeune, Towards out of distribution generalization for problems in mechanics, Computer Methods in Applied Mechanics and Engineering 400 (2022) 115569.

\bibitem{samek2021explaining}
W.~Samek, G.~Montavon, S.~Lapuschkin, C.~J. Anders, K.-R. M{\"u}ller, Explaining deep neural networks and beyond: A review of methods and applications, Proceedings of the IEEE 109~(3) (2021) 247--278.

\bibitem{zhong2022explainable}
X.~Zhong, B.~Gallagher, S.~Liu, B.~Kailkhura, A.~Hiszpanski, T.~Y.-J. Han, Explainable machine learning in materials science, npj Computational Materials 8~(1) (2022) 204.

\bibitem{OdenMoserGhattas2010I}
T.~Oden, R.~Moser, O.~Ghattas, {Computer Predictions with Quantified Uncertainty, Part I}, SIAM News 43~(9) (2010).

\bibitem{Babuska2008}
I.~Babuška, F.~Nobile, R.~Tempone, \href{https://www.sciencedirect.com/science/article/pii/S0045782507005142}{A systematic approach to model validation based on {B}ayesian updates and prediction related rejection criteria}, Computer Methods in Applied Mechanics and Engineering 197~(29) (2008) 2517--2539, validation Challenge Workshop.
\newblock \href {https://doi.org/https://doi.org/10.1016/j.cma.2007.08.031} {\path{doi:https://doi.org/10.1016/j.cma.2007.08.031}}.
\newline\urlprefix\url{https://www.sciencedirect.com/science/article/pii/S0045782507005142}

\bibitem{odenbabuska2017}
J.~T. Oden, I.~Babu{\v{s}}ka, D.~Faghihi, Predictive computational science: Computer predictions in the presence of uncertainty, Encyclopedia of Computational Mechanics Second Edition (2017) 1--26.

\bibitem{tan2022toward}
J.~Tan, B.~Liang, P.~K. Singh, K.~A. Farrell-Maupin, D.~Faghihi, Toward selecting optimal predictive multiscale models, Computer Methods in Applied Mechanics and Engineering 402 (2022) 115517.

\bibitem{yaseen2023quantification}
M.~Yaseen, X.~Wu, Quantification of deep neural network prediction uncertainties for vvuq of machine learning models, Nuclear Science and Engineering 197~(5) (2023) 947--966.

\bibitem{twomey1997validation}
J.~M. Twomey, A.~E. Smith, et~al., Validation and verification, Artificial neural networks for civil engineers: Fundamentals and applications (1997) 44--64.

\bibitem{mackay1995probable}
D.~J. MacKay, Probable networks and plausible predictions-a review of practical {B}ayesian methods for supervised neural networks, Network: computation in neural systems 6~(3) (1995) 469.

\bibitem{neal2012bayesian}
R.~M. Neal, Bayesian learning for neural networks, Vol. 118, Springer Science \& Business Media, 2012.

\bibitem{elsken2019neural}
T.~Elsken, J.~H. Metzen, F.~Hutter, Neural architecture search: A survey, The Journal of Machine Learning Research 20~(1) (2019) 1997--2017.

\bibitem{liu2021survey}
Y.~Liu, Y.~Sun, B.~Xue, M.~Zhang, G.~G. Yen, K.~C. Tan, A survey on evolutionary neural architecture search, IEEE transactions on neural networks and learning systems (2021).

\bibitem{wang2018hybrid}
B.~Wang, Y.~Sun, B.~Xue, M.~Zhang, A hybrid differential evolution approach to designing deep convolutional neural networks for image classification, in: AI 2018: Advances in Artificial Intelligence: 31st Australasian Joint Conference, Wellington, New Zealand, December 11-14, 2018, Proceedings 31, Springer, 2018, pp. 237--250.

\bibitem{ghosh2022designing}
A.~Ghosh, N.~D. Jana, S.~Mallik, Z.~Zhao, Designing optimal convolutional neural network architecture using differential evolution algorithm, Patterns 3~(9) (2022) 100567.

\bibitem{mendoza2016towards}
H.~Mendoza, A.~Klein, M.~Feurer, J.~T. Springenberg, F.~Hutter, Towards automatically-tuned neural networks, in: Workshop on automatic machine learning, PMLR, 2016, pp. 58--65.

\bibitem{hutter2019automated}
F.~Hutter, L.~Kotthoff, J.~Vanschoren (Eds.), Automated Machine Learning - Methods, Systems, Challenges, Springer, 2019.

\bibitem{optuna_2019}
T.~Akiba, S.~Sano, T.~Yanase, T.~Ohta, M.~Koyama, Optuna: A next-generation hyperparameter optimization framework, in: Proceedings of the 25th {ACM} {SIGKDD} International Conference on Knowledge Discovery and Data Mining, 2019.

\bibitem{faghihi2018fatigue}
D.~Faghihi, S.~Sarkar, M.~Naderi, J.~E. Rankin, L.~Hackel, N.~Iyyer, A probabilistic design method for fatigue life of metallic component, ASCE-ASME Journal of Risk and Uncertainty in Engineering Systems, Part B: Mechanical Engineering 4~(3) (2018).

\bibitem{tan2021}
J.~Tan, U.~Villa, N.~Shamsaei, S.~Shao, H.~M. Zbib, D.~Faghihi, A predictive discrete-continuum multiscale model of plasticity with quantified uncertainty, International Journal of Plasticity 138 (2021) 102935.

\bibitem{tan2022predictive}
J.~Tan, P.~Maleki, L.~An, M.~Di~Luigi, U.~Villa, C.~Zhou, S.~Ren, D.~Faghihi, A predictive multiphase model of silica aerogels for building envelope insulations, Computational Mechanics 69~(6) (2022) 1457--1479.

\bibitem{oden2013selection}
J.~T. Oden, E.~E. Prudencio, A.~Hawkins-Daarud, Selection and assessment of phenomenological models of tumor growth, Mathematical Models and Methods in Applied Sciences 23~(07) (2013) 1309--1338.

\bibitem{jha2020bayesian}
P.~K. Jha, L.~Cao, J.~T. Oden, Bayesian-based predictions of covid-19 evolution in texas using multispecies mixture-theoretic continuum models, Computational Mechanics 66~(5) (2020) 1055--1068.

\bibitem{liang2023bayesian}
B.~Liang, J.~Tan, L.~Lozenski, D.~A. Hormuth~II, T.~E. Yankeelov, U.~Villa, D.~Faghihi, Bayesian inference of tissue heterogeneity for individualized prediction of glioma growth, IEEE Transactions on Medical Imaging (2023).

\bibitem{lima2021bayesian}
E.~A. Lima, D.~Faghihi, R.~Philley, J.~Yang, J.~Virostko, C.~M. Phillips, T.~E. Yankeelov, Bayesian calibration of a stochastic, multiscale agent-based model for predicting in vitro tumor growth, PLoS Computational Biology 17~(11) (2021) e1008845.

\bibitem{prudencio2015}
E.~Prudencio, P.~Bauman, D.~Faghihi, K.~Ravi-Chandar, J.~Oden, A computational framework for dynamic data-driven material damage control, based on {B}ayesian inference and model selection, International Journal for Numerical Methods in Engineering 102~(3-4) (2015) 379--403.

\bibitem{farrell2015jcp}
K.~Farrell, J.~T. Oden, D.~Faghihi, A {B}ayesian framework for adaptive selection, calibration, and validation of coarse-grained models of atomistic systems, Journal of Computational Physics 295 (2015) 189--208.

\bibitem{oden2016reviewtumor}
J.~T. Oden, E.~A. Lima, R.~C. Almeida, Y.~Feng, M.~N. Rylander, D.~Fuentes, D.~Faghihi, M.~M. Rahman, M.~DeWitt, M.~Gadde, et~al., Toward predictive multiscale modeling of vascular tumor growth, Archives of Computational Methods in Engineering 23~(4) (2016) 735--779.

\bibitem{mackay1992bayesian}
D.~J. MacKay, Bayesian interpolation, Neural computation 4~(3) (1992) 415--447.

\bibitem{bishop1995neural}
C.~M. Bishop, Neural networks for pattern recognition, Oxford university press, 1995.

\bibitem{psaros2023uncertainty}
A.~F. Psaros, X.~Meng, Z.~Zou, L.~Guo, G.~E. Karniadakis, Uncertainty quantification in scientific machine learning: Methods, metrics, and comparisons, Journal of Computational Physics (2023) 111902.

\bibitem{YANG2021109913}
L.~Yang, X.~Meng, G.~E. Karniadakis, \href{https://www.sciencedirect.com/science/article/pii/S0021999120306872}{B-pinns: Bayesian physics-informed neural networks for forward and inverse pde problems with noisy data}, Journal of Computational Physics 425 (2021) 109913.
\newblock \href {https://doi.org/https://doi.org/10.1016/j.jcp.2020.109913} {\path{doi:https://doi.org/10.1016/j.jcp.2020.109913}}.
\newline\urlprefix\url{https://www.sciencedirect.com/science/article/pii/S0021999120306872}

\bibitem{olivier2021bayesian}
A.~Olivier, M.~D. Shields, L.~Graham-Brady, Bayesian neural networks for uncertainty quantification in data-driven materials modeling, Computer methods in applied mechanics and engineering 386 (2021) 114079.

\bibitem{LINKA2022115346}
K.~Linka, A.~Schäfer, X.~Meng, Z.~Zou, G.~E. Karniadakis, E.~Kuhl, \href{https://www.sciencedirect.com/science/article/pii/S0045782522004327}{Bayesian physics informed neural networks for real-world nonlinear dynamical systems}, Computer Methods in Applied Mechanics and Engineering 402 (2022) 115346, a Special Issue in Honor of the Lifetime Achievements of J. Tinsley Oden.
\newblock \href {https://doi.org/https://doi.org/10.1016/j.cma.2022.115346} {\path{doi:https://doi.org/10.1016/j.cma.2022.115346}}.
\newline\urlprefix\url{https://www.sciencedirect.com/science/article/pii/S0045782522004327}

\bibitem{MORA2023116207}
C.~Mora, J.~T. Eweis-Labolle, T.~Johnson, L.~Gadde, R.~Bostanabad, \href{https://www.sciencedirect.com/science/article/pii/S0045782523003316}{Probabilistic neural data fusion for learning from an arbitrary number of multi-fidelity data sets}, Computer Methods in Applied Mechanics and Engineering 415 (2023) 116207.
\newblock \href {https://doi.org/https://doi.org/10.1016/j.cma.2023.116207} {\path{doi:https://doi.org/10.1016/j.cma.2023.116207}}.
\newline\urlprefix\url{https://www.sciencedirect.com/science/article/pii/S0045782523003316}

\bibitem{kontolati2023influence}
K.~Kontolati, S.~Goswami, M.~D. Shields, G.~E. Karniadakis, On the influence of over-parameterization in manifold based surrogates and deep neural operators, Journal of Computational Physics 479 (2023) 112008.

\bibitem{meng2021multi}
X.~Meng, H.~Babaee, G.~E. Karniadakis, Multi-fidelity {B}ayesian neural networks: Algorithms and applications, Journal of Computational Physics 438 (2021) 110361.

\bibitem{buntine1991bayesian}
W.~Buntine, A.~Weigend, Bayesian back-propagation. technical report fia-91-22 (1991).

\bibitem{ritter2018scalable}
H.~Ritter, A.~Botev, D.~Barber, A scalable laplace approximation for neural networks, in: 6th International Conference on Learning Representations, ICLR 2018-Conference Track Proceedings, Vol.~6, International Conference on Representation Learning, 2018.

\bibitem{deng2022accelerated}
Z.~Deng, F.~Zhou, J.~Zhu, Accelerated linearized laplace approximation for bayesian deep learning, Advances in Neural Information Processing Systems 35 (2022) 2695--2708.

\bibitem{immer2021improving}
A.~Immer, M.~Korzepa, M.~Bauer, Improving predictions of bayesian neural nets via local linearization, in: International conference on artificial intelligence and statistics, PMLR, 2021, pp. 703--711.

\bibitem{immer2021scalable}
A.~Immer, M.~Bauer, V.~Fortuin, G.~R{\"a}tsch, K.~M. Emtiyaz, Scalable marginal likelihood estimation for model selection in deep learning, in: International Conference on Machine Learning, PMLR, 2021, pp. 4563--4573.

\bibitem{humt2019laplace}
M.~Humt, Laplace approximation for uncertainty estimation of deep neural networks, Ph.D. thesis, TUM (2019).

\bibitem{martens2015optimizing}
J.~Martens, R.~Grosse, Optimizing neural networks with kronecker-factored approximate curvature, in: International conference on machine learning, PMLR, 2015, pp. 2408--2417.

\bibitem{botev2017practical}
A.~Botev, H.~Ritter, D.~Barber, Practical gauss-newton optimisation for deep learning, in: International Conference on Machine Learning, PMLR, 2017, pp. 557--565.

\bibitem{qian2023biomimetic}
K.~Qian, A.~S. Liao, S.~Gu, V.~A. Webster-Wood, Y.~J. Zhang, Biomimetic iga neuron growth modeling with neurite morphometric features and cnn-based prediction, Computer Methods in Applied Mechanics and Engineering 417 (2023) 116213.

\bibitem{pouchard2023rigorous}
L.~Pouchard, K.~G. Reyes, F.~J. Alexander, B.-J. Yoon, A rigorous uncertainty-aware quantification framework is essential for reproducible and replicable machine learning workflows, Digital Discovery 2~(5) (2023) 1251--1258.

\bibitem{krishnanunni2022layerwise}
C.~Krishnanunni, T.~Bui-Thanh, Layerwise sparsifying training and sequential learning strategy for neural architecture adaptation, arXiv preprint arXiv:2211.06860 (2022).

\bibitem{muto2008}
M.~Muto, J.~L. Beck, Bayesian updating and model class selection for hysteretic structural models using stochastic simulation, Journal of Vibration and Control 14~(1-2) (2008) 7--34.

\bibitem{vila2000bayesian}
J.-P. Vila, V.~Wagner, P.~Neveu, Bayesian nonlinear model selection and neural networks: A conjugate prior approach, IEEE Transactions on neural networks 11~(2) (2000) 265--278.

\bibitem{chitsazan2015prediction}
N.~Chitsazan, A.~A. Nadiri, F.~T.-C. Tsai, Prediction and structural uncertainty analyses of artificial neural networks using hierarchical bayesian model averaging, Journal of Hydrology 528 (2015) 52--62.

\bibitem{shekhar2020hierarchical}
P.~Shekhar, A.~Patra, Hierarchical approximations for data reduction and learning at multiple scales, Foundations of Data Science 2~(2) (2020) 123--154.

\bibitem{shekhar2022forward}
P.~Shekhar, A.~Patra, A forward--backward greedy approach for sparse multiscale learning, Computer Methods in Applied Mechanics and Engineering 400 (2022) 115420.

\bibitem{huan2013simulation}
X.~Huan, Y.~M. Marzouk, Simulation-based optimal {B}ayesian experimental design for nonlinear systems, Journal of Computational Physics 232~(1) (2013) 288--317.

\bibitem{riis2022bayesian}
C.~Riis, F.~Antunes, F.~H{\"u}ttel, C.~Lima~Azevedo, F.~Pereira, Bayesian active learning with fully bayesian gaussian processes, Advances in Neural Information Processing Systems 35 (2022) 12141--12153.

\bibitem{dougherty2015}
R.~Dehghannasiri, B.-J. Yoon, E.~R. Dougherty, Efficient experimental design for uncertainty reduction in gene regulatory networks, in: BMC bioinformatics, Vol.~16, BioMed Central, 2015, pp. 1--18.

\bibitem{paquette2023optimal}
A.~Paquette-Rufiange, S.~Prudhomme, M.~Laforest, Optimal design of validation experiments for the prediction of quantities of interest, arXiv preprint arXiv:2303.06114 (2023).

\bibitem{williams1995bayesian}
P.~M. Williams, Bayesian regularization and pruning using a laplace prior, Neural computation 7~(1) (1995) 117--143.

\bibitem{rudy2021sparse}
S.~H. Rudy, T.~P. Sapsis, Sparse methods for automatic relevance determination, Physica D: Nonlinear Phenomena 418 (2021) 132843.

\bibitem{jaynes2003}
E.~T. Jaynes, Probability theory: The logic of science, Cambridge University press, 2003.

\bibitem{daxberger2021laplace}
E.~Daxberger, A.~Kristiadi, A.~Immer, R.~Eschenhagen, M.~Bauer, P.~Hennig, Laplace redux-effortless {B}ayesian deep learning, Advances in Neural Information Processing Systems 34 (2021) 20089--20103.

\bibitem{fenics}
M.~S. Aln{\ae}s, J.~Blechta, J.~Hake, A.~Johansson, B.~Kehlet, A.~Logg, C.~Richardson, J.~Ring, M.~E. Rognes, G.~N. Wells, The {FEniCS} project version 1.5, Archive of Numerical Software 3~(100) (2015).
\newblock \href {https://doi.org/10.11588/ans.2015.100.20553} {\path{doi:10.11588/ans.2015.100.20553}}.

\bibitem{ABLATE}
M.~McGurn, D.~Salac, {ABLATE} ablative boundary layers at the exascale, version 0.12.23, \url{https://ablate.dev} (2013).

\bibitem{sarkar2024carbon}
A.~Sarkar, P.~K. Singh, L.~Zhu, D.~Faghihi, S.~Ren, Carbon-sequestration straw cellulose-aerogel gradient thermal insulation material, ACS Applied Engineering Materials (2024).

\bibitem{an2023flexible}
L.~An, M.~Di~Luigi, J.~Tan, D.~Faghihi, S.~Ren, Flexible percolation fibrous thermal insulating composite membranes for thermal management, Materials Advances 4~(1) (2023) 284--290.

\bibitem{bhattacharjee2023integrating}
S.~Bhattacharjee, Integrating scientific machine learning and physics-based models for quantification of uncertainty in thermal properties of silica aerogel, Ph.D. thesis, State University of New York at Buffalo (2023).

\bibitem{maupin2018validation}
K.~A. Maupin, L.~P. Swiler, N.~W. Porter, Validation metrics for deterministic and probabilistic data, Journal of Verification, Validation and Uncertainty Quantification 3~(3) (2018) 031002.

\bibitem{georgalis2023combined}
G.~Georgalis, K.~Retfalvi, P.~E. DesJardin, A.~Patra, Combined data and deep learning model uncertainties: An application to the measurement of solid fuel regression rate, International Journal for Uncertainty Quantification 13~(5) (2023).

\bibitem{surina2022measurement}
G.~Surina~III, G.~Georgalis, S.~S. Aphale, A.~Patra, P.~E. DesJardin, Measurement of hybrid rocket solid fuel regression rate for a slab burner using deep learning, Acta Astronautica 190 (2022) 160--175.

\bibitem{raissi2019physics}
M.~Raissi, P.~Perdikaris, G.~E. Karniadakis, Physics-informed neural networks: A deep learning framework for solving forward and inverse problems involving nonlinear partial differential equations, Journal of Computational physics 378 (2019) 686--707.

\bibitem{jha2024}
P.~K. Jha, J.~T. Oden, \href{https://www.sciencedirect.com/science/article/pii/S0045782523007193}{Residual-based error corrector operator to enhance accuracy and reliability of neural operator surrogates of nonlinear variational boundary-value problems}, Computer Methods in Applied Mechanics and Engineering 419 (2024) 116595.
\newblock \href {https://doi.org/https://doi.org/10.1016/j.cma.2023.116595} {\path{doi:https://doi.org/10.1016/j.cma.2023.116595}}.
\newline\urlprefix\url{https://www.sciencedirect.com/science/article/pii/S0045782523007193}

\bibitem{li2023isogeometric}
A.~Li, Y.~J. Zhang, Isogeometric analysis-based physics-informed graph neural network for studying traffic jam in neurons, Computer Methods in Applied Mechanics and Engineering 403 (2023) 115757.

\bibitem{box1979}
G.~E. Box, Robustness in the strategy of scientific model building, Robustness in statistics 1 (1979) 201--236.

\bibitem{box2011bayesian}
G.~E. Box, G.~C. Tiao, Bayesian inference in statistical analysis, John Wiley \& Sons, 2011.

\bibitem{zhou2021effective}
X.~Zhou, W.~Zhang, H.~Xu, T.~Zhang, Effective sparsification of neural networks with global sparsity constraint, in: Proceedings of the IEEE/CVF Conference on Computer Vision and Pattern Recognition, 2021, pp. 3599--3608.

\bibitem{oLeary2022learning}
T.~O’Leary-Roseberry, X.~Du, A.~Chaudhuri, J.~R. Martins, K.~Willcox, O.~Ghattas, Learning high-dimensional parametric maps via reduced basis adaptive residual networks, Computer Methods in Applied Mechanics and Engineering 402 (2022) 115730.

\bibitem{oLeary2022derivative}
T.~O’Leary-Roseberry, U.~Villa, P.~Chen, O.~Ghattas, Derivative-informed projected neural networks for high-dimensional parametric maps governed by pdes, Computer Methods in Applied Mechanics and Engineering 388 (2022) 114199.

\bibitem{paris2021learning}
S.~Wang, H.~Wang, P.~Perdikaris, Learning the solution operator of parametric partial differential equations with physics-informed deeponets, Science advances 7~(40) (2021) eabi8605.

\bibitem{Karniadakis2019deeponet}
L.~Lu, P.~Jin, G.~E. Karniadakis, Deeponet: Learning nonlinear operators for identifying differential equations based on the universal approximation theorem of operators, arXiv preprint arXiv:1910.03193 (2019).

\bibitem{li2020fourier}
Z.~Li, N.~Kovachki, K.~Azizzadenesheli, B.~Liu, K.~Bhattacharya, A.~Stuart, A.~Anandkumar, Fourier neural operator for parametric partial differential equations, arXiv preprint arXiv:2010.08895 (2020).

\bibitem{Karniadakis2023physics}
S.~Goswami, A.~Bora, Y.~Yu, G.~E. Karniadakis, Physics-informed deep neural operator networks, in: Machine Learning in Modeling and Simulation: Methods and Applications, Springer, 2023, pp. 219--254.

\end{thebibliography}

\end{document}